\documentclass[onecolumn,prd,noshowpacs,nofootinbib,amsmath,amssymb,superscriptaddress]{revtex4}
\usepackage{graphicx}
\usepackage{amssymb}
\usepackage{amsmath}
\usepackage{amsfonts}
\usepackage{latexsym}
\usepackage{hyperref}
\usepackage[all]{xy}
\usepackage{slashed}
\usepackage{physics}
\usepackage{accents}
\usepackage{tensor}
\usepackage{afterpage}
\usepackage{booktabs,siunitx,dcolumn}
\usepackage{color}
\usepackage{cases}
\usepackage[dvipsnames]{xcolor}
\usepackage[mathscr]{euscript}
\usepackage{mathtools}
\usepackage{etoolbox}

\newtoggle{isdraft}
\togglefalse{isdraft}

\newcommand{\be}{\begin{equation}}
\newcommand{\ee}{\end{equation}}
\newcommand{\ba}{\begin{eqnarray}}
\newcommand{\ea}{\end{eqnarray}}
\newcommand{\nn}{\nonumber}
\def\d{\delta}
\def\del{\partial}
\def\x{{\bf x}}
\def\y{{\bf y}}

\def\k{{\bf k}}
\def\p{{\bf p}}
\def\q{{\bf q}}
\def\[{\left[}
\def\]{\right]}
\def\({\left(}
\def\){\right)}
\def\<{\langle}
\def\>{\rangle}
\def\q{\mathbf{q}}

\def\x{\mathbf{x}}

\def\O{\mathcal{O}}
\def\SS{\mathcal{S}}
\def\L{\mathcal{L}}
\def\H{\mathcal{H}}

\def\D{\mathcal{D}}
\def\F{\mathcal{F}}

\def\vph{\varphi}
\def\tg{\tilde{g}}

\def\eps{\epsilon}
\def\Mp{M_{\rm p}}
\def\kmin{k_{\rm min}}
\def\kmax{k_{\rm max}}
\def\xmin{x_{\rm min}}
\def\xmax{x_{\rm max}}
\def\rarr{\rightarrow}

\def\sqz{r}

\def\Sys{\mathscr{S}}
\def\Env{\mathscr{E}}
\def\Glob{\mathscr{H}}
\def\Frag{\mathscr{F}}
\def\Apt{\mathscr{A}}
\def\jstrut{\vphantom{\Big[}}

\def\nab{\nabla}

\def\App{Appendix}

\begin{document}

\title{Classical branches and entanglement structure in the wavefunction of cosmological fluctuations}

\author{Elliot Nelson}
\affiliation{Perimeter Institute for Theoretical Physics, Waterloo, Ontario, Canada}
\author{C.\ Jess Riedel}
\affiliation{Perimeter Institute for Theoretical Physics, Waterloo, Ontario, Canada}

\begin{abstract}
The emergence of preferred classical variables within a many-body wavefunction is encoded in its entanglement structure in the form of redundant classical information shared between many spatially local subsystems.    We show how such structure can be generated via cosmological dynamics from the vacuum state of a massless field, causing the wavefunction to branch into classical field configurations on large scales.  An accelerating epoch first excites the vacuum into a superposition of classical fields as well as a highly sensitive bath of squeezed super-horizon modes.  During a subsequent decelerating epoch, gravitational interactions allow these modes to collect information about longer-wavelength fluctuations. This information disperses spatially, creating long-range redundant correlations in the wavefunction. The resulting classical observables, preferred basis for decoherence, and system/environment decomposition emerge from the translation invariant wavefunction and Hamiltonian, rather than being fixed by hand. We discuss the role of squeezing, the cosmological horizon scale, and phase information in the wavefunction, as well as aspects of wavefunction branching 
for continuous variables and in field theories.
\end{abstract}

\maketitle

\tableofcontents

\section{Introduction}

Assuming a homogeneous initial state of the universe, the inhomogeneous primordial density perturbations we observe imprinted on the cosmic microwave background correspond to one member of a much larger ensemble of possibilities.  Further assuming a pure initial state $\ket{\Psi}$, such as the Bunch-Davies vacuum invoked in inflationary models, this ensemble can be identified with the states $\{\ket{\Psi_i} = P_i \ket{\Psi}\}$, where each orthogonal projector $P_i$ selects a joint configuration for a preferred set of commuting observables accessible to our instruments.  Unlike laboratory experiments on pristine microscopic systems, measurements of arbitrary cosmological observables cannot be made.  Rather, there is a quantum-classical transition at some point during the evolution of the early universe \cite{Kiefer:2008ku} which picks out a preferred set of observables, encompassing all measurements that might be made by any feasible experiment. In turn, this induces a preferred set of wavefunction \emph{branches} $\ket{\Psi_i}$.   This paper is motivated by a desire to identify these branches from first principles and, in particular, to eliminate dependence on certain assumptions that must be made with earlier treatments.  We build up only from spatial entanglement in the quantum state, a technique that can be extended outside the realm of cosmology to understand amplification and emergent classical observables in arbitrary many-body systems.

Given a particular subsystem distinguished from a larger environment, the theory of decoherence \cite{joos2003decoherence,zurek2003decoherence,schlosshauer2008decoherence} can provide a general method for identifying a preferred basis of the subsystem, and hence a preferred complete set of commuting observables, that is stable over times much longer than the subsystem-environment interaction timescale.  Phase information between states in this basis is irreversibly dispersed into the environment and hence cannot be accessed through measurements on the subsystem alone.  Decoherence is said to reduce the problem of finding a preferred \emph{basis} to the problem of finding a preferred \emph{subsystem}.  This is indeed a significant reduction, as the number of bases is much larger than the number of subsystems, and the subsystems that are feasibly accessible to experiment are much more salient than the bases.  Although this is sufficient for most practical purposes, the formal problem of identifying the preferred subsystems\footnote{Subsystems can equivalently be defined by a preferred set of observables \cite{zanardi2004quantum,viola2007entanglement}.} from first principles remains \cite{zurek1998decoherence,dugi2012parallel,jekni-dugi2014quantum}.

It is \emph{not} the case that all subsystems decohere in some basis. To a limited extent, the theory of decoherence has been inverted to find variables that are protected from interactions with the environment (e.g., decoherence-free subspaces \cite{palma1996quantum,lidar1998decoherence-free,lidar2003decoherence-free}) which has found uses in applications like quantum error-correcting codes.  Error correction, both classical and quantum, is ultimately grounded in the principle of locality; by distributing information across multiple, spatially disjoint degrees of freedom, local errors can detected and corrected before they corrupt the extended information.  Recently, it has been proposed that spatial locality and the associated entanglement structure may also single out a set of robust classical observables in non-equilibrium many-body states -- \emph{without} any reference to a preferred subsystem -- by the presence of ``redundant records'' \cite{riedel2017classical} (defined precisely in \S \ref{sec:prelim}).  Such records are commonly produced in decohering subsystems through a process known as quantum Darwinism \cite{zurek2000einselection,ollivier2005environment,ollivier2004objective,zurek2009quantum} in which information about the subsystem's preferred observables is copied into the environment, not just once, but redundantly into many disjoint locations.  This generates a characteristic type of classical correlations, interpretable as a \emph{classical} error correcting code, with so-called broadcast structure \cite{korbicz2014objectivity,horodecki2015quantum,brandao2015generic} as a special case.  The phenomenon has been explored in many concrete models \cite{blume-kohout2005simple,blume-kohout2006quantum,blume-kohout2008quantum,paz2009redundancy,zwolak2010redundant,riedel2010quantum,riedel2011redundant,riedel2012rise,korbicz2014objectivity,horodecki2015quantum} and has been proposed as a method for selecting a single preferred set of consistent histories \cite{riedel2016objective,riedel2017classical}. The set of all observables that are redundantly recorded must mutually commute -- except for a set of error-correcting counterexamples that become rarer and more pathological as the redundancy increases \cite{riedel2017classical} -- and so the observables induce a preferred decomposition of the state into branches, i.e., the projections of the state $\ket{\Psi}$ on the joint eigenspaces of the preferred observables.

In this paper we apply these ideas for the first time to the cosmological setting.  Decoherence has been extensively studied in this context \cite{kiefer2000entropy,lombardo2005decoherence,Kiefer:2006je,sharman2007decoherence,martineau2007decoherence,nambu2008entanglement,burgess2008decoherence,lim2015quantum,boddy2016how,nelson2016quantum}, but only by assuming a preferred separation between subsystem and environment, e.g., by showing that long-wavelength primordial perturbations are decohered by short-wavelength modes and other fields; the existence of alternate subsystems decohering and yielding incompatible branch structure was formally an open question.\footnote{Within the context of consistent histories, the absence of a principle for picking out preferred observables or branches manifests as the existence of incompatible sets of consistent histories (also known as \emph{alternate realms}).  This is sometimes called the \emph{set-selection problem} and solving it has been variously characterized as crucial \cite{paz1993environment-induced,goldstein1995linearly,dowker1996consistent,kent1996remarks,kent1997consistent,goldstein1998quantum,anastopoulos1998preferred,kent2000quantum,okon2015consistent}, merely preferable \cite{hartle1989quantum,gell-mann1990quantum,gell-mann1994equivalent}, and completely unnecessary \cite{griffiths1998choice,griffiths1998comment,griffiths2000consistent,wallace2010decoherence,griffiths2013consistent,griffiths2014new,griffiths2015consistent}.  Several incomplete solutions to the problem have been proposed \cite{kent1997quantum,anastopoulos1998preferred,wallden2014contrary,riedel2016objective,riedel2017classical}.}  Here, we establish, assuming only the special status of spacial locality, that long-wavelength perturbations are unconditionally preferred, essentially ruling out the possibility that an alternate choice of subsystem would reveal the decoherence of an incompatible (i.e., non-commuting) set of observables.  We do this by showing not just that these preferred modes decohere, but additionally that their configuration is recorded redundantly in many spatially disjoint regions.\footnote{Given the freedom to consider an arbitrary decomposition of the global Hilbert space into subsystems, there will exist incompatible observables that decohere and produce ``records'' with respect to that decomposition, at least instantaneously. But all of these pseudo-records, and the subsystems that contain them, will \emph{necessarily} overlap in space and hence will not be separately accessible to \emph{any} collection of spatially disjoint observers.  For an example with the decomposition associated with momentum space, see \App~\ref{sec:momentum-records}.}

We seek to understand the formation of branch structure when it initially develops, in the broadest possible cosmological setting.  Our model consists of a generic massless scalar field variable $\vph$ initialized in the vacuum state, which could describe scalar curvature fluctuations, tensor fluctuations, an inflaton field, or any other massless mode.
Due to its self-interactions, the field acts as both the system and decohering environment, with modes on a given scale being redundantly recorded by modes on shorter scales. The resulting wavefunction branches are peaked around long-wavelength classical field configurations. 
The dynamics will be those induced by a period of cosmological acceleration, which stretches the modes of the field to superhorizon scales, followed by a period of deceleration, in which the modes re-enter the cosmological horizon and regions of space come back into causal contact (and share information).
While the accelerating epoch occurs in an approximately de Sitter background, we emphasize that the crucial effect of this epoch -- the WKB classicality and squeezing of the modes in the superhorizon regime -- is not unique to inflation, and also occurs in bouncing cosmologies (see e.g., \cite{Battarra:2013cha}).
The primary reason for an inflationary epoch is that the field is effectively in the (Minkowski space) ground state at very early times, so we are able to generate classical branches from the most simple and symmetric initial state.

We build on significant earlier work studying decoherence and the quantum-to-classical transition in cosmological models.
As discussed in \cite{Guth:1985ya,Grishchuk:1989ss,Grishchuk:1990bj,albrecht1994inflation,Polarski:1995jg,Kiefer:1998qe} and related works, the Bunch-Davies vacuum becomes highly squeezed as modes redshift to super-Hubble scales, with the state in phase space having support only along the line $\pi\approx\pi_c(\vph)$ where the conjugate momentum is that of a field $\vph$ evolving classically. Furthermore, the field freezes at late times, rendering the velocity or conjugate momentum unobservably tiny.  Consequently, essentially all feasible measurements (if they were performed) would yield a classical distribution; observing non-classical results would require infeasible measurements of nonlocal observables \cite{kiefer1998quantum--classical,kiefer1998emergence}. Our goal is identify, as abstractly as possible, the properties of the cosmological wavefunction that produce these results, with the intention of applying them in the future to arbitrary non-cosmological system, ultimately generalizing these sorts of restrictions on the feasibility of real-world measurements.
Classicality in this sense is a necessary condition for the formation of spatially redundant records and wavefunction branching as we define them.

Our model is universal in the sense that, even in the absence of matter fields, gravity alone is capable of generating the necessary interactions \cite{Maldacena:2002vr,Chen:2006nt}.  Of course, matter self-interactions or multiple fields can introduce stronger (but still perturbative) interactions, and hence earlier decoherence, but these are model dependent.  We expect that such modifications would induce branching that is similar in most important qualitative ways to the generic case studied here.
We refer the reader to \cite{lombardo2005decoherence,Sakagami:1987mp,Brandenberger:1990bx,Prokopec:2006fc,Mazur:2008wa} for discussions of decoherence from matter self-interactions and other non-minimal couplings, and note that some earlier works \cite{Martineau:2006ki,Calzetta:1995ys,Franco:2011fg} have also studied gravitational interactions.

Our model of objective wavefunction branching for a translation-invariant scalar field is general enough to be of interest outside cosmology.  A long-term goal for future work is the identification of robust universal criteria that define provably-irreversible wavefunction branches in all systems including, as a special case, laboratory measurement.  Our current reliance on an unambiguous, a-priori concept of spatial locality will, of course, prevent it from being immediately applied in the context of quantum gravity, especially outside the perturbative regime.  
However, holographic approaches to gravity may use the idea of redundant records to identify preferred states in the boundary theory that correspond to quasiclassical spacetime geometries, recovering a notion of locality in the bulk \cite{nomura2017classical}.   
Locality might also be built up from more abstract properties of the Hamiltonian \cite{cotler2017locality}.

The identification of mathematical principles for the identification of wavefunction branch structure is valuable quite generally, but let us briefly list some reasons, both fundamental and practical, for why the inflationary early universe is a particularly good venue: (1) The early universe strains the conventional formulation of quantum mechanics as a strictly operational theory \cite{hartle1989quantum,wallace2016what}.  In particular, assigning a quantum state to the universe \cite{hartle1983wave} is unavoidable and, since there are no observers in the past, the measurable observables are pre-determined (rather than chosen by an experimentalist) and cannot be repeated. (2) Inflation is the most popular cosmological model of this time period. (3) Linear inflationary evolution is highly symmetric and exactly solvable, and perturbation theory from this solution is relatively well understood.  (4) Most study of the quantum-classical transition occurs in the nonrelativistic setting, but a fully satisfactory understanding must extend into the actual arena of fundamental physics: relativistic field theory.  (5) The inflationary state is out of equilibrium, a necessary condition for the time-asymmetric proliferation of branches.

The paper is structured as follows.
In Section \ref{sec:summary} we summarize our results.
In Section \ref{sec:prelim} we review preliminary concepts including decoherence, redundant records, wavefunction branches, branching in field theory, the background cosmology in our model, and the wavefunction description of inflationary fluctuations.
In Section \ref{sec:linear} we review the linear theory for fluctuations in a de Sitter background, including the phase space representation and squeezing of the quantum state.
In Section \ref{sec:inflation} we study the effects of gravitational interactions during inflation, first considering the simplified case of coupling to an infinite-wavelength background, and then evolving the wavefunction and quantifying the information recorded in a given localized mode of the field.
In Section \ref{decelerating-section} we study the behavior of the wavefunction in a decelerating era following inflation, and show that the correlations between modes that were created during inflation are amplified so that localized modes in many spatial regions contain redundant information about the field on superhorizon scales.
We discuss our results in Section \ref{sec:discussion}.

For notation, $\k$ and $\q$ will denote spatial wavenumbers in three dimensions, with $\q$ reserved for long-wavelength modes that are decohering, and $\k$ primarily for shorter-wavelength modes acting as an environment.  The magnitudes of vectors are denoted as $k = |\k|$.
We will use the Fourier decomposition convention $X(\x)\equiv\int\!\frac{\dd^3\k}{(2\pi)^3}X_\k e^{i\k\cdot\x}$ 
for any quantity $X=\zeta,\vph$, etc., and the integration of wavenumbers will often be shortened to $\int\!\frac{\dd^3\k}{(2\pi)^3}X_\k \equiv\int_\k X_\k$.
We will denote conformal time in an inflationary epoch as $\tau$, and conformal time in a post-inflation decelerating epoch as $\eta$.
Overdots will denote derivatives with respect to cosmic time $t$, and primes with respect to conformal time $\tau$ (or $\eta$).
For spatial derivatives of fields, we use the following conventions: $(\nab\vph)^2=\d^{ij}\del_i\vph\del_j\vph$, $\nab^2=\d^{ij}\del_i\del_j$, and $\nab^{-2}\vph\equiv\int\!\frac{\dd^3\k}{(2\pi)^3}(-k^{-2})\vph_\k e^{i\k\cdot\x}$.
A bar over two operators denotes the symmeterized product with subscript indices exchanged: $\overline{X_\k Y_\p} = (X_\k Y_\p + Y_\k X_\p)/2$.
A prime on a correlation function will denote the omission of a momentum-conserving delta function, e.g., $\<\vph_\k\vph_{\k'}\>=(2\pi)^3\d^3(\k+\k')\<\vph_\k\vph_{\k'}\>'$.
The notation $\Psi[\vph](t)$ will denote the wave functional evaluated at a field configuration $\vph(\x)$, and at time $t$.
The Hilbert spaces of systems $\Sys$, environments $\Env$, and the relevant universe $\Glob$ are written in an upright script.
Actions $\SS$, Lagrangians $\L$, and Hamiltonians $\H$ are written in calligraphic script while the Hubble constant $H$ is in Roman.
We distinguish the Lagrangian (or Hamiltonian) density from the Lagrangian (or Hamiltonian) by adding a spatial argument, e.g., $\L=\int d^3\x\L(\x)$.
Most Hamiltonians will be taken with respect to conformal time $\tau$ (or $\eta$), with Schr\"{o}dinger equation $i\partial_{\tau} \Psi = \H \Psi$ (or $i\partial_{\eta} \Psi = \H \Psi$). 
Lastly, note that we will use $\d\phi$ for fluctuations of the inflaton, $\zeta$ for the curvature perturbation on uniform-density hypersurfaces, and $\vph$ for a generic field variable in a more general class of models. (Within that class of generic models, there are particular choices of coupling constants such that $\vph$ can be interpreted as a rescaling of $\zeta$ or of $\d\phi$.)

Throughout the figures in this paper we use the Wigner function in phase space to depict quantum states.\footnote{The Wigner function of a single variable $x$ is a real-valued function of phase space given by the Fourier transform of the off-diagonal direction of the variable's density matrix $\rho$: $W(x,p)\equiv(2\pi)^{-1}\int \!\dd\Delta x\,\exp(-i p \Delta x) \langle x+\Delta x/2 \vert\rho \vert x-\Delta x/2\rangle =(2\pi)^{-1}\int \!\dd\Delta p \exp(-i x \Delta p)\langle p+\Delta p/2 \vert\rho \vert p-\Delta p/2\rangle$.  It satisfies $\int \! \dd x \, W(x,p)= \langle p\vert\rho\vert p\rangle$, $\int \! \dd p \, W(x,p)= \langle x\vert\rho\vert x\rangle$ and hence $\int \! \dd x \dd p\, W(x,p)= 1$. The Wigner function of a Gaussian state is Gaussian (and hence positive) with the same variances in $x$ and $p$.  By virtue of the invertible nature of the Fourier transform, two distinct states $\rho$ and $\rho'$ have distinct Wigner functions.}  Slightly abusing terminology, we refer to the region where the Wigner function is neither zero nor exponentially suppressed as the \emph{support}.  Since the states we consider are Gaussian or nearly Gaussian, their support is well represented by an ellipse whose boundary is a contour containing most of the Wigner probability mass.  Squeezed states thus appear as narrow, tilted ellipses.

\section{Summary of Results}
\label{sec:summary}

In this paper, we show that the wavefunction of a massless field undergoes evolution with the schematic form
\begin{subequations}\label{long_short_records}
\begin{alignat}{3}
\ket{\Psi} & = \(\sum_{\vph_S} \psi_S(\vph_S)|\vph_S\>\)
\(\sum_{\vph_M} \psi_M(\vph_M)|\vph_M\>\)
\(\sum_{\vph_L} \psi_L(\vph_L)|\vph_L\>\) \label{product_state_SML} \\
& \rarr \sum_{\vph_{S,M,L}} \psi_S^{(\vph_M)}(\vph_S)|\vph_S\>
\cdot\psi_M^{(\vph_L)}(\vph_M)|\vph_M\>
\cdot\psi_L(\vph_L)|\vph_L\> \label{entangled_state_SML}
\end{alignat}
\end{subequations}
Here, ``S'', ``M'', and ``L'' denote short-, medium-, and long wavelength modes; $|\vph\>$ indicates approximate eigenstates of the field operator $\hat{\vph}$, sharply peaked around a classical field configuration; $\psi(\vph)$ denotes a wavefunction over configurations $\vph$; and $\psi^{(\tilde\vph)}(\vph)$ denotes a wavefunction over the configuration $\vph$ of a shorter wavelength mode  \emph{conditional} on the configuration $\tilde\vph$ of a longer wavelength mode. For sufficiently different values $\tilde\vph$ and $\tilde{\vph}'$ of the longer wavelength configuration, the conditional states are orthogonal: $\int \! d \vph\, \psi^{(\tilde{\vph}')}(\vph)^\dagger \psi^{(\tilde\vph)}(\vph) \approx 0$.
In the initial (vacuum) state, Eq. \eqref{product_state_SML}, modes on different scales are unentangled.
After evolving under the cosmological dynamics, the field is in a highly entangled state, Eq. \eqref{entangled_state_SML}, with the property that the conditional states of short (medium) wavelength modes contain redundant information about the field on medium (long) wavelengths.
We include medium modes as well as short and long modes in the schematic division into momentum bands in order to emphasize that the modes of the field act \textit{both} as an environment recording information (reflected in the conditional wavefunction $\psi_M^{(\vph_L)}(\vph_M)$) \textit{and} as a system being decohered (reflected in the entanglement of field eigenstates $|\vph_M\>$ with the conditional wavefunctions $\psi_S^{(\vph_M)}(\vph_S)$ of short modes).

\begin{figure}
\includegraphics[scale=0.45]{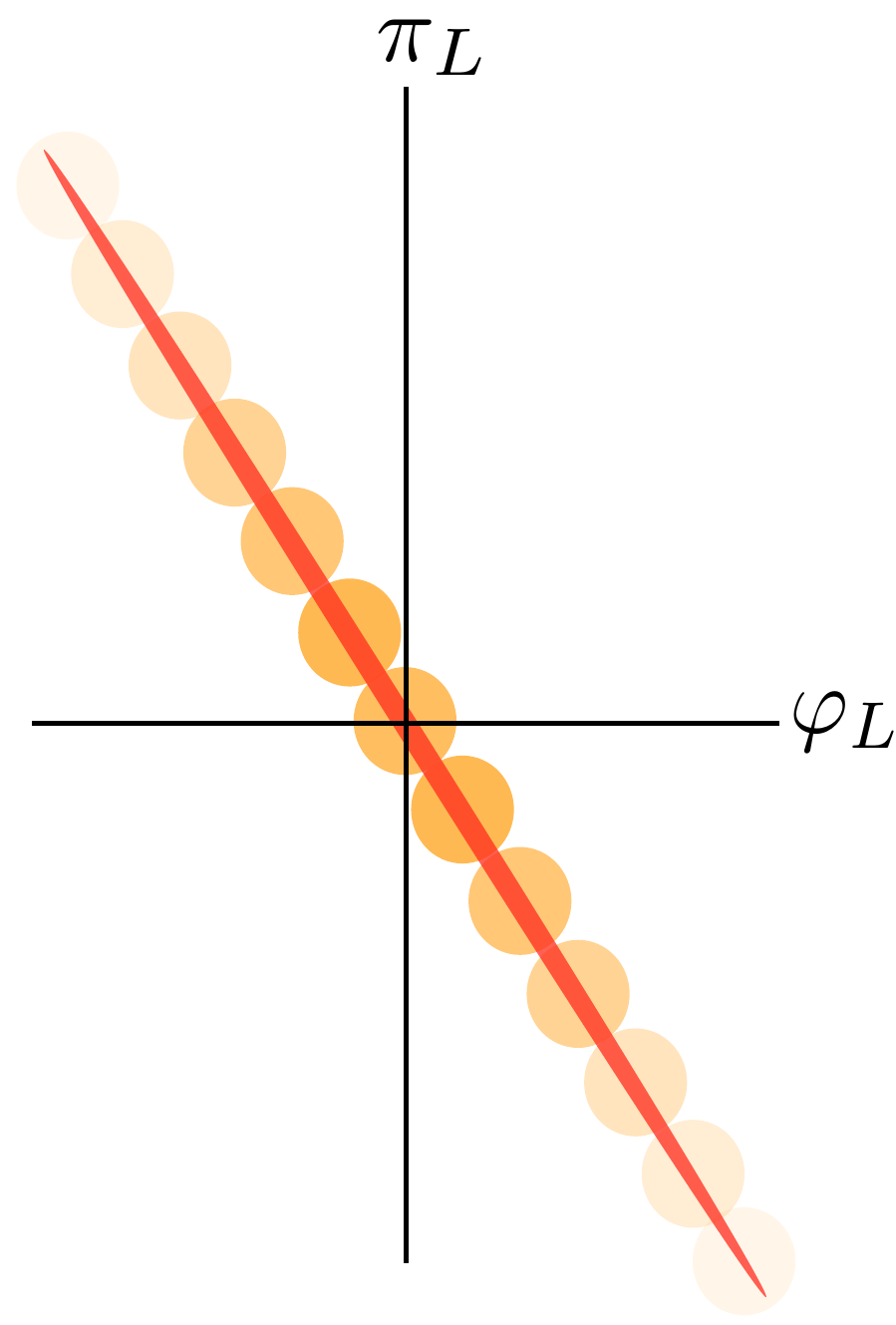} \hspace{4cm}\includegraphics[scale=0.45]{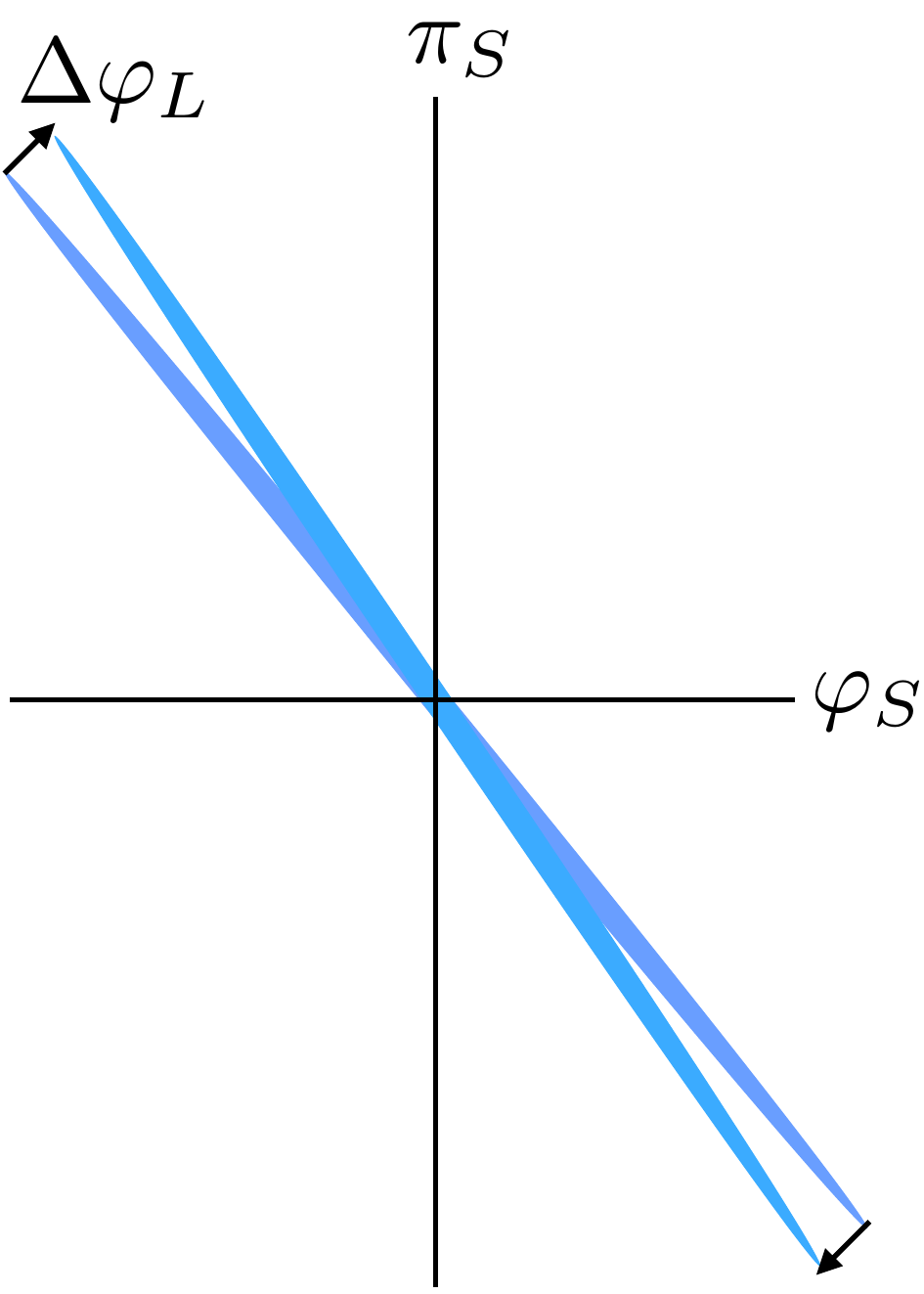}
\caption{LEFT: After redshifting to super-horizon scales, a mode of the field evolves into an increasingly squeezed state, which is shown here as a thin red ellipse in phase space of area $\sim\hbar$, roughly corresponding to a contour of the Wigner function associated with that pure state.  This may be viewed as a superposition of coherent states corresponding to classical field values, with corresponding classical field velocities. In the presence of interactions, this extended superposition fails to remain isolated and becomes entangled with many other modes of the field.  The induced decoherence of the original mode causes its state to become more mixed, and hence spread over an area in phase space much larger than $\hbar$.  This can be decomposed as an incoherent mixture of coherent states with much less squeezing, depicted here by their roughly circular Wigner contours.  The opacity of the circles indicates their relative probability mass. RIGHT: A highly squeezed state is easily perturbed into a completely orthogonal state by a tiny shift to its angle of orientation in phase space, induced by a change in the background field, $\Delta\vph_L$. Here, $\vph_S = \int_{\k} g(\k) \vph_\k$ is a localized mode constructed from Fourier modes that have wavelengths short relative to the background $\vph_L$ but have redshifted during inflation to superhorizon scales ($|k\tau|\ll1$) in order to become squeezed. In a decelerating era after inflation, the modes become highly correlated with background fluctuations $\vph_L$ on the scale of the cosmological horizon, recording long-wavelength information in phase space like the hand of a clock recording the time.}
\label{fig:superposition-and-clocks}
\end{figure}

During an inflationary epoch, the modes redshift to scales much larger than the cosmological horizon, and are excited into highly squeezed quantum states, which -- as indicated in the left panel of Figure~\ref{fig:superposition-and-clocks} -- are superpositions of coherent states peaked around classical field values.
Furthermore, a squeezed state is easily displaced to a completely orthogonal state by any perturbation to its parameters that induces a slight rotation in phase space. This allows the modes to act as highly sensitive measuring devices, able to respond to other degrees of freedom like the hands on a very precise clock, as in the right panel of Figure \ref{fig:superposition-and-clocks}.
Consequently, the accelerating epoch excites the field into a state which is simultaneously a delicate superposition of classical outcomes, and a decohering environment which is highly sensitive to disturbances introduced by coupling modes.
In particular, highly squeezed (superhorizon) modes are potentially capable of recording precise information about comparatively longer-wavelength background fluctuations, which have evolved into a superposition of many classical states.

Minimal gravitational interactions during inflation introduce an entanglement between modes, through which a long-wavelength background field can shift the parameters describing shorter-wavelength modes (for instance, by shifting the local Hubble expansion rate, or the local scale factor). However, we will see that due to the scale-invariance of the inflationary power spectrum, a given mode is entangled with modes on all scales in the same way, so that any recorded information about the long-wavelength field is lost due to the noise from its entanglement with other (shorter) modes. Furthermore, any records formed after modes cross the horizon during inflation are unable to propagate spatially: Once a region of space inflates to superhorizon scales, it acts as a separate universe, and no quanta can propagate out of that region in the finite amount of conformal time allowed during inflation (and more generally, while the mode remains outside the horizon).
Consequently, any information about the classical field in a given region remains fixed in space, and inaccessibly entangled between many modes.

\begin{figure}
\includegraphics[scale=0.5]{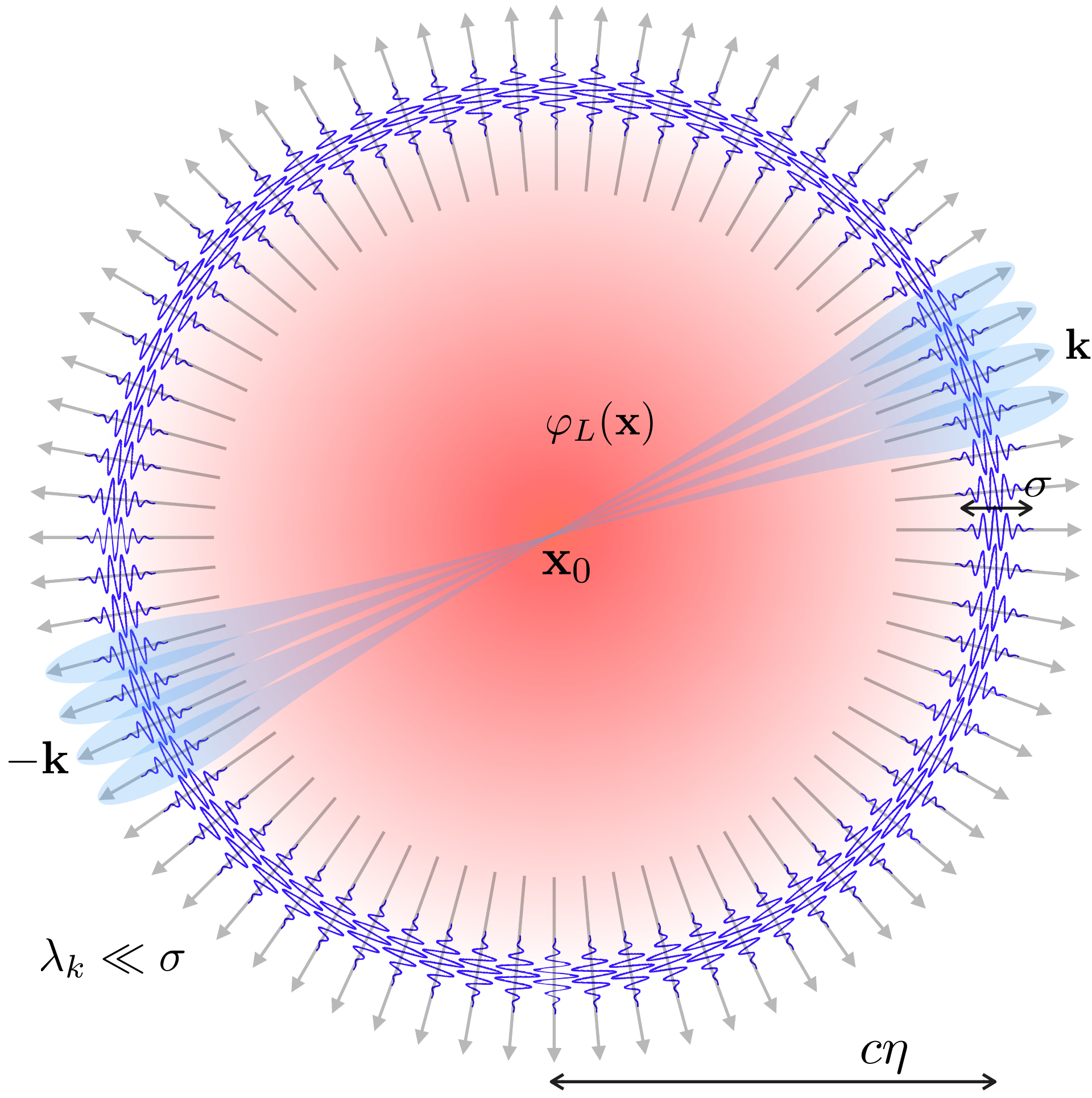}
\caption{
During an accelerating epoch, modes with (approximate) wavenumber $\pm\k$ localized to a region of size $\sigma\gg\lambda_k=2\pi/|\k|$ at point $\x_0$ become extremely squeezed (as in Figure~\ref{fig:superposition-and-clocks}), and thus capable of recording sensitive information about the long-wavelength field $\vph_L(\x)$.  In a subsequent decelerating epoch, the quanta in these squeezed modes propagate in opposing directions $(\k,-\k)$ in the form of entangled EPR pairs.  Gravitational interactions allow modes to record variation in the long-wavelength background $\vph_L$ on the scale of the cosmological horizon $(c\eta)$ as they propagate, encoded in their orientation in phase space.
Although the recorded information is encoded in EPR correlations between the left- and right-moving components at antipodal points on the shell, and thus could only be revealed by a joint measurement of both components, many such records are produced for EPR-entangled short modes propagating in approximately the same direction $\pm\k$, generating redundant records of the same long-wavelength variable in many disjoint spatial regions (indicated as multiple blue EPR-entangled modes).	This process occurs in all spatial regions and directions, and at all scales, first for smaller volumes that re-enter the horizon earlier.}  
\label{fig:records_propagate}
\end{figure}

Once inflation ends, however, and modes later re-enter the cosmological horizon during a decelerating epoch, they are able to propagate and distribute information spatially, as illustrated in Figure~\ref{fig:records_propagate}.
Furthermore, we will see that the decaying of the (massless) modes after re-entering the horizon effectively enhances the entanglement between a given short mode and long-wavelength fluctuations on the scale of the horizon (relative to its entanglement with shorter modes), allowing for very precise records to form.
This is described in \S~\ref{sec:records_decel}, with the recorded information (the sensitivity of the ``clock'' in the right panel of Figure \ref{fig:superposition-and-clocks}) quantified in Eq. \eqref{twopoint_inequality}, and the long-wavelength variable being recorded given in Eq. \eqref{vph_khat_def}.
We will see that in order for these records to form and to propagate into many disjoint spatial regions, the preceding inflationary epoch must last for a sufficiently long time, and interactions must be strong enough:
\be
g_{\rm total}\(\frac{a_f}{a_i}\)_{\rm inf}^4\gg1.
\ee
Here, $g_{\rm total}$ quantifies the net effect of the relevant gravitational interactions, and $a_f/a_i$ is the amount of expansion during inflation. Because the scale factor grows exponentially during inflation, even an extremely small interaction is sufficient to generate wavefunction branching after inflation.

The formation of redundant records (and hence branches) in the wavefunction is, in this particular cosmological system, captured by the phase information in the wavefunction $\Psi[\vph]$ in the field configuration basis. (The conditional states of Eq. \eqref{long_short_records} carry information in their growing phases,
\be
\psi^{(\vph_L)}(\vph)\sim e^{i g_0\cdot a(t)\vph_L}\psi^{(0)}(\vph),
\ee
where the cosmological expansion from $a(t)$ becomes large compared to the small interaction strength $g_0$.) The squeezing of the modes, corresponding growth of phase oscillations in the wavefunction, and decoherence in the superhorizon regime prepares the wavefunction in a state that will inevitably branch into (locally distinguishable) classical fields once the modes fall back within the horizon scale after the accelerating era ends.

\section{Preliminary Concepts}
\label{sec:prelim}

\subsection{Branches and Records}
In this paper we seek to contribute toward an ambitious long-term goal: the identification of mathematical criteria that define the branch structure of the wavefunction of large many-body systems, up to and including the entire universe \cite{riedel2017classical}.  
In other words, we seek, from first (information-theoretic) principles, a more-or-less unique method for decomposing the global, Schr\"odinger-picture wavefunction into orthogonal components, $\ket{\Psi} = \sum_i \ket{\Psi_i}$, where each component $\ket{\Psi_i}$ has an unambigous physical interpretation as a macroscopically distinct configuration corresponding to a possible classical outcome.  We furthermore expect branches to have these intuitive properties: 
(1) ``Collapse'': a superposition of branches $\ket{\Psi}=\sum_i \ket{\Psi_i}$ cannot feasibly be distinguished from the corresponding incoherent mixture $\rho =\sum_i \ket{\Psi_i}\bra{\Psi_i}$ with any reasonable measurement. 
(2) ``Irreversibility'': the branch decomposition $\ket{\Psi} = \sum_i \ket{\Psi_i}$ at an given time is a \emph{coarse-graining} of the decomposition $\ket{\Psi} = \sum_j \ket{\Psi_j^\prime}$ at all later times,\footnote{Strictly speaking, branching cannot continue forever \cite{dowker1995properties,kent1996quasiclassical,halliwell1999somewhere,woo1999consistent-histories,buniy2005hilbert,zeh2012role}.  Even if Hilbert space is formally infinite dimensional, reasonable IR and UV cutoffs mean that there is a maximum number of orthogonal branches. However, this would be unobservable if branching is effectively irreversible in a closed system until it thermalizes, at which point observers and experimentation cannot persist.} 
in the sense that $\ket{\Psi_i} = \sum_{j \in J_i} \ket{\Psi_j^\prime}$ where $\{J_i\}$ is some partition of the range of $j$.

At the least, formal criteria for branches ought to recover, as a special case, the standard decoherence story when a system $\Sys$ is measured by an apparatus $\Apt$ which is subsequently decohered by large many-body environment $\Env$:
\ba \label{eq:decoh-toy}
\ket{\Psi(t_0)} = \left[\sum_i \ket{S_i}\right]\ket{A_0}\ket{E_0} \quad \longrightarrow \quad \ket{\Psi(t_1)} = \sum_i \ket{\Psi_i(t_1)} = \sum_i \ket{S_i}\ket{A_i}\ket{E_i}
\ea
Here, $\ket{S_0} = \sum_i \ket{S_i}$, $\ket{A_0}$, and $\ket{E_0}$ are the initial states of the system, apparatus, and environment, respectively, and $\ket{A_i}$ and $\ket{E_i}$ are the conditional states of the apparatus and environment following the measurement (satisfying $\braket{S_i}{S_j} = \braket{A_i}{A_j} = \braket{E_i}{E_j} = 0$ for $i \neq j$).  We do not necessarily expect criteria to identify (for instance) the difference between the measured system, the artificial apparatus, and the natural environment, as is encoded in the tensor decomposition $\Glob = \Sys \otimes \Apt \otimes \Env$ of Hilbert space.  
Indeed, most large systems undergoing dynamics similar to the form in Eq.~\eqref{eq:decoh-toy} are not experiments in laboratories, but rather are natural amplification processes such as the decoherence of macroscopic fields by the nearby medium \cite{anglin1996decoherence}, the decoherence of hydrodynamic variables by the microscopic constituents of a fluid \cite{brun1996decoherence,halliwell1998decoherent,brun1999classical,anastopoulos1998preferred}, the decoherence of small particles by ambient radiation \cite{joos1985emergence,schlosshauer2008decoherence,riedel2010quantum,riedel2011redundant,korbicz2014objectivity}, or the decoherence of the long-wavelength primordial fluctuations by shorter wavelengths (and by other fields and particles) \cite{Kiefer:1998qe,lombardo2005decoherence,Kiefer:2006je,Burgess:2014eoa,nelson2016quantum,Sakagami:1987mp,Brandenberger:1990bx,Prokopec:2006fc,Mazur:2008wa}. 
In these naturally occurring examples, the ``measured system'' is often a collective variable, e.g., the average magnetization of a region, the average pressure within a volume, or the center-of-mass of a macroscopic object.  
Even when a tensor structure decomposition ($\Sys \otimes \Env_1 \otimes \Env_2 \otimes \cdots$) can be formally assigned, it will generically be unstable in time and underdetermined.  
Therefore, the primary goal is simply to identify the time-dependent set of orthogonal branches $\{\ket{\Psi_i(t)}\}$.

Although sufficient criteria for fully describing the complete branch structure of a many-body wavefunction is not currently known, in this work we will be satisfied if we can identify a non-trivial coarse-graining of these branches in the early universe without assuming any systems or observables as preferred a priori.
Any given decomposition $\{\ket{\Psi_i}\}$ corresponds to the set $\{\Omega\}$ of all mutually commuting observables of the form $\Omega = \sum_i \omega_i \ket{\Psi_i}\bra{\Psi_i}$, so intuitively we would like to derive at least some large subset $\{\Omega\}' \subset \{\Omega\}$ of preferred classical variables from first principles.

Of course, we will need to have access to \emph{some} additional structure in order to distinguish different states in Hilbert space from each other at all.  Here, we are investigating what can be derived based only on the distinguishability of different spatial regions \cite{riedel2017classical}. In non-relativistic many-body physics, this can take the form of a preferred tensor decomposition of Hilbert space into microscopic lattice sites: $\Glob = \bigotimes_{\x} \Glob_{\x}$ where, for instance, each local system $\Glob_{\x}$ might be a qubit or more generally a qudit.  (Relativistic field theories are discussed in the next section.)  Building on the intuition developed from studies of quantum Darwinism, we tentatively assert that different branches should be not just orthogonal but \emph{locally orthogonal} at many places. 
By this we mean that any two different branches $\ket{\Psi_i(t)}$ and $\ket{\Psi_j(t)}$ should be distinguishable by a hypothetical measurement on some region (subset of the lattice), $\Frag = \bigotimes_{\x \in F} \Glob_{\x} \subset \Glob$, and that furthermore this should be true for many disjoint regions ($\Frag^{(1)}$, $\Frag^{(2)}$, $\Frag^{(3)}$, etc.).\footnote{There is no strict threshold for ``many'' but, in a sense that can be made precise, the branch decomposition will be unique excepting counterexamples that become more and more pathalogical as the number of orthogonal regions increases \cite{riedel2017classical}.} This condition can be expressed as
\ba \label{eq:records-def}
\ket{\Psi_i} = \Pi^{\Frag^{(1)}}_i \ket{\Psi} = \Pi^{\Frag^{(2)}}_i \ket{\Psi} = \Pi^{\Frag^{(3)}}_i \ket{\Psi} = \cdots
\ea
where, for each $\Frag^{(a)}$, $\{\Pi^{\Frag^{(a)}}_i\}_i$ is a set of orthogonal projectors: $\Pi^{\Frag^{(a)}}_i = (\Pi^{\Frag^{(a)}}_i)^2 = (\Pi^{\Frag^{(a)}}_i)^\dagger$ and $\sum_i \Pi^{\Frag^{(a)}}_i = I^{\Frag^{(a)}}$, where $I^{\Frag^{(a)}}$ denotes the identity operator on ${\Frag^{(a)}}$.  An equivalent condition is that 
\ba \label{eq:ortho-conditional-states}
\Tr\[\rho^{\Frag^{(a)}}_{\Frag^{(b)}:i} \, \rho^{\Frag^{(a)}}_{\Frag^{(b)}:j}\]=0 ,\qquad i \neq j
\ea
where 
\ba
\rho^{\Frag^{(a)}}_{\Frag^{(b)}:i} \equiv \Tr_{\,\overline{\Frag^{(a)}}}\[\Pi_i^{\Frag^{(b)}}\ket{\Psi}\bra{\Psi}\Pi_i^{\Frag^{(b)}}\]
\ea
is the state on region $\Frag^{(a)}$ conditional on region $\Frag^{(b)}$ occupying the subspace labeled by $i$.
Here $\overline{\Frag} = \bigotimes_{\x \notin F} \Glob^{\x}$ denotes the complement of any region $\Frag$, so $\Glob = {\Frag} \otimes \overline{\Frag}$.  When this condition holds, we say that any observable $\Omega^{\Frag^{(a)}}$ local to a region $\Frag^{(a)}$ has a \emph{record} when its eigenspaces correspond to the local projectors: $\Omega^{\Frag^{(a)}} = \sum_i \omega_i \Pi^{\Frag^{(a)}}_i$ (for distinct eigenvalues, $\omega_i \neq \omega_j$ for $i \neq j$).  Note that Eq.~\eqref{eq:records-def} defines a relation between records that is manifestly symmetric and transitive, avoiding the need to distinguish any particular record as preferred over the others.

Thus, we will characterize branches by a sort of multipartite entanglement which is most cleanly exemplified by a generalized GHZ state of $N$ qudits and $d$ alternatives:
\ba \label{eq:generalized-GHZ-branch}
\ket{\Psi} = \sum_{i=1}^d \ket{i}^{\otimes N} = \ket{1}_1\ket{1}_2\cdots\ket{1}_N + \ket{2}_1\ket{2}_2\cdots\ket{2}_N + \cdots + \ket{d}_1\ket{d}_2\cdots\ket{d}_N
\ea
Indeed, the equivalent conditions Eqs.~\eqref{eq:records-def} and \eqref{eq:ortho-conditional-states} are 
automatically satisfied in the special case of Eq.~\eqref{eq:decoh-toy} when the environment has many spatially disjoint parts ($\Env = \Env^{(1)}\otimes\Env^{(2)}\otimes\Env^{(3)}\otimes\cdots$), 
each of which are in an 
orthogonal state conditional on the state of the system 
$\Sys$, that is, $\ket{E_i} = \vert E_i^{(1)}\rangle\vert E_i^{(2)}\rangle\vert E_i^{(3)}\rangle\cdots$ where $\langle E_i^{(a)}\vert E_j^{(a)}\rangle=0$ for $i\neq j$.  
But this condition can be satisfied much more generally.  For instance, the state $\ket{\Psi} = \ket{\Psi_1} + \ket{\Psi_2}$ where
\begin{align}\begin{split}\label{eq:first-example-branches}
\ket{\Psi_1} &= \big(\ket{0}\ket{0}+\ket{1}\ket{1}\big)\big(\ket{0}\ket{0}+\ket{1}\ket{1}\big)\cdots\big(\ket{0}\ket{0}+\ket{1}\ket{1}\big), \\
\ket{\Psi_2} &= \big(\ket{0}\ket{0}-\ket{1}\ket{1}\big)\big(\ket{0}\ket{0}-\ket{1}\ket{1}\big)\cdots\big(\ket{0}\ket{0}-\ket{1}\ket{1}\big)
\end{split}\end{align}
contains redundant records on neighboring pairs of qudits, but not on any individual qudit.  To see this, we choose the projectors $\Pi_1^{(n,n+1)} = \ket{+}_{n,n+1}\!\bra{+}$ and $\Pi_2^{(n,n+1)} = \ket{-}_{n,n+1}\!\bra{-}$ jointly on the $n$-th and $n\! +\! 1$-th qudits, where $\ket{\pm}_{n,n+1} = (\ket{0}_{n}\ket{0}_{n+1} \pm \ket{1}_{n}\ket{1}_{n+1})/\sqrt{2}$, and observe that $\Pi_i^{(n,n+1)}\ket{\Psi} = \ket{\Psi_i}$, $i=1,2$, for any odd $n$.  On the other hand, branching states with redundant records may possess entanglement between different recording regions even within a single branch, such as the state $\ket{\Psi} = \ket{\Psi_1} + \ket{\Psi_2}$ where now
\begin{align}\begin{split}\label{eq:second-example-branches}
\ket{\Psi_1} &= \big(\ket{0}\ket{0}+\ket{1}\ket{1}\big)\big(\ket{0}\ket{0}+\ket{1}\ket{1}\big)\cdots\big(\ket{0}\ket{0}+\ket{1}\ket{1}\big), \\
\ket{\Psi_2} &= \big(\ket{2}\ket{2}+\ket{3}\ket{3}\big)\big(\ket{2}\ket{2}+\ket{3}\ket{3}\big)\cdots\big(\ket{2}\ket{2}+\ket{3}\ket{3}\big).
\end{split}\end{align}
In this case, we could choose the projectors $\Pi_1^{(n)} = \ket{0}_n\!\bra{0} + \ket{1}_n\!\bra{1}$ and $\Pi_2^{(n)} = \ket{2}_n\!\bra{2} + \ket{3}_n\!\bra{3}$ on the $n$-th qudit to see that each qudit has a record of the branch structure even though, conditional on a particular branch $\ket{\Psi_i}$, neighboring qudits are partially entangled and therefore are not in a pure state.

There exists weak but suggestive evidence that the existence of highly redundant records in the above sense is necessary and sufficient to formalize our intuitive notion of branches.  
Redundant records appear \emph{necessary} for describing variables we might call quasiclassical for the simple reason that (a) all reasonable measuring devices create redundant records by virtue of their primary purpose of amplification, and (b) it would be impossible for multiple observers to even \emph{know} the value of an observable without their internal mental states carrying a record.  (Consider, for instance, the many records that exist about the outcome of any laboratory measurement, or the position of any macroscopic object, on account of the tiny subset of photons that are sufficient to reconstruct the variable through passive observation \cite{riedel2010quantum,riedel2011redundant}.) 
Redundant records may also be \emph{sufficient} in light of the fact that a unique branch decomposition is induced, except in the case of error-correcting counterexamples, by the set of all redundantly recorded observables \cite{riedel2017classical}. 
However, at this point it is unclear whether redundant records can guarantee irreversibility (although the proliferation of a macroscopic number of records is certainly suggestive of it).

\subsection{Branching of Fields}

When considering field theories, the need for branch criteria that do not depend on a preferred subsystem is brought into stark relief.   Without particle conservation, subsets of particles grouped into macroscopic objects become plainly ephemeral.  It has been suggested that preferred subsystems might be obtained from the distinction between particle species (e.g., photons and non-photons \cite{kent2014solution}), but even this is suspect in light of our expectation of invariance under local field redefinitions that mix species \cite{apfeldorf2001field}.  
It is likely that currently known fields are only indirectly related to more fundamental field to be discovered in the future. In contrast, spatial locality becomes more fundamental in the relativistic setting, and we focus on criteria the depend only on the entanglement between disjoint spatial regions.  Local field redefinitions are analogous to local changes of basis (i.e., local unitaries) on individual sites of a discrete lattice, and they leave spatial entanglement invariant subject to the following subtleties.

In relativistic\footnote{The Reeh-Schlieder theorem and related properties also hold in many non-relativistic theories \cite{requardt1982spectrum}.} field theories, no precisely local records can exist because no finite-energy state is an eigenstate of any observable that can be exactly localized to a bounded spatial region \cite{redhead1995more,summers2008yet}.  
Thus, Eq.~\eqref{eq:records-def} cannot hold exactly.\footnote{Furthermore, adjacent patches of space cannot, for any finite-energy state, be completely disentangled from each other.  This signals that there is no sensible notion of a tensor decomposition of space, $\bigotimes_x\Glob^{(x)}$ or $\mathclap{\,\,\,\,\,\int}{\otimes}\, \dd \Glob^{(x)} $, in a quantum field theory.}  This is a consequence of the Reeh-Schlieder theorem,\footnote{The (generalized) Reeh-Schlieder theorem states that for any two finite-energy state $\ket{\Psi}$ and $\ket{\chi}$ (including the vacuum), and for every bounded spatial region $\Frag$, there exists a (not necessarily normalized or Hermitian) operator $\Lambda^{\Frag}$ on $\Frag$ such that $\Lambda^{\Frag}\ket{\Psi} = \ket{\chi}$. 
If there was a finite-energy branch $\ket{\Psi_1}$ that is preserved by a projector local to a region $\Frag$ (i.e., $\Pi_i^{\Frag}\ket{\Psi_1} = \ket{\Psi_1}$) then for any state $\ket{\chi}$ there would be an operator $\Lambda^{\Frag'}$ on the disjoint region $\Frag'$ such that $\Pi_i^{\Frag}\ket{\chi} = \Pi_i^{\Frag} \Lambda^{\Frag'}\ket{\chi}= \Lambda^{\Frag'}\Pi_i^{\Frag} \ket{\Psi_1}= \Lambda^{\Frag'} \ket{\Psi_1} = \ket{\chi}$.  But this would be mean that every $\ket{\chi}$ is preserved by $\Pi_i^{\Frag}$ and hence the projector must be trivial: $\Pi_i^{\Frag}=I^{\Frag}$.} which similarly implies that no detector confined to a finite spatial region -- as all physical detectors are -- can ever record the presence of a particle-eigenstate with perfect accuracy \cite{redhead1995more}.  
This might be a signal that the concept of spatially disjoint records will eventually need to be generalized in the relativistic domain, or simply that branches will only ever be well-defined asymptotically (as are particle in-states and out-states).
In this paper our focus is on investigating how much we can learn using only local records; we will largely side-step the Reeh-Schlieder property by simply requiring only that records be localized in disjoint regions up to exponentially tiny tails.  More precisely, the localized observables we consider will be of Gaussian form 
\ba
\vph(\x_0,\k_0,\sigma) = \int_{\k} e^{-(\k-\k_0)^2\sigma^2/2+i(\k-\k_0/2)\cdot\x_0} \vph_\k = \int \!\dd{\x}\, e^{-(\x-\x_0)^2/(2\sigma^2)-i\k_0\cdot(\x-\x_0/2)} \vph_\x
\ea
where the triplet $(\x_0,\k_0,\sigma)$ denotes the central position, central momentum, and characteristic spatial extent of the localized variable.
These variables are as spatially disjoint as \emph{any} physical quantum system in a field theory can be.  
We will consider two observables $\vph(\x_0,\k_0,\sigma)$ and $\vph(\x_0',\k_0',\sigma)$ to be essentially spatially disjoint when $|\x_0 -\x_0^\prime| \gg \sigma$.   The spatial entanglement within the vacuum state implied by the Reeh-Schlieder property is known to fall off exponentially with distance for massive fields and as a power law for massless ones \cite{summers1987bells,summers1987bells-a}.  We will find below that, for the massless field $\vph$, records in localized Gaussian modes will be accurate only up to errors that fall like a power law with the relevant distance scales, although the connection to masslessness is not yet clear.

Lastly, note that, unlike the simple examples of discrete branching in 
Eqs.~(\ref{eq:generalized-GHZ-branch}-\ref{eq:second-example-branches}), 
we will be concerned below with branching of continuous variables such that individual wavefunction branches will be associated with finite ranges of the variable.   In some cases this will be associated with a trade-off between the number of disjoint records defining the branches and the accuracy with which they record the relevant variable.  See \App~\ref{sec:continuous-branching} for details. The local variables $\vph(\x_0,\k_0,\sigma)$ will be distributed as Gaussians and, as described in detail below, we will consider them to have a record distinguishing between two branches of interest when the distributions of $\vph(\x_0,\k_0,\sigma)$ \emph{conditional} on either branch are easily distinguishable.  Proving rigorously that the branch uniqueness theorem \cite{riedel2017classical} can be extended to include approximately localized records with approximate accuracy, although plausible due to tiny errors, is not proved here but is left to future work.

\subsection{Background Cosmology}
\label{sec:cosmology_review}

In this paper, we will impose a fixed background cosmology, and ask how the wavefunction of cosmological fluctuations $\Psi[\vph]$ evolves in that background.
The (classical) background cosmology is of the Friedmann-Robertson-Walker form
\be
ds^2 = -dt^2 + a^2(t) d\x^2.
\ee
We will first consider an inflationary epoch with an approximately de Sitter background, $a(t) \propto e^{Ht}$, with $H(t)\equiv\dot{a}/a\approx{\rm const}$.
During the inflationary era, we will use conformal time $\tau$, defined by $d\tau = dt/a$ and satisfying
\be
a(\tau)=-1/H\tau.
\ee
Note that $\tau$ runs from $-\infty$ to $0$, and we will be mainly interested in the late time limit $\tau\rarr0$ (or $a\rarr\infty$).

After a sufficiently long (to be made precise) period of inflation, ending at a small but finite $\tau_f<0$, we will modify the background expansion and introduce a decelerating epoch, characterized by equation of state $p=\rho/3$, that is, radiation.
In this era, the scale factor evolves as $a(t)\sim\sqrt{t}$. We will introduce a new conformal time coordinate $\eta$ for this epoch, again defined by $d\eta= dt/a$, in terms of which the scale factor can be written as
\be
a(\eta) = a(0)+a'(0)\eta
\ee
where $a(0)$ and $a'(0)$ are fixed by imposing continuity of $a'$ (or of the Hubble radius) across the transition from inflation to radiation.

We will assume that the Hubble rate is changing very slowly, quantified by the slow-roll parameter $\eps\equiv-\dot{H}/H^2 \ll 1$, so that the dynamics of the fluctuations can be approximated as dynamics in de Sitter space. This slow-roll variation affects the amplitude of metric fluctuations, as discussed in Appendix \ref{app:review_metric}, as well as their gravitational interactions, as discussed in \S \ref{sec:long_modes_examples} and Appendix \ref{sec:cubic_zeta}.  We will absorb this dependence into the rescaled field $\vph$ and into generic coefficients for cubic interactions.

\subsection{Phase Information in the Inflationary Wavefunction}
\label{sec:WFinflation}

The formation of wavefunction branches, and effective wavefunction collapse, is grounded in the general phenomenon of amplification, and so only appears in closed quantum systems that explicitly break time-reversal symmetry.\footnote{As always, the fundamental dynamics are Lorentz invariant and time-reversal symmetry is broken only by the special initial conditions of the universe.}  More specifically, we identify branches by the increasing amount of space-like entanglement found on consecutive Cauchy surfaces. Therefore it will often be useful to work in the Schr\"odinger picture, where the wavefunction of the universe is a time-dependent functional of the field(s) on consecutive spatial slices.

During inflation, the relevant degrees of freedom are scalar and tensor metric fluctuations, $\zeta(\x)$ and $\gamma_{ij}(\x)$. These are defined and reviewed in Appendix~\ref{app:review_metric}. The scalar mode $\zeta$ is a dynamical degree of freedom in all inflationary models, appearing as the Goldstone boson associated with breaking the time translation invariance of de Sitter space by introducing a small variation in the Hubble rate \cite{Cheung:2007st,Cheung:2007sv}. Its self-interactions and couplings to the tensor modes (gravitons) are universal gravitational interactions.
The wavefunction for scalar and tensor modes, $\Psi[\zeta(\x),\gamma_{ij}(\x)](t)$, satisfies the Schr\"{o}dinger equation
\be
i\hbar\frac{d}{dt}\Psi[\zeta,\gamma](t) = \H(t) \Psi[\zeta,\gamma](t)
\ee
where $\H=\H[\zeta(\x),\gamma_{ij}(\x),\pi_\zeta(\x),\pi_{\gamma,ij}(\x)]$ is the Hamiltonian, which acts on the wave functional as an operator via its dependence on the functional derivatives $\pi_\zeta(\x)=-i\hbar\d/\d\zeta(\x)$ and $\pi_{\gamma,ij}(\x)=-i\hbar\d/\d\gamma_{ij}(\x)$.

Most of the literature on inflationary cosmology has focused on observable correlation functions\footnote{The procedure of identifying branches can be used to distinguish which observables $\hat{\O}$ will take robust classical values -- namely, the operators whose eigenstates are the branches -- from the larger set of observables that do not because they involve non-local phases (and hence are infeasible to measure).} of $\zeta$ (or $\gamma$), which become time-independent in the $t\rarr\infty$ limit as the field freezes out. While these calculations usually use the Heisenberg picture, they can of course be reproduced from the amplitude of the wavefunction:
\be\label{field_correlators}
\<\hat{\O}[\zeta,\gamma](t)\> = \int\!\D\zeta\D\gamma\,\, \O[\zeta,\gamma]|\Psi[\zeta,\gamma](t)|^2.
\ee

At late times (in the super-Hubble regime), the wavefunction takes the WKB form of a rapidly oscillating phase given by the action, and a constant amplitude. This is because the conjugate momentum dependence of the Hamiltonian is suppressed as the fields freeze out ($\dot{\zeta},\dot{\gamma}_{ij}\rarr0$), so the Hamiltonian just acts on the wavefunction multiplicatively.  
Suppressing the tensor arguments,
\be
\H[\zeta,\pi_\zeta]=\H[\zeta,-i\d/\d\zeta] \,\,\overset{t \to \infty}{\approx}\,\, \H[\zeta]. \nn
\ee
Furthermore, since $\dot{\zeta}\rarr0$ after horizon crossing, we have $\H\rarr-\L=-\int\! d^3\x\,\L(\x)$. The Schr\"{o}dinger equation is therefore solved by the WKB form
\be\label{psi_WKB}
\Psi[\zeta,\gamma](t)\rarr e^{i\SS[\zeta,\gamma](t)}|\Psi[\zeta,\gamma]| \ \ \ \ \ (\text{super-Hubble regime})
\ee
when all modes have crossed the horizon. Here $\SS(t)=\int^t \! dt' d^3\x \, \L(\x)$, and the amplitude $|\Psi|$ becomes time-independent.

The phase information in Eq.~\eqref{psi_WKB} is not relevant for correlation functions of the fields. It only affects correlation functions involving the conjugate momenta or field velocities, which leave no observable effect since $\dot{\zeta},\dot{\gamma}\rarr0$ during inflation.
However, this phase information is crucial for decoherence \cite{Nelson:2016kjm}, the formation of redundant records, and wavefunction branching.

In particular, the rapid inflationary growth of the (free theory) action in Eq.~\eqref{psi_WKB} leads to the squeezing of the quantum state for each mode. Including interactions, we will see that this squeezing makes the modes of the field very sensitive as a decohering environment or measurement apparatus for longer-wavelength degrees of freedom.
While inflationary dynamics will not be sufficient to generate redundant records of the fields in many \textit{spatial} regions, we emphasize that the WKB classicality of the inflationary wavefunction, and corresponding squeezing of the modes, prepares a state which is highly susceptible to branching (much like Schr\"{o}dinger's cat). Decoherence and branching occur once the modes re-enter the horizon and propagate spatially.
The resulting long-range redundant correlations, which are indicative of branching, will be encoded in the phase information generated by interactions, or equivalently, in correlation functions which depend on both the fields and their conjugate momenta.

\section{Linear Inflationary Dynamics}
\label{sec:linear}

In this section we give an elementary review of linear inflationary dynamics for the benefit of readers with a background in quantum information rather than cosmology or field theory.  Other readers can safely skip to \S~\ref{sec:wigner} after noting our choice of notation and the form of the Hamiltonian, Eq.~\eqref{H_phi}.

\subsection{Field definition and Hamiltonian}

The Lagrangian for a massless scalar field in an approximate de Sitter background spacetime, $a(\tau)=-1/H\tau$, defined with respect to conformal time $\tau$, is
\ba
\L &=& \frac{1}{2}a^2(\tau) H^2 \int d^3\x \[ (\vph')^2 - (\nabla\vph)^2 \] \\
&=&\frac{1}{2\tau^2}\int_\k\(\vph'_\k\vph'_{-\k} - k^2\vph_\k\vph_{-\k}\),
\ea
where the prime indicates $\dd/\dd\tau$, and we deploy the notation $\int_\k \equiv \int\!\frac{\dd^3\k}{(2\pi)^3}$. By including a factor of $H^2$, we have implicitly defined $\vph$ to be a dimensionless field variable with fluctuations of order unity in the late-time limit,
\be
\left.\<\vph_\k\vph_{\k'}\>\right|_{\tau\rarr0} = (2\pi)^3\d^3(\k+\k')\frac{1}{2k^3},
\ee
instead of a dimensionful field with fluctuations of order the Hubble scale.
The canonically conjugate variable is $\pi(\x) = \partial_{\vph'(\x)} \L = \vph'(\x)/\tau^2$, or equivalently\footnote{Note that the momentum-space variable itself does not pick up a minus sign because \begin{align}
\pi_\k =  \int \dd^3\x p(\x)e^{-i\k \cdot \x} = \tau^{-2} \int \dd^3\x \vph'(\x)e^{-i\k \cdot \x} = \tau^{-2} \frac{\dd}{\dd \tau} \int \dd^3 \x \vph(\x)e^{-i\k \cdot \x} = \tau^{-2} \frac{\dd}{\dd \tau}\vph_\k =  \tau^{-2} \vph'_\k.
\end{align}  You can use the same elementary Fourier manipulations to show that $p(\x) = \partial_{\vph(\x)} \L$ is equivalent to $\pi_{\k} = \partial_{-\vph_\k} \L$.  Likewise for Hamilton's equations, $\vph'_\k = \partial_{\pi_{-\k}} \H$, $p'_\k = -\partial_{\vph_{-\k}} \H$ exhibits the sign flip in momentum space, leading to the correct equations of motion \eqref{eq:neweom} without the sign flip.} $\pi_\k =\vph'_\k/\tau^2$. The resulting Hamiltonian is\footnote{Note that $\H+\L=\int \dd^3 \x \vph'(\x)p(\x) = (2\pi)^{-3}\int \dd^3\k \vph'_{\k}\pi_{-\k} = \frac{1}{2}(2\pi)^{-3} \int \dd^3\k (\vph'_{\k}\pi_{-\k} + \vph'_{-\k}\pi_{\k})$.  Furthermore, $ \vph'_{\k}\pi_{-\k} \neq  \pi_{-\k} \vph'_{\k}$, so we are implicitly choosing an operator ordering when we quantize.}
\begin{align}\begin{split}\label{H_phi}
\H&= \left[\int_\k \vph'_{\k}\pi_{-\k} \right] - \L \\
&= \frac{1}{2} \int_\k \[\frac{1}{a^2H^2}\pi_\k \pi_{-\k} + a^2 H^2 k^2 \vph_\k\vph_{-\k}\] \\
&= \frac{1}{2} \int_\k \[\tau^2\pi_\k \pi_{-\k} + \frac{k^2}{\tau^2} \vph_\k\vph_{-\k}\]. \\
\end{split}\end{align}

\subsection{Heisenberg-Picture Dynamics}

The classical equations of motion are 
\begin{align}
\label{eq:neweom}
\vph'_\k = \tau^2 \pi_{\k}, \qquad
p'_\k = -\frac{k^2}{\tau^2} \vph_{\k}
\end{align}
or equivalently,
\be
\vph''_\k - \frac{2}{\tau}\vph'_\k + k^2\vph_\k = 0.
\ee
The classical solution is given by the linear combination
\begin{align}\begin{split}
\label{eq:newvarsolutions}
\vph_\k(\tau) &= \frac{1}{\sqrt{2k^3}} \[(1+ik\tau)e^{-ik\tau}a_\k+(1-ik\tau)e^{ik\tau}a^\dagger_{-\k}\], \\
\pi_\k(\tau) &= \frac{1}{\sqrt{2k^3}} \frac{k^2}{\tau} \[e^{-ik\tau}a_\k + e^{ik\tau}a^\dagger_{-\k}\],
\end{split}\end{align}
with mode coefficients $a_\k$, $a^\dagger_\k$. While the conjugate momentum blows up, the physical velocity goes to zero:
\be\label{vph_dot}
\dot{\vph}_\k \equiv \frac{d\vph_\k}{dt} = -H\tau\vph'_\k(\tau) = -H(k\tau)^2  \frac{1}{\sqrt{2k^3}} \[e^{-ik\tau}a_\k + e^{ik\tau}a^\dagger_{-\k}\].
\ee
This freezing is fast enough that the classical mode approaches a constant value. That is, $\int^\infty dt \dot{\vph}=\int^0 d\tau\vph'$ converges, as is already clear from the $\tau\rarr0$ limit of $\vph_\k(\tau)$.

We quantize by promoting the mode coefficients to time-independent raising and lowering operators satisfying
\be
[a_\k,a^\dagger_{\k'}]=(2\pi)^3\d^3(\k-\k'),
\ee
which is equivalent\footnote{Remember that $k = \abs{\k}$, so $\k \to -\k$ does not flip the sign of $k$.} 
to the canonical commutation relations 
\begin{align}
[\vph_\k(\tau),\pi_{\k'} (\tau)] = [\vph_\k,\pi_{\k'} ] = i(2\pi)^3\delta^3(\k+\k'),
\end{align}
where the Heisenberg picture operators are defined as $\vph_\k(\tau) = \exp\(i\int^\tau d\tau' \H\)\vph_\k \exp\(-i\int^\tau d\tau' \H\)$.
One can then check that Eq.~\eqref{eq:newvarsolutions} solve the von Neumann equations of motion 
\begin{align}
\frac{\dd}{\dd\tau}\vph_\k(\tau) = -i[\H,\vph_\k(\tau)], \qquad \frac{\dd}{\dd\tau}\pi_\k(\tau) = -i[\H,\pi_\k(\tau)].
\end{align}

Lastly, it will be useful to introduce ``rotated'' canonical variables $(\Phi_\k(\tau),\Pi_\k(\tau))$:
\be\label{wigner_rotate}
\( \begin{array}{c}
\Phi_\k \cdot \sqrt{k^3} \\
\Pi_\k / \sqrt{k^3}
\end{array} \) = 
\( \begin{array}{cc}
   \cos(\theta(k\tau))  &  -\sin(\theta(k\tau))  \\
    \sin(\theta(k\tau))  &  \cos(\theta(k\tau))
\end{array} \)
\( \begin{array}{c}
\vph_\k \cdot \sqrt{k^3} \\
\pi_\k / \sqrt{k^3}
\end{array} \),
\ee
which are defined to be uncorrelated, $\langle\Phi_\k(\tau)\Pi_{\k'}(\tau)\rangle=0$ at all times.
From this one can determine that the time-dependent angle $\theta$ must be
\be
\theta(k\tau) = k\tau - \frac{1}{3}(k\tau)^3 + \O(k\tau)^5 \approx k\tau
\ee
in the $k\tau\rarr0$ limit. From this, we can find that in the same limit,
\ba
\<\Phi_\k\Phi_{\k'}\> &=& \frac{1}{2k^3}(k\tau)^2 (2\pi)^3\d^3(\k+\k'), \nn \\
\<\Pi_\k\Pi_{\k'}\> &=& \frac{1}{2} k^3 \frac{1}{(k\tau)^2} (2\pi)^3\d^3(\k+\k'). \label{Phi_Pi_two_pt}
\ea

\subsection{Schr\"{o}dinger Wavefunction, Wigner Function, and Phase Space Dynamics}
\label{sec:wigner}

The time evolution can be described in the Heisenberg picture in terms of operators, as in Eq.~\eqref{eq:newvarsolutions}, but we now construct the wavefunction in the Schr\"{o}dinger picture.
The solution for the wavefunction for a pair of modes $(\vph_\k,\vph_{-\k})$ 
is of the Gaussian form\footnote{Recall that for the real field $\vph(\x)$, the momentum modes $\vph_\k$ are complex, with the position-space reality condition $\vph_\x = \vph_\x^\dagger$ enforced by $\vph_\k = \vph_{-\k}^\dagger$, so that $|\vph_\k|^2 = \vph_\k^\dagger \vph_\k = \vph_{- \k} \vph_\k$.  For the purpose of taking functional derivatives, we can treat the fields $\vph_\k$ and $\vph_\k^\dagger$ as ``independent'' in the same way as for complex scalars $z = a+ib$ and  $z^* = a-ib$, i.e., by defining $\partial_z \equiv (\partial_a -i \partial_b)/2$ and $\partial_{z^*} \equiv (\partial_a +i \partial_b)/2$ so that $\partial_z z = 1 = \partial_{z^*} z^*$ and $\partial_z z^* = 0 = \partial_{z^*} z$.}
\be\label{psi_k}
\psi(\vph_\k,\vph_{-\k}) \equiv \psi(\vph_\k) = {\rm{(norm.)}} \exp[-|\vph_\k|^2A(k,\tau)],
\ee
and satisfies the Schr\"{o}dinger equation,
\be
i \frac{\dd}{\dd\tau}\psi(\vph_\k) = \H_\k(\vph_\k,-i\dd/\dd\vph_\k)\psi(\vph_\k),
\ee
where we have defined the single-mode Hamiltonian $\H_\k$ by $\H\equiv\sum_\k \H_\k$.
Using Eq. \eqref{H_phi}, this leads to a first-order, nonlinear differential equation for $A(k,\eta)$ \cite{Burgess:2014eoa,boddy2016how}:
\be\label{A_eom_inflation} 
i\frac{d}{d\tau}A(k,\tau) = \tau^2 A^2(k,\tau) - k^2/\tau^2.
\ee
The nonlinear $A^2$ term comes from the kinetic term in the Hamiltonian; two factors of $\pi=-id/d\vph$ yield two factors of $A(k,\eta)$ from the exponent of the wavefunction. The $k^2$ term comes from the gradient term in the Hamiltonian, and is dominant when the mode is frozen outside the horizon.
The boundary condition $\lim_{\tau\rarr-\infty} A(k,\tau)$ is set using the canonical choice of the Bunch-Davies vacuum in the infinite past.\footnote{The Bunch-Davies state is constructed by placing each mode in the ground state of the Hamiltonian in the limit $\tau \rarr -\infty$. This is most easily done by performing a time-dependent field redefinition to the Mukhanov coordinate, $v=-\vph/\tau$, so that the Hamiltonian for each mode becomes time-independent in the infinite past:  $\lim_{\tau \rarr -\infty} \H_{\k}^{(v)}(\tau) = (k^2 |v_{\k}|^2 + |\pi_{\k}^{(v)}|^2)/2$.  This ground state is well-defined in any coordinates at any finite time, and the mode is not excited out of this state until it crosses the horizon.}
With this initial condition the solution is\footnote{See \cite{Polarski:1995jg} for related discussion of the Schr\"{o}dinger-picture evolution. Another way to derive $A(k,\tau)$ is to write the wavefunction as a path integral, $\Psi[\vph]|_\tau=\int^\vph\D\tilde{\vph} e^{i\SS[\tilde{\vph}]}\Psi_0[\tilde{\vph}]$, where $S=\frac{1}{2} \int d^3\x d\tau \tau^{-2} ((\vph')^2 - (\nab\vph)^2)$. The path integral is a Gaussian integral.}
\be\label{A_k}
A(k,\tau) = k^3\frac{1-i/k\tau}{1+k^2\tau^2}.
\ee
(Note that we have suppressed a factor of the total volume in which the field is being decomposed into $\k$ modes, which arises from discretizing the integral in the Hamiltonian, $(2\pi)^{-3}\int d^3\k\rarr V^{-1}\sum_\k$. In the remainder of this section, we will implicitly include this factor in the canonical variables: $V^{-1/2}(\vph,\pi)\rarr(\vph,\pi)$.)

\begin{figure}
\hspace{-1.0cm}
\includegraphics[scale=0.45]{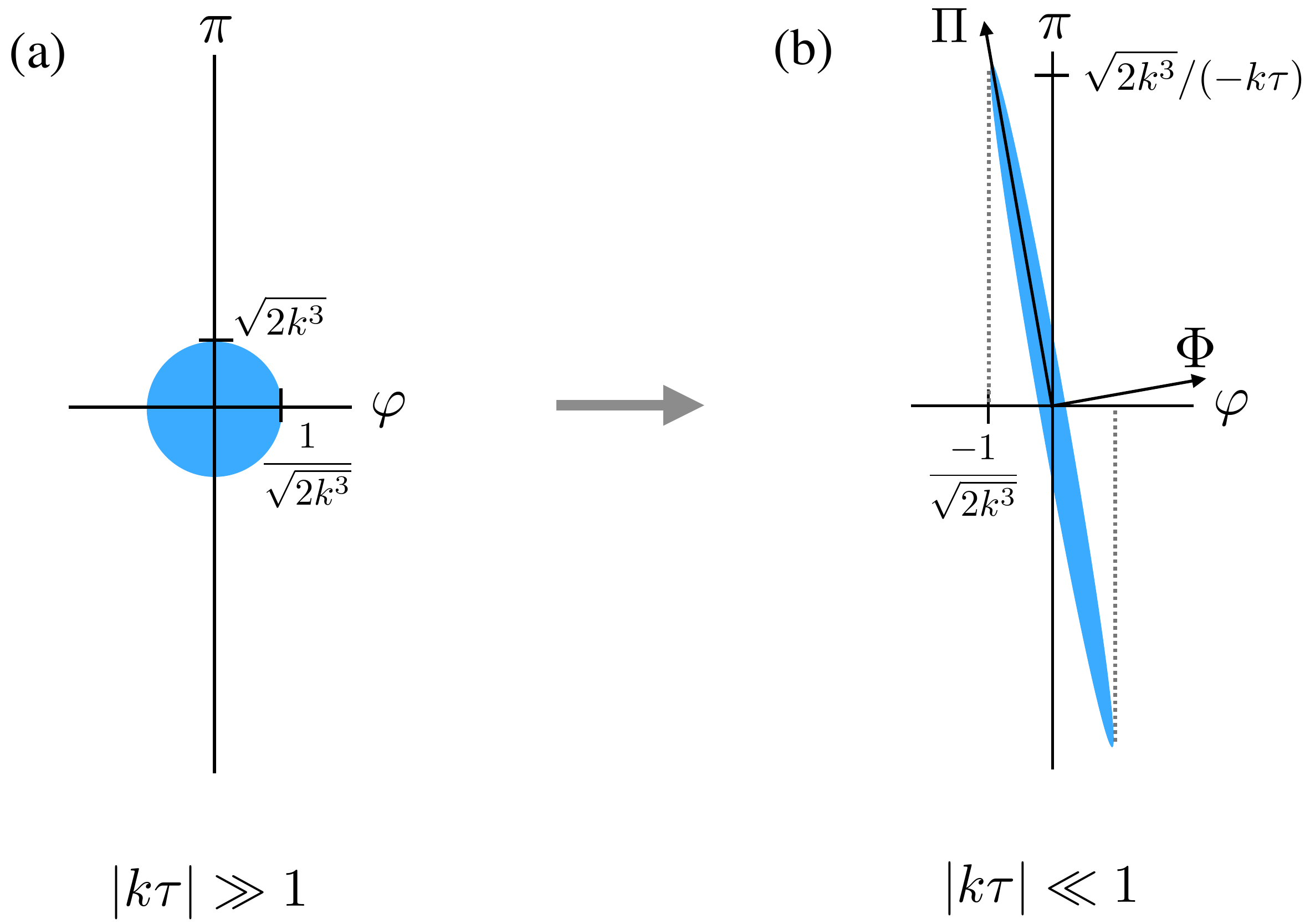}
\caption{The shaded region is the phase space region in which $r^2|\Phi_\k|^2 + r^{-2}|\Pi_\k|^2\lesssim1$, where the Wigner function $W_0(\vph_\k,\pi_\k;k,\tau)$ is not (close to) zero. On the axes, we suppress the wavenumber $\k$, and are also implicitly plotting either the real or imaginary part $\vph_\k$, which are two independent modes in the same quantum state. (So, $\vph=\vph^{\rm(Re,Im)}_\k$, $\pi=\pi^{\rm(Re,Im)}_\k$, $\Phi=\Phi^{\rm(Re,Im)}_\k$, and $\Pi=\Pi^{\rm(Re,Im)}_\k$.) At early times (a) the mode is deep inside the horizon and is just a harmonic oscillator in its ground state. At late times (b), the state becomes highly squeezed. The conjugate momentum grows as $1/|k\tau|$, while the spread in the $\Phi_\k$ direction is suppressed, going as $|k\tau|$. Note that $\pi_\k$ is anticorrelated with $\vph_\k$ due to the relative factor of $1/k\tau<0$.}
\label{fig:wigner}
\end{figure}

The corresponding Wigner function is \cite{Polarski:1995jg}
\be
W_0(\vph_\k,\pi_\k;k,\tau) = \int d^2\Delta\vph_\k e^{-i(\pi_\k^*\Delta\vph_\k+\pi_\k\Delta\vph^*_\k)} \rho_0(\vph_\k-\Delta\vph_\k/2,\vph_\k+\Delta\vph_\k/2),
\ee
where
\be\label{rho_0}
\rho_0(\vph_\k-\Delta\vph_\k/2,\vph_\k+\Delta\vph_\k/2) = \psi(\vph_\k-\Delta\vph_\k/2)\psi^*(\vph_\k+\Delta\vph_\k/2)
\ee
is the (pure state) density matrix. Using Eq.~\eqref{psi_k}, we find that the Wigner function in terms of the uncorrelated canonical variables of Eq.~\eqref{wigner_rotate} is\footnote{Note that the Wigner function is a phase space density, so changing variables $(\vph_\k,\pi_\k) \to (\Phi_\k,\Pi_\k)$ requires re-normalizing.  But we're not worrying about normalization right now.}
\be\label{W_0}
W_0(\Phi_\k,\Pi_\k;k,\tau) \propto \exp[-\sqz_0^2(2k^3)|\Phi_\k|^2 - \sqz_0^{-2}(2k^{-3})|\Pi_\k|^2],
\ee
where the squeezing ratio is 
\be\label{sqz}
\sqz_0 \equiv -1/k\tau
\ee
up to corrections in $k\tau$.
We show the contour of the Wigner function in Figure \ref{fig:wigner}. At late times, $\Pi_\k\approx \pi_\k$, and $\Pi_\k$ has a spread that scales as $1/|k\tau|$, as is clear also from Eq.~\eqref{eq:newvarsolutions}. On the other hand, the spread of $\Phi_\k$ is suppressed by $1/k\tau$ relative to the spread in $\vph_\k$, revealing the squeezing of the state.

As mentioned in \S~\ref{sec:WFinflation}, the squeezing is a consequence of the phase oscillations in the wavefunction, Eq.~\eqref{psi_k}, which come from the action.  Indeed, evaluating the action, $S=\frac{1}{2} \int \! d^3\x d\tau\, \tau^{-2} ((\vph')^2 - (\nab\vph)^2)$, for a single pair of modes $(\vph_\k,\vph_{-\k})$ in the superhorizon regime where $\vph_\k\rarr{\rm const.}$ and the gradient term dominates, we have
\be\label{action_phase} 
i S_{\k,-\k}(\vph_\k)\approx i (-k^2)|\vph_\k|^2\int^\tau\frac{d\tau}{\tau^2} = -k^3 \(\frac{-i}{k\tau}\)|\vph_\k|^2,
\ee
This is the phase in the wavefunction, Eq.~\eqref{psi_k}, and is enhanced relative to the real part of the exponent by the squeezing factor $1/|k\tau|$.

\section{Inflationary Epoch with Gravitational Interactions}
\label{sec:inflation}

In this section we will study the evolution of the wavefunction during inflation, including the entanglement between different scales generated by gravitational interactions.  After looking at the oversimplified case of redundant records of an infinite-wavelength background fluctuation in \S~\ref{sec:long_modes_inflation}, and reviewing the physical meaning of a long-wavelength fluctuation for different field variables in \S~\ref{sec:long_modes_examples}, we will perturbatively evolve the wavefunction in \S~\ref{sec:WF_nonlin_inflation} and show in \S~\ref{sec:no_records_inflation} that a long-wavelength fluctuation is not recorded by any one spatially localized short mode (even though short modes collectively do contain long-wavelength information), and is not recorded in more than one spatial region.

We will use a generic model for the leading gravitational interactions in an inflationary background:
\be\label{cubic_model} 
\L_3(\x) \equiv H^2 a^2(\tau) \(g\vph(\nab\vph)^2 + \tilde{g} \vph\vph'^2 \),
\ee
defined with respect to conformal time $\tau$.
The small dimensionless couplings $(g,\tilde{g})$ are left as free parameters. We will implicitly specify them when we specify $\vph$ as a field variable in a particular gauge in \S~\ref{sec:long_modes_examples}, which determines the couplings in terms of the inflationary background. 
The factor of $H^2$ ensures that the dimensionless couplings do indeed quantify the relative amplitude of the interactions in comparison to the free Lagrangian. (In particular, we define $(g,\tilde{g})$ to include any parametric suppression in the action from $H/\Mp$ as well as from slow roll parameters.)

This model captures the leading gravitational interactions in the case of slow-roll inflation, whether we consider $\vph$ to be a rescaled version of the adiabatic curvature $\zeta$ in the uniform density gauge, or of inflaton fluctuations $\d\phi$ in the spatially flat gauge.
Moreover, the two (massless) graviton modes can be described with the same model at the linear level, and, as noted in \S~\ref{sec:long_modes_examples}, gravitational interactions involving tensor modes are of the same form as Eq.~\eqref{cubic_model}.

\subsection{Cubic Interactions and Long-wavelength Fluctuations}
\label{sec:long_modes_inflation}

Consider a constant background fluctuation, $\vph_L=\rm const$. The quadratic Lagrangian for the modes will be shifted via their coupling to the $\vph_L$ in Eq.~\eqref{cubic_model}:
\be\label{L_shift_vph} 
\L_2(\x)+\Delta\L_2(\x) = \frac{1}{2\tau^2} \[(1+2\tilde{g}\vph_L)\vph'^2 - (1-2g\vph_L)(\nab\vph)^2 \].
\ee
Now, recall that in the Schr\"{o}dinger evolution of the Gaussian wavefunction, the kinetic and gradient terms appear respectively as the first and second terms on the RHS of Eq.~\eqref{A_eom_inflation}. In the superhorizon regime ($|k\tau|\ll1$), the phase of the wavefunction, ${\rm Im}A$, is sourced only by the gradient term:
\be 
\frac{d}{d\tau}{\rm Im}A(k,\tau) = k^2/\tau^2 + \O(\tau^0).
\ee 
and will thus be modified by the coupling to $\vph_L$:
\be
\frac{d}{d\tau}{\rm Im}A(k,\tau) \rarr (1-2g\vph_L)k^2/\tau^2 + \O(\tau^0) \ \ \ \ \text{for} \ \ g\vph_L\neq0.
\ee
Therefore, in the presence of a long-wavelength background, the phase grows at a slightly different rate:
\be\label{ImA_shift}
{\rm Im}A(k,\tau) \rightarrow (1-2g\vph_L) {\rm Im}A(k,\tau).
\ee

A small fractional shift in the phase is enough to shift the wavefunction to a completely orthogonal state: As we saw in \S~\ref{sec:wigner}, a large phase is equivalent to a high degree of squeezing, and even a small change in the squeezing factor alters the state to a non-overlapping state. This can be seen by computing the overlap of two wavefunctions, conditioned on two different long-wavelength backgrounds $\vph_L$ and $\vph_L+\Delta\vph_L$,
\be
\tensor[_{\vph_L}] {\bra{\psi_\k(\tau)}\ket{\psi_\k(\tau)}}{_{\vph_L+\Delta\vph_L}} =  \int d^2\vph_\k \psi(\vph_\k;\tau|\vph_L)\psi^*(\vph_\k;\tau|\vph_L+\Delta\vph_L),
\ee
where $\psi(\vph_\k;\tau|\vph_L)$ is evaluated with the shifted phase, Eq.~\eqref{ImA_shift}.
(The real part, ${\rm Re}A$, is also slightly perturbed for nonzero $g$ and $\tilde{g}$. However, since both $\<\vph\vph\>$ and $\<\pi\pi\>$ scale as $1/{\rm Re}A$, this does not affect the angle of orientation of the Wigner function, and thus cannot shift the state to an orthogonal state, as in Figure \ref{fig:superposition-and-clocks}.)
The overlap falls to zero for an increasingly small difference $\Delta\vph_L$ in the late-time limit. 
From Eqs.~\eqref{psi_k}, \eqref{A_k}, and \eqref{ImA_shift}, we obtain\footnote{It is important that $|g\vph_L|\ll1$ here.}
\be\label{time_shift_1mode} 
\frac{|\tensor[_{\vph_L}] {\bra{\psi_\k(\tau)}\ket{\psi_\k(\tau)}}{_{\vph_L+\Delta\vph_L}}|^2}{|\bra{\psi_\k(\tau)}\ket{\psi_\k(\tau)}|^2} \approx \[1+\frac{g^2\Delta\vph_L^2}{k^2\tau^2} \]^{-1}\rarr \(\frac{k\tau}{g\Delta\vph_L}\)^2 \ \ \text{for} \ \ |k\tau|\ll |g\Delta\vph_L|.
\ee
So we see that -- as shown earlier in Figure \ref{fig:superposition-and-clocks} -- even a small shift $g\Delta\vph_L\ll1$ tilts the squeezed state to an orthogonal state, once $|k\tau|\lesssim g\Delta\vph_L$.
In other words, the mode becomes sensitive to a long-wavelength background $\vph_L$ once it has redshifted $\sim\ln(1/g\vph_L)$ $e$-folds beyond the horizon. For a typical long-wavelength background $\vph_L=\O(\<\vph_L^2\>^{1/2})$, this is $\sim\ln(1/g)$ $e$-folds.
Because there are many modes which become squeezed, the coupling to a constant background $\vph_L$ generates many records of $\vph_L$ in momentum space.\footnote{Note that while modes that have just crossed the horizon, $|k\tau|\sim1$, are only slightly squeezed and individually contain very little information, there are vastly more such modes (in three spatial dimensions). These Hubble-scale modes are the dominant contribution to decoherence of $\vph_L$ \cite{Nelson:2016kjm}, and \textit{collectively} contain many more records of $\vph_L$. However, in this paper we focus (for simplicity and clarity) on records in modes that have evolved far outside the horizon and become highly squeezed.}

In this section we have so far considered the oversimplified case of an interaction which couples modes only to an infinite-wavelength background, ignoring interactions between finite-wavelength modes. In reality, local interactions couple all momentum-conserving combinations of modes. We will see that the entanglement between modes in a scale-invariant spectrum compromises a single mode's sensitivity to a long-wavelength background. In \S \ref{sec:no_records_inflation} below we will evolve the inflationary wavefunction and compute the (lack of) recorded long-wavelength information.

In \S~\ref{decelerating-section}, we will see that the post-inflationary evolution of the wavefunction in a decelerating epoch modifies the entanglement in a way that generates records similar to the simplified momentum-space records described above by simultaneously amplifying the correlation with long-wavelength modes and distributing records into many spatial regions.

\subsection{Long-wavelength Fluctuations: Effects on Local Geometry}
\label{sec:long_modes_examples}

In this section we review the physical meaning of Eqs.~\eqref{cubic_model} and \eqref{L_shift_vph} for gravitational interactions (which fix $g$ and $\tilde{g}$ in terms of the background cosmology) and for different field variables during inflation.

\textit{Scalar Curvature Perturbation $\zeta$.}

For the scalar curvature perturbation $\zeta$ as defined in the comoving (uniform density) gauge, with hypersurfaces of uniform matter density, a constant background fluctuation shifts the quadratic action as
\be\label{L2_shift_zeta}
\L^{(\zeta)}_{2,t}(\x)+\Delta \L^{(\zeta)}_{2,t}(\x)= \eps a^3 \(1+3\zeta_L\) \dot{\zeta}^2 - \eps a(1+\zeta_L)(\nab\zeta)^2,
\ee
as first derived in \cite{Maldacena:2002vr}, and reviewed in \App~\ref{sec:cubic_zeta}.
The long-wavelength background $\zeta_L$ only appears as a perturbation to the local scale factor, $a(t)\rarr(1+\zeta_L)a(t)$.
(In this case, we have written the Lagrangian with respect to cosmic time (hence the label $t$), for which it is easier to see this effect.)
Consequently, the conditional wavefunction for a short-wavelength mode with wavenumber $\k$ in the presence of a constant background will be shifted in conformal time:
\be\label{conditional_state}
\ket{\psi_\k(\tau)}_{\zeta_L} \approx \ket{\psi_\k(\tau(1-\zeta_L))},
\ee
at leading order in $\zeta_L$.
This shift backward or forward in time increases or decreases the uncertainty in the time derivative $\dot{\zeta}_\k$ for shorter modes, which is decaying to zero as the modes freeze out after horizon crossing. In the superhorizon regime, $\dot{\zeta}$ can be written as an operator in terms of $\zeta$ \cite{Assassi:2012et},
\be\label{zeta_dot_assassi}
\dot{\hat{\zeta}} = H^{-1} a^{-2}e^{-2\hat{\zeta}} \(\nab^2\hat{\zeta} + \frac{1}{2}(\nab\hat{\zeta})^2 \) + \O(\eps,\nab^4),
\ee
where the corrections include the parts that do not commute with $\zeta$, and we use hats to emphasize that the equality is at the operator level.  We see that when a long-wavelength background $\zeta_L$ is turned on,
\be
\dot{\hat{\zeta}} \rarr e^{-2\zeta_L}\dot{\hat{\zeta}},
\ee
indicating the long-wavelength modulation of freezing out of short modes.
From Eq.~\eqref{zeta_dot_assassi}, the squeezed-limit ($q\ll k,k'$) three-point function  between a long-wavelength $\zeta$ mode and two shorter-wavelength $\dot{\zeta}$ modes is
\be\label{zdotdot_squeezed_maintext}
\<\zeta_\q\dot{\zeta}_\k\dot{\zeta}_{\k'}\> \approx -4 \<\zeta_\q\zeta_{-\q}\>\<\dot{\zeta}_\k\dot{\zeta}_{-\k}\>, \hspace{1cm} (q\ll k \ll aH=-1/\tau)
\ee
up to corrections of order $q/k$.

The cubic interactions shown in Eq.~\eqref{L2_shift_zeta} do not lead to observable correlations in the late-time three-point function. In particular, the resulting squeezed-limit three-point function $\<\zeta_{\q}\zeta_\k\zeta_{\k'}\>$ %
is zero because the two interactions cancel each other. (As discussed in \App~\ref{sec:cubic_zeta}, the full three-point function (at $\tau\rarr0$) is suppressed by slow-roll parameters, and receives no contributions from the cubic interactions shown above.)
As the modes freeze out as $|\Psi[\zeta]|^2\rarr{\rm const.}$, correlation functions involving only $\zeta$ become time-independent, and are unaffected by the effective shift in time from $\zeta_L$ in Eq.~\eqref{L2_shift_zeta}.
However, because the conjugate momentum (or field velocity) is highly time-dependent, correlation functions involving the conjugate momentum or velocity -- such as the three-point function $\<\zeta\dot{\zeta}\dot{\zeta}\>$ (or $\<\zeta\pi\pi\>$) -- and are sensitive to the shift in time from $\zeta_L$.
These correlation functions probe the additional phase information not contained in the classical distribution $|\Psi[\zeta]|^2$.

\textit{Inflaton Perturbations $\d\phi$.}

In the spatially flat gauge, the metric is unperturbed and the scalar fluctuations are described in the matter sector, which we will temporarily assume to be the source of inflation via a potential $V(\phi)$ of a scalar field $\phi$. We will see that a long-wavelength inflaton background $\d\phi_L$ acts as a shift in the energy scale or Hubble rate.

From the Friedmann equation, $H^2=V(\bar{\phi}+\d\phi)/3\Mp^2$, along with the slow-roll parameter $\eps=-\dot{H}/H^2=\frac{1}{2}\Mp^2(V'/V)^2$, infinitesimal changes in $H$ and $V$ from a shift $\d\phi_L$ to the background field are related as\footnote{We assume that $\dot{\phi}>0$, so $\d\phi>0$ is associated with a shift forward in time and down the potential.} 
\be\label{phi_gauge_H_shift}
\frac{\Delta H}{H} = \frac{1}{2}\frac{\Delta V}{V} = - \eps\frac{\d\phi_L}{\sqrt{2\eps}\Mp}.
\ee
Furthermore, an increase in $H$ slightly shrinks the horizon scale, which effectively expands the modes relative to the horizon and thus shifts the modes forward in time or scale factor:
\be\label{time_H_shift}
\(\frac{\Delta a}{a}\)_{\rm eff} = \frac{\Delta H}{H}.
\ee
From the cubic interactions in the spatially flat gauge, Eq.~(3.8) in \cite{Maldacena:2002vr}, it is straightforward to check how the long-wavelength background shifts the quadratic Lagrangian \footnote{As in $\zeta$ gauge, we keep only the cubic terms with a factor of $\d\phi$ without derivatives. Other terms can be neglected in the limit of a constant background field.}:
\be
\L_{2,t}^{(\d\phi)}(\x)+\Delta\L_{2,t}^{(\d\phi)}(\x) = \frac{1}{2}a^3\(1 - \frac{\eps\d\phi_L}{\sqrt{2\eps}\Mp}\)\dot{\d\phi}^2 - \frac{1}{2}a\(1 + \frac{\eps\d\phi_L}{\sqrt{2\eps}\Mp}\)(\nab\d\phi)^2.
\ee
From Eqs.~\eqref{phi_gauge_H_shift} and \eqref{time_H_shift} we can write this as:
\be
\L_{2,t}^{(\d\phi)}(\x)+\Delta\L_{2,t}^{(\d\phi)}(\x) = \(\frac{H+\Delta H}{H}\)^{-2} \[\frac{1}{2}(a+\Delta a_{\rm eff})^3\dot{\d\phi}^2 - \frac{1}{2} (a+\Delta a_{\rm eff}) (\nab\d\phi)^2 \].
\ee
This is exactly what we expect from thinking of $\d\phi_L$ as a shift in $H$ (and hence in the effective ``time'' $a_{\rm eff}$): The overall factor in parentheses shifts the amplitude of inflaton fluctuations, $\<\d\phi^2\>\sim H^2 \rarr (H+\Delta H)^2$, which is the same as shifting the Hubble rate. In addition, the scale factor is shifted, which will advance or retard the modes in their evolution. %

\textit{Tensor Modes.}

The leading cubic interactions 
between two scalar and one graviton mode, and one scalar and two graviton modes, are respectively \footnote{In Eqs.~\eqref{L_zzg} and \eqref{L_ggz} we set $\Mp\equiv1$ and have neglected derivative and slow-roll suppressed interactions, as for the scalar mode self-interactions. To obtain the $\gamma\gamma\zeta$ interactions in Eq.~\eqref{L_ggz}, one must integrate by parts the interactions in \cite{Maldacena:2002vr} several times, in the same way as for the $\zeta\zeta\zeta$ interactions discussed in \App~\ref{sec:cubic_zeta}.} \cite{Maldacena:2002vr}
\ba
\L_3^{(\zeta\zeta\gamma)}(\x) &\approx& \eps a \gamma_{ij}\del_i\zeta\del_j\zeta, \label{L_zzg} \\
\L_3^{(\gamma\gamma\zeta)}(\x) &\approx& \frac{1}{8}\(3a^3\zeta\dot{\gamma}_{ij}\dot{\gamma}_{ij} - a\zeta(\nab\gamma_{ij})^2 \). \label{L_ggz}
\ea
The resulting effect from Eq.~\eqref{L_zzg} of a long-wavelength tensor background fluctuation $\gamma_{ij,L}$ on short-wavelength scalar modes is to create a slightly anisotropic background spacetime which distorts the scalar modes $\zeta_\k$ by shifting their wavenumbers, $k^2\rarr k^2 - \gamma_{ij,L}k_i k_j$, which shifts the angle of orientation of the Wigner function. Note that scalar modes propagating in the same direction as the gravitational wave background satisfy $\gamma_{ij,L}k_i k_j=0$, and are not affected.
Similarly, the effect from Eq.~\eqref{L_ggz} of a long-wavelength scalar curvature fluctuation $\zeta_L$ on short-wavelength tensor modes is once again an effective shift in the scale factor or time.

\subsection{Nonlinear Evolution of the Wavefunction}
\label{sec:WF_nonlin_inflation}

We would like to quantify the recorded information about long-wavelength fluctuations in terms of the entanglement structure of the wavefunctional $\Psi[\vph(\x)](\tau)$. In this section, we review the time evolution of the wavefunction in the presence of small interactions, which can be solved perturbatively in the interaction strength.

Any interactions will generate a non-Gaussian part of the wavefunction, which we parametrize as
\be
\Psi[\vph]\equiv\Psi_G[\vph]\cdot\Psi_{NG}[\vph],  \nn
\ee
where the Gaussian part is
\be
\Psi_G[\vph](\tau) = {\rm (norm.)} \exp\[-\frac{1}{2}\int_\k \vph_\k\vph_{-\k}A(k,\tau) \],
\ee
and $\Psi_{NG}$ captures the effects of interactions and has the form
\be\label{Psi_NG}
\Psi_{NG}[\vph](\tau) \approx \exp\[\int_{\q,\k,\k'}(2\pi)^3\d^3(\q+\k+\k')\vph_\q\vph_\k\vph_{\k'}\F_{\q,\k,\k'}(\tau) \].
\ee
The complex, time-dependent function $\F$ is isotropic, depending only on the wavenumber magnitudes $q,k,k'$, and
\be
\int_{\q,\k,\k'}\equiv\int\frac{d^3\q}{(2\pi)^3}\frac{d^3\k}{(2\pi)^3}\frac{d^3\k'}{(2\pi)^3}, \nn
\ee
with the delta function enforcing translation invariance ($\Psi[\vph(\x)]=\Psi[\vph(\x+\x_0)]$). The evolution of $\F$ is determined by the Schr\"{o}dinger equation
\be
i\frac{d}{d\tau} \Psi[\vph](\tau) = \( \H[\vph,-i\d/\d\vph](\tau) \) \Psi[\vph](\tau),
\ee
with $\pi=-i\d/\d\vph$.
Expanding the Schr\"{o}dinger equation perturbatively, one obtains an equation of motion for $\F$ \cite{Nelson:2016kjm},
\be\label{F_eom}
\F'(\tau) - i\alpha(\tau)\F(\tau) = -i \H^{(\rm int)}(\tau),
\ee
where we have suppressed the wavenumber dependence. The source function $\H^{(\rm int)}_{\k,\k',\q}(\tau)$ is defined by the action of the interacting Hamiltonian with respect to conformal time\footnote{Note that this Hamiltonian differs by a scale factor from the cosmic time Hamiltonian used in \cite{Nelson:2016kjm}.}, $\H^{\rm int}=\int\! d^3\x \,\H^{\rm int}(\x) = -\int\! d^3\x\,\L^{\rm int}(\x)$ \footnote{The Hamiltonian density $\H=\pi\vph'-\L$ receives quadratic corrections to the momentum $\pi$, which generate cubic terms in both $\pi\vph'$ and the free Lagrangian. However, these cancel, so we are left with $\H^{\rm int}=-\L^{\rm int}$ at cubic order.}, on the Gaussian wavefunction:
\be\label{H_int_source_def}
\(\H^{\rm int}[\vph,-i\d/\d\vph]\Psi_G\)\big|_\tau \equiv \[ \int_{\k,\k',\q}(2\pi)^3 \d^3(\q+\k+\k') \vph_\k\vph_{\k'}\vph_{\q} \(\H^{(\rm int)}_{\k,\k',\q}(\tau)\) \] \Psi_G.
\ee
Eq.~\eqref{F_eom} can be solved explicitly, yielding
\be\label{F_gen_solution}
\F_{k,k',q}(\tau) =-i \int_{-\infty}^\tau d\tau' \H^{(\rm int)}_{\k,\k',\q}(\tau')\exp\[i\int_{\tau'}^\tau d\tau''\alpha_{k,k',q}(\tau'')\].
\ee
Here, the function
\be
\alpha_{k,k',q}(\tau) \equiv - \tau^2 (A(k,\tau) + A(k',\tau) + A(q,\tau))
\ee
is determined by the linear dynamics, where $A(k,\tau)$ was introduced in Eq.~\eqref{A_k} and defines the Gaussian wavefunction $\Psi_G$.

The calculation of $\F$ for the $\vph(\nab\vph)^2$ interaction is given in \cite{Nelson:2016kjm}. The real part of $\F$ asymptotes to a constant function as $\tau\rarr0$, and determines the late-time three-point function $\<\vph\vph\vph\>$, which is hoped to be detectable as primordial non-Gaussianity in post-inflationary observables.
On the other hand, the imaginary part ${\rm Im}\F$ is sourced by the real part of $\H^{(\rm int)}_{\k,\k',\q}$, which grows with the scale factor. It contributes to the WKB phase, shown in Eq.~\eqref{psi_WKB} in terms of the curvature perturbation $\zeta$.
In the limit where all modes are in the superhorizon regime, the solution is
\be 
{\rm Im}\F_{k,k',q}(\tau) = - \frac{g}{3\tau} (k^2+k'^2+q^2), \hspace{1cm} (q,k,k'\ll 1/|\tau|)
\label{ImF_inflation}
\ee
which is essentially the cubic interaction in Fourier space.
The effect of ${\rm Im}\F$ for a constant background, $q\rarr0$, is a shift to the Gaussian phase ${\rm Im}A$, which easily deforms the state to a non-overlapping state.

Note that the $\tg\vph\vph'^2$ interaction does not generate a rapidly growing phase during inflation. This is because the modes freeze out, with $\vph'_\k(\tau)\propto k\tau\rarr0$, so the interaction does not grow in time.\footnote{Despite the freezing of individual modes on super-Hubble scales, the $\tg\vph\vph'^2$ does contribute to decoherence of long-wavelength fluctuations due to the growing bath of Hubble-scale modes, which collectively (although not individually) contain an increasing amount of information.}
The absence of a phase from the $\tg\vph\vph'^2$ interaction recovers the result of \S~\ref{sec:long_modes_inflation} that the effect of $\tg$ on the quadratic Lagrangian in the presence of a long-wavelength background, in Eq.~\eqref{L_shift_vph}, does not shift the short modes to an orthogonal state. %
However, the $\tg\vph\vph'^2$ will become relevant after inflation when modes re-enter the horizon and unfreeze.

\subsection{Localized Gaussian Modes}

In order to quantify spatially localized records in the evolving wavefunction, we introduce localized field modes, defined by convolving the field with a Gaussian window function:
\begin{align}\begin{split}\label{phi_pi_wavepacket}
\vph^{(\sigma)}(\k,\x) \equiv \int_{\k'} W(\k'|\k,\x) \vph_{\k'}, \hspace{1cm}
\pi^{(\sigma)}(\k,\x) \equiv \int_{\k'} W(\k'|\k,\x) \pi_{\k'},
\end{split}\end{align}
where the convolution function, 
\be
W(\k'|\k,\x) = (2\sqrt{\pi}\sigma)^{3/2} \times\exp\[-\frac{1}{2}\sigma^2|\k'-\k|^2 + i\k'\cdot\x\],
\ee
picks out Fourier modes near wavenumber $\k$, and spatial modes $\vph(\y)=\int_\k e^{i\k\cdot\y}\vph_\k$ within a distance $\sim\sigma$ of $\x$.  
The normalization factor is chosen so that $\int_{\k'} |W(\k'|\k,\x)|^2 = 1$, which fixes the canonical commutation relation
\begin{align}
\[\vph^{(\sigma)}(\k,\x),\pi^{(\sigma)}(\p,\y) \] &= \int_{\k',\p'} W(\k'|\k,\x)W(\p'|\p,\y) \[\vph_{\k'},\pi_{\p'}\] \\
&= i\hbar \int_{\k'} W(\k'|\k,\x)W(-\k'|\p,\y) \\
&= i\hbar\exp\[-\frac{1}{4\sigma^2}|\x-\y|^2-\frac{\sigma^2}{4}|\k+\p|^2+\frac{i}{2}(\k-\p)\cdot(\x-\y)\],
\end{align}
using $[\vph_{\k'},\pi_{\p'}] = i\hbar(2\pi)^3 \delta^3(\k'+\p')$.
In particular, this reduces to $\[\vph^{(\sigma)}(\k,\x),\pi^{(\sigma)}(-\k,\x) \]=i\hbar$ for $\x=\y$ and $\k=-\p$.
For reference, we also define the Fourier transformed window function
\be\label{Wtilde}
\tilde{W}(\y|\k,\x) \equiv \int_{\k'}e^{-i\k'\cdot\y}W(\k'|\k,\x) = \frac{1}{\pi^{3/4}\sigma^{3/2}}\exp\[-\frac{|\x-\y|^2}{2\sigma^2}+i\k\cdot(\x-\y)\].
\ee
Like the Fourier modes $\vph_{-\k} = \vph_{\k}^\dagger$, the localized modes obey $\vph^{(\sigma)}(-\k,\x) = \vph^{(\sigma)}(\k,\x)^\dagger$ because $W(-\k'|-\k,\x) = W(\k'|\k,\x)^*$.  They can similarly be broken into two degrees of freedom, the real and imaginary parts
\begin{eqnarray}\begin{split}
\vph_{S,r} &\equiv& \frac{1}{\sqrt{2}} \(\vph^{(\sigma)}(\k,\x) + \vph^{(\sigma)}(-\k,\x) \),  \\
\vph_{S,i} &\equiv& \frac{1}{\sqrt{2}i} \(\vph^{(\sigma)}(\k,\x) - \vph^{(\sigma)}(-\k,\x) \). \label{vph_S_def}
\end{split}\end{eqnarray}
$S=S(\k,\x,\sigma)$ refers to short wavelengths, since we will be interested in the limit where the peak wavelength $2\pi/k$ is much shorter than the spatial scale $\sigma$ over which the localized mode is spread, which will in turn be shorter than the long-wavelength modes being recorded. At the same time, we assume the peak wavelength is much longer than the horizon scale, $|k\tau|\ll1$, so that the mode is highly squeezed.

With these definitions, we will be able to quantify the amount of information recorded by modes of the field in spatially localized regions.

\subsection{Long-wavelength Influence on Localized Short-wavelength Modes}
\label{sec:no_records_inflation}

In this section, we will quantify the amount of long-wavelength information recorded in spatially localized modes of the field on superhorizon scales.
It will be convenient to use the rotated canonical variables $(\Phi,\Pi)$, for which we once again define localized Gaussian variables,
\begin{align}\begin{split}
\Phi^{(\sigma)}(\k,\x,\tau) &\equiv \int_{\k'} W(\k'|\k,\x) \Phi_{\k'}(\tau), \\
\Pi^{(\sigma)}(\k,\x,\tau) &\equiv \int_{\k'} W(\k'|\k,\x) \Pi_{\k'}(\tau). \label{Phi_Pi_wavepackets}
\end{split}\end{align}
Just as in Eq.~\eqref{vph_S_def}, the corresponding real and imaginary parts are
\begin{align}\begin{split}\label{Phi_Pi_S_def}
\Phi_{S,r} &\equiv \frac{1}{\sqrt{2}} \(\Phi^{(\sigma)}(\k,\x) + \Phi^{(\sigma)}(-\k,\x) \), \qquad\qquad \Phi_{S,i} \equiv \frac{1}{\sqrt{2}i} \(\Phi^{(\sigma)}(\k,\x) - \Phi^{(\sigma)}(-\k,\x) \), \\
\Pi_{S,r} &\equiv \frac{1}{\sqrt{2}} \(\Pi^{(\sigma)}(\k,\x) + \Pi^{(\sigma)}(-\k,\x) \), \qquad\qquad \Pi_{S,i} \equiv \frac{1}{\sqrt{2}i} \(\Pi^{(\sigma)}(\k,\x) - \Pi^{(\sigma)}(-\k,\x) \).
\end{split}\end{align}
It is straightforward to check from Eqs.~\eqref{Phi_Pi_two_pt} and \eqref{Phi_Pi_wavepackets} that 
\ba
\<\Phi_{S}^2\>(k,\tau) \equiv \<\Phi_{S,r}^2\>(k,\tau) &=& \<\Phi_{S,i}^2\>(k,\tau) = \frac{1}{2} \frac{(k\tau)^2}{k^3} + \O(1/\sigma k), \nn \\
\<\Pi_{S}^2\>(k,\tau) \equiv \<\Pi_{S,r}^2\>(k,\tau) &=& \<\Pi_{S,i}^2\>(k,\tau) = \frac{1}{2}\frac{1}{(k\tau)^2} k^3  + \O(1/\sigma k), \label{Phi_Pi_2pt_linear}
\ea
and by definition,
\be
\<\overline{\Phi_S\Pi_S}\> \equiv \<\overline{\Phi_{S,r}\Pi_{S,r}}\> = \<\overline{\Phi_{S,i}\Pi_{S,i}}\> = 0.
\ee
Here we have defined these quantities without the $r$ and $i$ subscripts, since they are the same in either case. We will omit these subscripts in general, with the understanding that expressions hold for both of the two modes.
Note that $\<\Phi_S^2\>\<\Pi_S^2\>=\<|\Phi_\k|^2\>\<|\Pi_\k|^2\>+\O(1/\sigma k)$, with the $\O(1/\sigma k)$ corrections capturing the mixing due to entanglement between spatial regions. This vanishes in the $\sigma\rarr\infty$ limit, in which case we recover the case of pure, unentangled Fourier modes, saturating the uncertainty principle.

We will quantify the amount of long-wavelength information carried by a localized short mode in terms of its conditional two-point functions. In particular, the overlap between the reduced state for a localized short mode \textit{conditioned} on a particular background of long-wavelength fluctuations, and the reduced state for the same mode conditioned on a different background (as depicted in Figure \ref{fig:superposition-and-clocks}), is determined by their respective two-point functions. 

\begin{figure}
\includegraphics[scale=0.45]{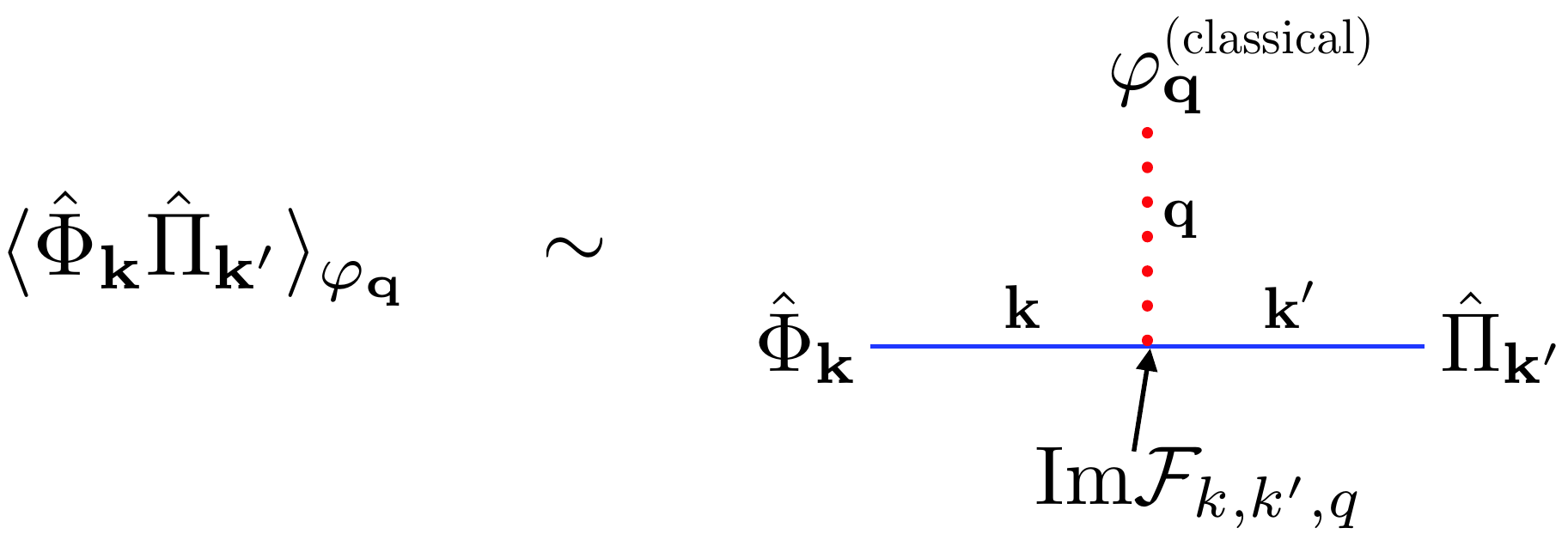}
\caption{A Feynman-like diagram for the two-point function $\<\Phi_\k\Pi_{\k'}\>_{\varphi_L}(\tau)$, conditioned on a classical configuration of long-wavelength modes $q\ll k,k'$. The short-wavelength modes become correlated via the classical mode, represented as an external leg with soft momentum $\q=-\k-\k'$. The vertex function is given by the phase (imaginary part) of the cubic part of the wavefunction, ${\rm Im}\F_{k,k',q}(\tau)$. (We use hats in this figure to emphasize that $\Phi$ and $\Pi$ are quantum operators in contrast to the classical mode being conditioned on.)}
\label{fig:conditional_2pt_diagram}
\end{figure}

We will start by computing the conditional correlator $\<\overline{\Phi_S \Pi_S} \>_{\vph_L}\equiv\frac{1}{2}\<(\Phi_S\Pi_S+\Pi_S\Phi_S)\>_{\vph_L}$ for a localized short-wavelength mode.
It is easiest to first find the corresponding Fourier space correlator, $\<\overline{\Phi_\k\Pi_{\k'}}\>_{\varphi_L} \equiv \frac{1}{2} \<(\Phi_\k\Pi_{\k'}+\Pi_\k\Phi_{\k'})\>_{\varphi_L}$, and then convolve to localized Gaussian modes.
In the Schr\"{o}dinger picture,
\be
\<\hat{\O}[\vph,\pi]\> = \int D\vph \Psi^*[\vph]\O[\vph,-i\d/\d\vph]\Psi[\vph]
\ee
for any well-defined operator $\O$.
To condition on a long-wavelength configuration, we treat all modes $\vph_\q$ for $q<q^\star$ for a cutoff scale\footnote{We would like to ask whether short modes $k$ respond \textit{only} to sufficiently long-wavelength modes $q\lesssim1/\sigma$.  Letting $q^\star>\sigma^{-1}$ and asking whether the dependence on conditional modes is sensitive to $q^\star$ will allow us to determine this.} $q^\star\gg\sigma^{-1}$ in the wavefunction as having fixed classical values, so that they can be factored outside of expectation values over the shorter modes.
Using Eq.~\eqref{A_k} for the Gaussian part $\Psi_G$, and ($\Phi,\Pi$) as defined in Eq.~\eqref{wigner_rotate}, a straightforward calculation gives the leading $\O(g)$ contribution.
For $|\k+\k'|\ll k,k'$, this is
\ba
\<\overline{\Phi_\k\Pi_{\k'}}\>_{\varphi_L}(\tau) &=& \vph_\q \frac{-3}{k^3}{\rm Im}\F_{k,k',q}(\tau) + \O(q/k,(k\tau)^0,g^2), \hspace{1cm} (\q\equiv-\k-\k') \\
&=& \vph_\q \frac{g}{k\tau} + \O(q/k,(k\tau)^0,g^2). \label{Phi_Pi_kspace}
\ea
In the second line, we have used Eq.~\eqref{ImF_inflation}.
We see that the two short modes $\k$ and $\k'$ are correlated via their coupling to one of the long-wavelength modes being conditioned on. Due to momentum conservation, $\k+\k'+\q=0$, information about long-wavelength modes (small $q$) is shared between modes oriented in nearly opposite directions. We will see that because the field is excited into a two-mode squeezed state, with pairs of entangled quanta occupying $\k$ and $-\k$ modes, this information is shared between EPR pairs of particles which propagate in opposite directions in the post-inflationary epoch.
Note also that contributions from the real part ${\rm Re}\F$ have cancelled, leaving only the phase ${\rm Im}\F$. This is because $\<\Phi\Pi\>$ captures the effect of long modes on the tilt in phase space of the Wigner function for the short mode; this depends only on the shift to the phase of the wavefunction (see \S~\ref{sec:long_modes_inflation}), which comes from ${\rm Im}\F$.
The factor of $g/k\tau$ in Eq.~\eqref{Phi_Pi_kspace} is suppressed by the weak inflationary interactions, but exponentially enhanced by the squeezing factor. This enhances the sensitivity to long-wavelength modes and prepares the wavefunction in a state susceptible to the formation of a spatially redundant records after inflation.
Eq.~\eqref{Phi_Pi_kspace} is represented diagrammatically in Figure \ref{fig:conditional_2pt_diagram}: the classical background mode $\vph_\q$ correlates the two short modes by attaching as an external leg with soft momentum, with the vertex function given by ${\rm Im}\F$.

Finally, convolving from Fourier modes to quasi-local Gaussian wavepackets, we find that 
\be 
\<\overline{\Phi_{S}\Pi_{S}}\>_{\varphi_L}(k,\tau) 
 = \frac{g}{k\tau} \vph_\sigma(\x) + \O\(q/k,(k\tau)^0\),
\label{Phi_Pi_final}
\ee
where we have defined the coarse-grained field
\be\label{vph_sigma}
\vph_\sigma(\x) \equiv \int_\q \vph_\q e^{i\q\cdot\x-\frac{1}{4}\sigma^2q^2},
\ee
which is essentially independent of the upper limit $q^\star>1/\sigma$.
Note once again that the long-wavelength information is shared between entangled $\k$ and $-\k$ modes (now localized to a finite region), as indicated in Eq.~\eqref{Phi_Pi_S_def}.

The conditional two-point function $\<\Phi_\k\Phi_{\k'}\>_{\varphi_L}$ can be calculated in the same way, leading to the result
\ba 
\label{Phi_Phi_kspace}
\<\Phi_\k\Phi_{\k'}\>_{\varphi_L}(\tau) &\approx& \frac{(k\tau)^2}{k^3} \int^{q^\star}_{\q,\q'} \vph_\q \vph_{\q'} (2\pi)^3\d^3(\k+\k'+\q+\q')\times\frac{18}{k^6} {\rm Im}\F_{k,|\k+\q|,q}(\tau){\rm Im}\F_{k',|\k'+\q'|,q'}(\tau) + \<\Phi_\k\Phi_{\k'}\>_0(\tau) \nn \\
&\approx& \frac{2g^2}{k^3} \int^{q^\star}_{\q,\q'} \vph_\q \vph_{\q'} (2\pi)^3\d^3(\k+\k'+\q+\q') +  \<\Phi_\k\Phi_{\k'}\>_0(\tau),
\ea
where we have again used Eq.~\eqref{ImF_inflation} in the second line, the upper limit denotes $|\q|,|\q'|<q^\star$, and $\<\Phi_\k\Phi_{\k'}\>_0=\<\Phi_\k\Phi_{\k'}\>_{\vph_L=0}$ is the contribution from entanglement with other short modes which are not being conditioned on.
Integrating over Fourier modes to obtain the two-point function for Gaussian modes, we find
\ba 
\<\Phi_{S}^2\>_{\varphi_L}(k,\tau)
&\approx& \frac{2g^2}{k^3} \int^{q^\star}_{\q,\q'}\vph_\q \vph_{\q'} e^{i(\q+\q')\cdot\x-\frac{1}{4}\sigma^2|\q+\q'|^2} + \<\Phi_{S}^2\>_0(k,\tau). \\
&\equiv& \frac{2g^2}{k^3} \(\vph^2(\x) \)_\sigma + \<\Phi_{S}^2\>_0(k,\tau), \label{Phi_Phi_final_inflation}
\ea
where
\be
\(\vph^2(\x) \)_\sigma \equiv \int_\p \(\vph^2\)_\p e^{i\p\cdot\x-\frac{1}{4}\sigma^2p^2} = \int^{q^\star}_{\q,\q'} \vph_\q \vph_{\q'} e^{i(\q+\q')\cdot\x-\frac{1}{4}\sigma^2|\q+\q'|^2}
\ee
is a coarse-grained version of the squared field, $\vph^2(\x)$. (In the middle expression, $(\vph^2)_\p$ denotes a Fourier mode of this field.)

Lastly, $\<\Pi_S^2\>$ is unaffected by interactions at leading order, and is approximately given by the free theory expression, Eq.~\eqref{Phi_Pi_two_pt}.

\begin{figure}
\includegraphics[scale=0.55]{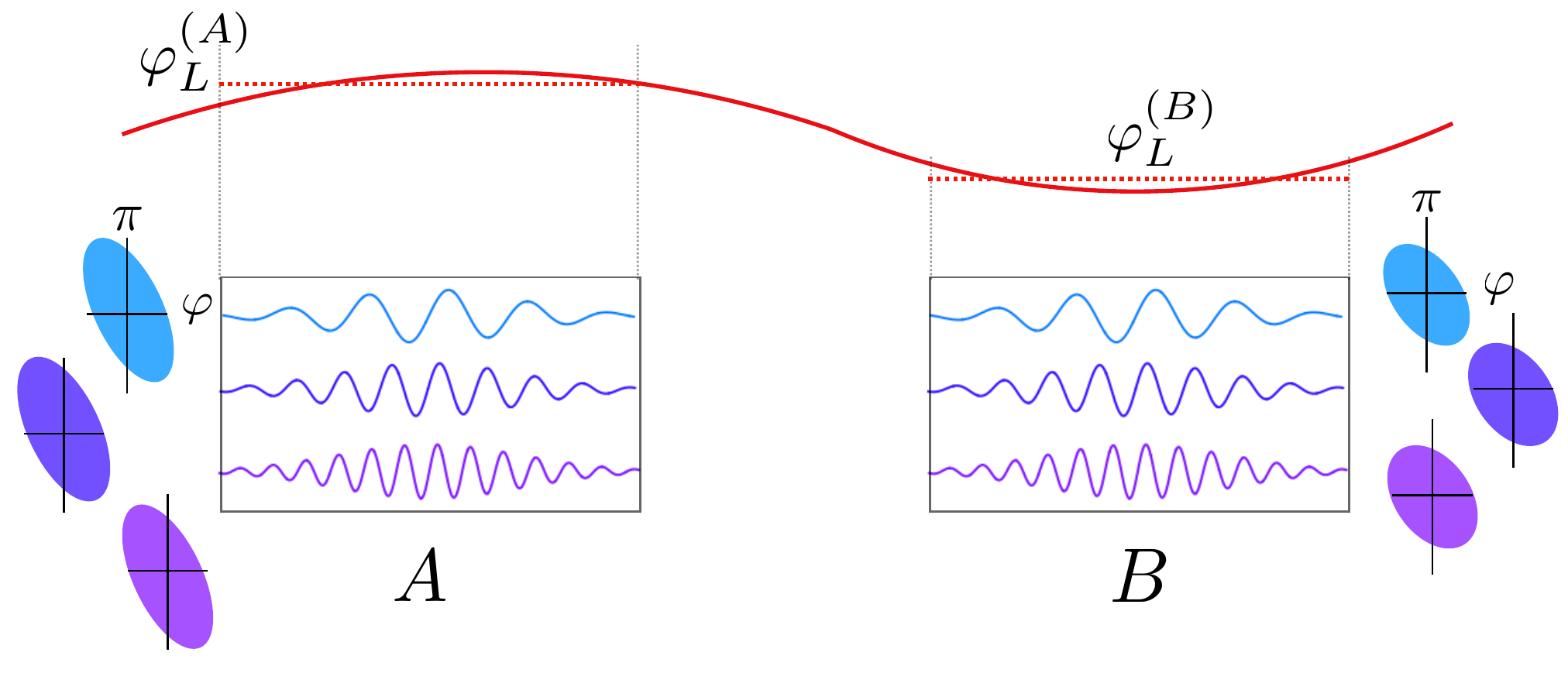}
\caption{A schematic depiction of entanglement of superhorizon modes during inflation. Modes localized to regions $A$ and $B$ become correlated with the respective long-wavelength backgrounds in each region after redshifting beyond the horizon, but cannot propagate spatially after this point, and thus cannot distribute information into spatially disjoint regions. Furthermore, the modes in region $A$ are entangled with each other (as are those in $B$). This increases the mixedness of their reduced states, which blurs out the corresponding Wigner functions in phase space, reducing their sensitivity to the long-wavelengh background. The information about $\vph_L^{(A)}$ or $\vph_L^{(B)}$ therefore remains inaccessibly entangled between the short modes in each respective region, and cannot be obtained from a single mode.}
\label{fig:realspace_no_records}
\end{figure}

In order to have a record, the Gaussian modes should be in distinguishable (i.e., orthogonal) states when conditioned on sufficiently different values of the background ${\varphi_L}$.  This is only possible if the area of support in phase space for a conditioned state, $\<\Phi_{S}^2\>_{\varphi_L}\<\Pi_{S}^2\>_{\varphi_L}-\<\overline{\Phi_{S}\Pi_{S}}\>_{\varphi_L}^2$, is much smaller than the area of support for the \emph{unconditioned} state $\<\Phi_{S}^2\>\<\Pi_{S}^2\>$.  (Recall $\<\overline{\Phi_{S}\Pi_{S}}\> = 0$.)  Achieving this requires a delicate cancellation between $\<\Phi_{S}^2\>_{\varphi_L}\<\Pi_{S}^2\>_{\varphi_L}$ and $\<\overline{\Phi_{S}\Pi_{S}}\>_{\varphi_L}^2$. However, from Eqs.~\eqref{Phi_Pi_final} and \eqref{Phi_Phi_final_inflation} we see that these two terms depend on different long-wavelength variables: the smoothed squared field, $(\vph^2(\x))_\sigma$, and the squared smoothed field, $\vph^2_\sigma(\x)$. (Relatedly, the contribution $\<\Phi_{S}^2\>_0$ from entanglement with (sufficiently short) unconditioned modes is not small compared to the contribution from conditioned modes.) For a scale-invariant power spectrum, these variables will be very different, as can be seen from their expectation values:
\ba
\<(\vph_\sigma)^2\> &=& \int_\q \frac{1}{2q^3} e^{-\frac{1}{2}\sigma^2 q^2}, \\
\<(\vph^2)_\sigma\> &=& \int^{q^\star}_\q \frac{1}{2q^3}.
\ea
While $\<(\vph_\sigma)^2\>$ only depends on modes longer than $\sigma$, $\<(\vph^2)_\sigma\>$ depends on modes at all scales up to the cutoff $q^\star$. This is because any two short-wavelength modes $\vph_\q$ and $\vph_{\q'}$ can contribute to $\<(\vph^2)_\sigma\>$ as long as $|\q+\q'|\lesssim\sigma^{-1}$. Since squaring generates long-wavelength pieces from short modes, squaring and smoothing do not commute. Therefore, we find that \textit{for a scale-invariant power spectrum}, $q^3\<\vph_\q\vph_{-\q}\>\approx{\rm const.}$, localized short-wavelength modes \textit{cannot} distinguish between different long-wavelength backgrounds. In short, the spatial locality of the cubic interaction results in any momentum-conserving triplet of modes ($\k+\k'+\q=0$) becoming entangled, and it is impossible to isolate the entanglement between very long and very short modes, since the latter are entangled with other short modes. Consequently, the short modes only resolve the background with a very large error due to the 
fractional difference between $(\vph_\sigma)^2$ and $\vph_\sigma^2$, and do not record any precise information about the background.\footnote{Interestingly, we found that if the cubic interaction was replaced in an ad hoc manner with a \textit{nonlocal} cubic interaction which removed by hand the coupling between all modes except those separated by a hierarchy of scales, then short modes do indeed record the background with high precision. Furthermore, a post-inflationary decelerating epoch with \textit{linear} evolution allows these momentum-space records to propagate into many disjoint spatial regions, once the short modes re-enter the cosmological horizon at late times. However, for local interactions and slow-roll inflation, the scale-invariance of the power spectrum results in modes at all scales being entangled. As a result, it is impossible to isolate long-wavelength information in spatial regions after inflation, since long-wavelength information formed during inflation is shared between many entangled modes, all propagating in different directions.} This is depicted schematically in Figure \ref{fig:realspace_no_records}.

We will now move on to a post-inflationary era of deceleration, which will remedy this problem by changing the power spectrum once modes re-enter the cosmological horizon, and allowing information to be distributed into many spatial regions.

\section{Post-Inflation Decelerating Epoch}
\label{decelerating-section}

We impose an end to inflation at some finite conformal time $\tau_f <  0$, at which all the modes of interest are far outside the horizon and thus very squeezed, and study the evolution of the wavefunction in a post-inflationary decelerating epoch. This will be a radiation-dominated epoch, $a \sim \sqrt{t}$.  The scale factor can be written in terms of (post-inflationary) conformal time $\eta$, satisfying $d\eta = dt/a$, which runs from $0$ to $\infty$:
\be\label{a_radiation}
a(\eta) = \frac{-1}{H\tau_f} \(1 + \frac{\eta}{-\tau_f} \),
\ee
where $H$ is the Hubble rate during the preceding period of inflation. Here, we have fixed the initial scale factor $a(0) =a(\tau = \tau_f)= -1/H\tau_f$ and rate of change $a'(0) = a'(\tau = \tau_f)=1/H\tau_f^2$ so that the scale factor and its derivative evolve continuously through the transition.

We choose to consider a radiation era because the resulting evolution of the wavefunction is relatively simple to solve, but we expect that any well-behaved decelerating epoch in which the modes eventually re-enter the horizon will result in the same qualitative behavior: decay of the modes upon horizon re-entry, along with spatial propagation of the many quanta occupying a mode.

In the following subsections, we will describe the linear evolution of the wavefunction in the radiation era, and then repeat the calculation of \S \ref{sec:inflation} in order to quantify the amount of recorded information. The main result is given in \S \ref{sec:records_decel}, where we contrast it with the inflationary case (\S \ref{sec:no_records_inflation}), and describe the propagation of redundant information into disjoint spatial regions.

\subsection{Linear Evolution of the Wavefunction}
\label{linear_WF_rad}

The linear evolution of the Gaussian wavefunction for $\vph$,
\be
\Psi_G[\vph](\eta) = {\rm (norm.)} \exp\[-\frac{1}{2}\int_\k \vph_\k\vph_{-\k}A(k,\eta) \],
\ee
is described by the continued evolution of $A(k,\eta)$, which we distinguish from the corresponding function during inflation by using $\eta$ instead of $\tau$ for conformal time. At the end of inflation, $\tau=\tau_f$, we have
\be
\label{linear-deceleration}
A(k,\tau=\tau_f) = k^3 \frac{1-i/k\tau_f}{1+k^2\tau_f^2}.
\ee
We evolve $A(k,\eta)$ during the decelerating epoch with this initial condition at $\eta=0$.
The free Hamiltonian for $\vph$ with respect to conformal time was introduced in Eq.~\eqref{H_phi}. In a generic post-inflationary cosmology $a(\eta)$ with conformal time $\eta$, we have
\be
\H[\vph,\pi] = \frac{1}{2}\int_\k\(\frac{1}{a^2H^2}\pi_\k\pi_{-\k} + a^2 H^2 k^2 \vph_\k\vph_{-\k} \),
\ee
where the factors of $H=H_{\rm inflation}$ result from defining $\vph$ to be dimensionless, with $2k^3\<|\vph_\k|^2\>=1$ at the end of inflation. The Schr\"{o}dinger equation,
\be
i\frac{d}{d\eta}\Psi_G[\vph] = \H[\vph,-i\d/\d\vph]\Psi_G[\vph]
\ee
determines the dynamical equation of motion for $A(k,\eta)$, which is \cite{boddy2016how,Burgess:2014eoa}
\be\label{A_eom} 
i\frac{d}{d\eta}A(k,\eta) = \frac{A^2(k,\eta)}{a^2(\eta)H^2} - a^2(\eta) H^2 k^2.
\ee
This is the same as Eq.~\eqref{A_eom_inflation}, with $-1/H\tau=a(\tau)$ replaced with a generic post-inflationary scale factor $a(\eta)$. As discussed below Eq.~\eqref{A_eom_inflation}, the two terms on the RHS of Eq.~\eqref{A_eom} come from the kinetic and gradient terms in the Hamiltonian. In the superhorizon regime, the $A^2$ (kinetic) term is small because the modes are frozen out. Once the modes re-enter the horizon and start to oscillate, the two terms are comparable.
(Note that in Minkowski space, we can set $a=1$, and the two terms cancel for the time-independent vacuum solution $A(k) \propto k$.)

Using Eq.~\eqref{a_radiation} for $a(\eta)$, we can write Eq.~\eqref{A_eom} as
\be\label{A_eom_radiation}
i\frac{d}{d\eta}A(k,\eta) = \tau_f^2(1-\eta/\tau_f)^{-2}A^2(k,\eta) - \tau_f^{-2}(1-\eta/\tau_f)^2 k^2.
\ee
The solution is controlled by the small parameter $|k\tau_f|$, which is the inverse of the amount of squeezing in the superhorizon mode $k$ at the end of the inflationary era. In particular, while the real part of $A$ (which controls the amplitude of perturbations) is not sensitive to $\tau_f$, the imaginary part scales as $1/|k\tau_f|^4$ in the decelerating epoch. Taking Eq.~\eqref{linear-deceleration} as the initial condition, an exact solution can be found\footnote{The exact solution is $A(k,\eta) = \[2 k^5 (\eta -\tau_f)^2\]\[2 k^4 \tau_f^4+\left(2 k^2 \tau_f^2-1\right) \cos (2k\eta)-2 k \tau_f \sin (2k\eta)+1\]^{-1}+i\Big[(\eta -\tau_f)  \Big(2 k^4 \tau_f^4 \allowbreak - \big(k\eta + k\tau_f -2 k\eta (k\tau_f)^2+2(k\tau_f)^3\big) \sin (2k\eta)+\left(2 k\eta\cdot k\tau_f-1\right) \cos (2k\eta)+1\Big)\Big] \[\tau_f^4 \left(2 k^4\tau_f^4+\left(2 k^2 \tau_f^2-1\right) \cos (2k\eta)-2 k \tau_f \sin (2k\eta)+1\right)\]^{-1}$, which recovers the initial condition \eqref{linear-deceleration} as $\eta \to 0$.} which is approximated by 
\ba\label{A_rad} 
A(k,\eta) &=& k^3 (k\eta)^2 \[\frac{1}{\sin^2(k\eta)} - i \frac{1}{(k\tau_f)^4} \(\cot(k\eta) - \frac{1}{k\eta}  \) \]  + \O(k\tau_f) \nn \\
&=& k^3 \frac{(k\eta)^2}{\sin^2(k\eta)} \[1 - i \frac{1}{(k\tau_f)^4} \(\sin(k\eta)\cos(k\eta) - \frac{1}{k\eta}\sin^2(k\eta) \) \] + \O(k\tau_f).
\ea
From the real part of $A$ we see that as the modes re-enter the horizon at $k\eta\sim1$, they oscillate and begin to decay in amplitude, with variance $\<|\vph_\k(\eta)|^2\>\propto1/{\rm Re}A(k,\eta)\propto1/(k\eta)^2$. We will see in \S~\ref{sec:phase_space_decel} that the phase behavior describes a simple rotation of a squeezed state in phase space, with the $1/(k\tau_f)^4$ enhancement of the phase increasing the squeezing of the state.

\subsection{Linear Evolution in Phase Space}
\label{sec:phase_space_decel}

\begin{figure}
\includegraphics[scale=0.45]{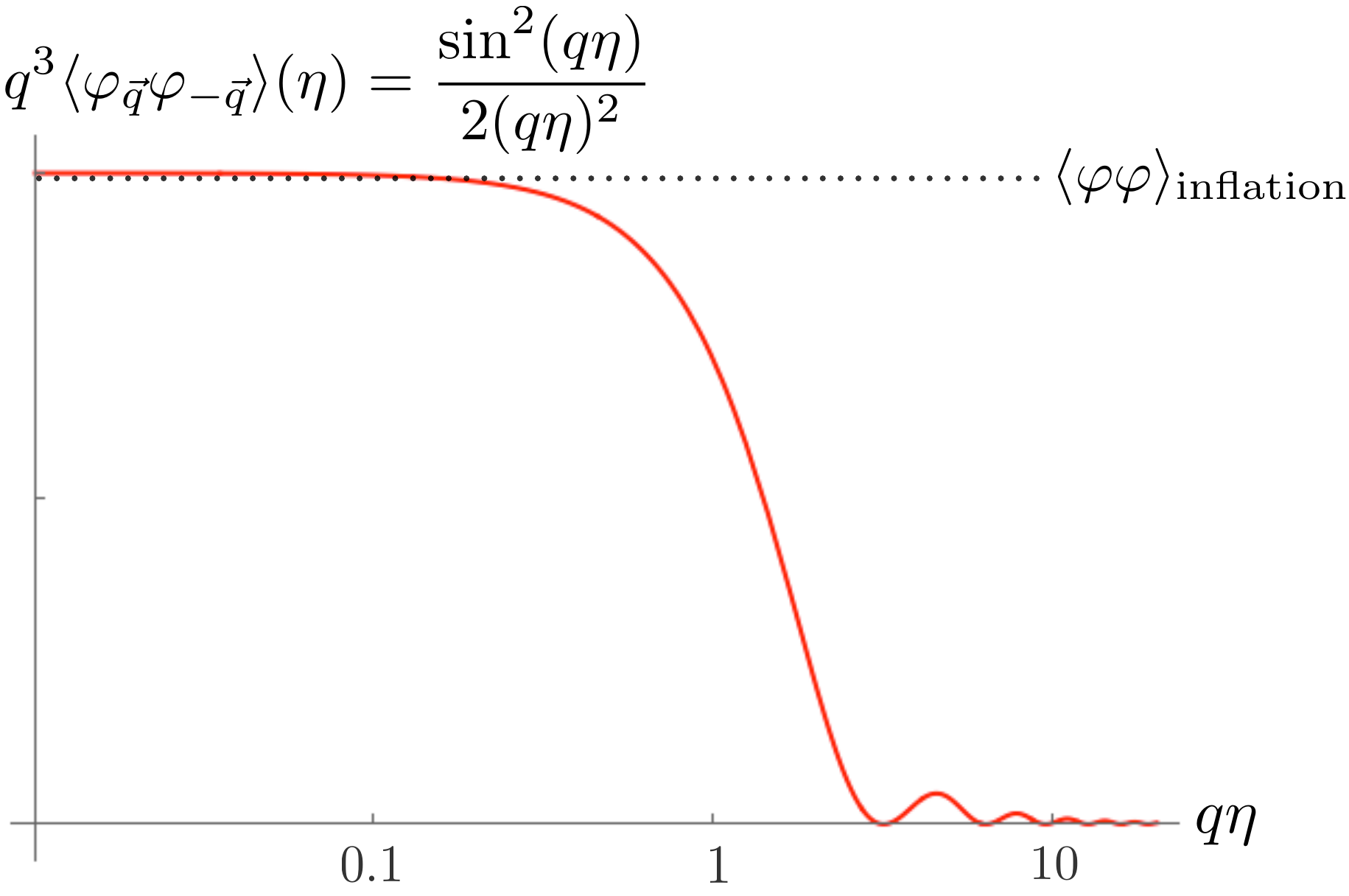}
\caption{The power spectrum of the field is a scale-invariant spectrum (set by inflation) on superhorizon scales, $q\eta\ll1$, and decays as $1/(q\eta)^2$ on sub-horizon scales, $q\eta\gg1$. The decay of the field on sub-horizon scales will result in an enhancement of coupling to superhorizon modes, allowing for more precise records to form.}
\label{fig:decay}
\end{figure}

The two-point functions of the field and its canonical momentum can be straightforwardly computed using Eq.~\eqref{A_rad}, and are \footnote{Recall that we use a prime on a correlation function to denote the omission of a momentum-conserving delta function, e.g., $\<\vph_\k\vph_{\k'}\>=(2\pi)^3\d^3(\k+\k')\<\vph_\k\vph_{\k'}\>'$.}:
\ba 
\<\vph_\k\vph_{\k'}\>'(\eta) &=& \frac{1}{2k^3}\frac{\sin^2(k\eta)}{(k\eta)^2} + \O(k\tau_f).
\label{power_spectrum_rad} \\
\<\overline{\vph_\k\pi_{\k'}}\>'(\eta) &=& \frac{1}{2} \frac{1}{(k\tau_f)^4} \(\sin(k\eta)\cos(k\eta) - \frac{1}{k\eta}\sin^2(k\eta) \) + \O((k\tau_f)^{-3}) \nn \\
&\rarr&  \frac{1}{2} \frac{1}{(k\tau_f)^4} \sin(k\eta)\cos(k\eta)
\hspace{1cm} (k\eta\gg1) \\
\<\pi_\k\pi_{\k'}\>'(\eta) &=& \frac{1}{(k\tau_f)^8}\frac{1}{2}k^3\(k\eta\cos(k\eta)+\sin(k\eta)\)^2 + \O((k\tau_f)^{-7}) \nn \\
&\rarr& \frac{1}{2} \frac{1}{(k\tau_f)^8}k^3(k\eta)^2\cos^2(k\eta)
\hspace{0.8cm} (k\eta\gg1)
\ea
Note that the corrections in $k\tau_f$ are important for $\<\pi\pi\>$ and $\<\vph\pi\>$ in the $\eta\rarr0$ limit, since the leading order terms from the $k\eta\gg1$ regime vanish here.
We see from the power spectrum, $\<\vph_\k\vph_{\k'}\>'(\eta)$, that the field decays in amplitude as $1/k\eta$ upon re-entering the horizon (Figure \ref{fig:decay}). We will see that this is crucial for the sensitivity of short modes to the super-horizon field, and resulting formation of redundant records.
Note that the relative phase in the second line grows as $(k\eta)^3$ for $k\eta\ll1$, describing the continued increase in squeezing in the superhorizon regime, until the mode re-enters the horizon with a maximum squeezing level at $k\eta\sim1$.
 
The evolution of the two-point functions also describes the rotation of the Wigner function in phase space. This can be clearly seen by working with the rotated variables $(\Phi,\Pi)$, which were introduced in Eq.~\eqref{wigner_rotate} for the inflationary phase and can be extended into the decelerating phase. In the sub-horizon regime, $k\eta\gg1$, and neglecting subleading corrections in $k\tau_f$, we have 
\ba
\Phi_\k(\eta) &\equiv& k\eta\cos(k\eta)\vph_\k - \frac{(k\tau_f)^4}{k\eta}\sin(k\eta)\frac{1}{k^3}\pi_\k + {\rm h.o.}, \nn \\
\frac{1}{k^3}\Pi_\k(\eta) & \equiv & \frac{k\eta}{(k\tau_f)^4}\sin(k\eta)\vph_\k + \frac{1}{k\eta}\cos(k\eta)\frac{1}{k^3}\pi_\k + {\rm h.o.},
\label{Phi_Pi_rad} 
\ea
where ``h.o.'' denotes higher order terms in the small parameters $1/k\eta$ and $k\tau_f$ that ensure that $\<\Phi_\k\Pi_{\k'}\> = 0$ exactly, 
and that canonical commutation is preserved exactly, $[\Phi_\k,\Pi_{\k'}]=[\vph_\k,\pi_{\k'}]$. 
The non-trigonometric factors of $k\eta$ in Eq.~\eqref{Phi_Pi_rad} are a matter of convention, and remove the overall scaling with $k\eta$ in $\vph$ (which decays as $1/k\eta$) and $\pi$ (which grows as $k\eta$). %
It is straightforward to check that the rotated momentum has the time-independent uncertainty
\be\label{PiPi_rad} 
\<\Pi_\k\Pi_{\k'}\>' = \frac{1}{2}\frac{1}{(k\tau_f)^8} k^3 + {\rm h.o.}, \hspace{1cm} (k\eta\gg1)
\ee
well after the mode re-enters the horizon.
Correspondingly, the uncertainty in $\Phi$ is zero up to corrections in $k\tau_f$, so the $\Phi$ direction is the direction in phase space in which the state is tightly squeezed, just as in the inflationary era. We can infer from the uncertainty principle (which is saturated for $(\Phi,\Pi)$ since each mode is in a pure state and $\<\Phi\Pi\>=0$) that
\be\label{PhiPhi_rad} 
\<\Phi_\k\Phi_{\k'}\>' = \frac{(k\tau_f)^8}{2k^3} + {\rm h.o.}, \hspace{1cm} (k\eta\gg1)
\ee
without actually calculating these corrections.

Let us briefly comment on the enhanced $1/(k\tau_f)^4$ dependence of mode squeezing, which is even stronger than the inflationary squeezing, $\propto 1/k\tau_f$.  During inflation, a given $k$-mode is initial excited out of its ground state at a time $\tau_0 \sim -1/k$ and the degree of squeezing achieved by the end of inflation $\tau = \tau_f$ is determined by the ratio of conformal times $\tau_0/\tau_f \sim 1/|k\tau_f|$.  But by the time the mode re-enters the horizon ($k\eta \sim 1$), the scaling becomes $1/(k\tau_f)^4$. This is because (i) $\Im dA/d\eta$ is initially parametrically enhanced by the scale factor $a^2(0)\propto (k\tau_f)^{-2}$, so the long-wavelength modes rapidly acquire additional relative squeezing during the transient period at the end of inflation, and (ii) by introducing a decelerating epoch, we have allowed for a large passage of conformal time during which $a^2$ can be further enhanced by $(\eta/\tau_f)^2$. In short, the squeezing is enhanced by the continued expansion of the universe while the mode remains frozen outside the horizon.

\subsection{Linear Evolution of Localized Modes}
\label{Phi_Pi_wavepacket}

Once again, we make use of localized Gaussian modes in order to quantify the amount of information recorded locally.
Since the modes remain highly squeezed, and furthermore rotate in phase space after re-entering the horizon, we once again use the rotated $(\Phi,\Pi)$ coordinates. 
As in the inflationary era, we localize the rotated canonical variables with a Gaussian convolution,
\be
\Phi^{(\sigma)}(\k,\x,\eta) \equiv \int_{\k'} W(\k'|\k,\x) \Phi_{\k'}(\eta), \hspace{1.5cm} \Pi^{(\sigma)}(\k,\x,\eta) \equiv \int_{\k'} W(\k'|\k,\x) \Pi_{\k'}(\eta).
\label{Phi_Pi_wavepackets_rad}
\ee
Crucially, once a mode with peak wavenumber $k\gg1/\sigma$ re-enters the horizon in the decelerating era, $\Phi^{(\sigma)}(\k,\x,\eta)$ and $\Pi^{(\sigma)}(\k,\x,\eta)$ \textit{depend only on the field $\vph$ and its conjugate momentum in the spatial regions of size $\sigma$ centered at the points $\x_\pm = \x\pm\hat{\k}\eta$ on opposite sides of $\x$, in the direction picked out by the peak wavenumber $\k$.} 
To see this, note that the sines and cosines in the definitions of $\Phi$ and $\Pi$, when expressed as exponentials, can be combined with the phase in the  convolution $W(\k'|\k,\x)$, leading to a dependence only on the field near $\x_\pm$, rather than the field near $\x$.
In terms of $\vph(\y)$ and $\pi(\y)$ at spatial points, we can integrate over $\k'$ to obtain
\ba
\Phi^{(\sigma)}(\k,\x,\eta) &=& \int d^3\y \Big[ \vph(\y) \frac{k\eta}{2} \(\tilde{W}(\y+\hat{\k}\eta|\k,\x)+\tilde{W}(\y-\hat{\k}\eta|\k,\x)\)  \nn \\
&& \hspace{0.8cm} - \pi(\y) \frac{1}{2i}\frac{(k\tau_f)^4}{k^3(k\eta)} \(\tilde{W}(\y-\hat{\k}\eta|\k,\x) - \tilde{W}(\y+\hat{\k}\eta|\k,\x)\) \Big]+\O(1/\sigma k) \nn \\
k^{-3}\Pi^{(\sigma)}(\k,\x,\eta) &=& \int d^3\y \Big[ \vph(\y) \frac{1}{2i}\frac{k\eta}{(k\tau_f)^4} \(\tilde{W}(\y-\hat{\k}\eta|\k,\x)-\tilde{W}(\y+\hat{\k}\eta|\k,\x)\)  \nn \\
&& \hspace{0.8cm} + \pi(\y) \frac{1}{2}\frac{1}{k^3(k\eta)} \(\tilde{W}(\y+\hat{\k}\eta|\k,\x) + \tilde{W}(\y-\hat{\k}\eta|\k,\x)\) \Big] +\O(1/\sigma k).
\label{Phi_Pi_Gaussian_realspace}
\ea
From the definition of $\tilde{W}$ in Eq.~\eqref{Wtilde} we see that the integrand only has support at points $\y$ within a distance $\sim\sigma$ of $\x\pm\hat{\k}\eta$.
Thus, the phase space rotation of $\Phi_\k$ and $\Pi_\k$ over a time $\eta\gg1/\sigma$ captures the spatial propagation of quanta localized in a region of size $\sigma$ away from their initial location at $\eta=0$.

It is important to note that, while the leading terms in Eq.~\eqref{Phi_Pi_Gaussian_realspace} have compact support in $\sigma$-sized regions at all times $\eta$, the $1/\sigma k$ corrections capture the spreading of the Gaussian mode at large times. This occurs for $\sigma/\eta\lesssim\sigma^{-1}/k$, or $\eta\gtrsim\sigma^2 k $, which is when quanta propagating in slightly different directions around the peak wavenumber $\k$ have dispersed outside a $\sigma$-sized region.
We will not be concerned with the spatial distribution of modes that have begun to spread out, since as long as all the relevant modes became highly squeezed during inflation, there will exist many short modes with $k\gtrsim\eta/\sigma^2$ for choices of $\sigma\ll\eta$, all containing the same redundant information localized in regions of size $\sigma$.
Modes which have spread outside of a $\sigma$-sized region may of course contain additional information, increasing the (already large) total redundancy of records for a given long-wavelength variable.

Before moving on, we write down the two-point functions for Gaussian modes.
Again defining $\Phi_{S,r}$ and $\Phi_{S,i}$ as in Eq.~\eqref{Phi_Pi_S_def}, it is straightforward to check from Eqs.~\eqref{vph_S_def}, \eqref{Phi_Pi_wavepackets_rad}, \eqref{PiPi_rad}, and \eqref{PhiPhi_rad} that during the radiation era,
\ba 
\<\Phi_{S}^2\>(k) \equiv \<\Phi_{S,r}^2\>(k) &=& \<\Phi_{S,i}^2\>(k) = \frac{(k\tau_f)^8}{2k^3} + \O(1/\sigma k), \nn \\
\<\Pi_{S}^2\>(k) \equiv \<\Pi_{S,r}^2\>(k) &=& \<\Pi_{S,i}^2\>(k) = \frac{1}{2}\frac{1}{(k\tau_f)^8} k^3  + \O(1/\sigma k), \label{Phi_Pi_2pt_linear_rad}
\ea
and by definition,
\be
\<\Phi_S\Pi_S\> \equiv \<\Phi_{S,r}\Pi_{S,r}\> = \<\Phi_{S,i}\Pi_{S,i}\> = 0.
\ee
We will again omit the $r$ or $i$ subscripts because the two modes behave identically.

Note that the information contained in these two-point functions comes from field degrees of freedom $\{\vph(\y),\pi(\y)\}$ located near $\x+\hat{\k}\eta$ \textit{and} degrees of freedom near $\x-\hat{\k}\eta$, as indicated in Eq. \eqref{Phi_Pi_Gaussian_realspace}. This is because the translation invariant inflationary dynamics created entangled pairs of quanta in $\k$ and $-\k$ modes in a two-mode squeezed state of the particle basis form $\sum_n c_n |n_\k;n_{-\k}\>$. Once $\eta\gtrsim\sigma$, the quanta in a localized mode propagate in opposite directions in entangled EPR pairs, carrying information in the entanglement between them.

We now have all the needed pieces from the evolution of the Gaussian part of the wavefunction in place, and will move on to calculating effects from mode coupling in the initial state on these two-point functions.

\subsection{Nonlinear Evolution of the Wavefunction}

Having worked through the linear theory, we now compute the evolution of the leading non-Gaussian part of the wavefunction during the radiation epoch.
Recall that the leading mode couplings appear in the cubic part of the wavefunction,
\ba\label{WF_cubic_radiation}
\Psi[\vph](\eta) &\approx& {\rm (norm.)} \exp\[-\frac{1}{2}\int_\k \vph_\k\vph_{-\k}A(k,\eta) + \int_{\q,\k}\vph_\q\vph_\k\vph_{\k'}\F_{q,k,k'}(\eta) \] \nn \\
&\equiv& \Psi_G[\vph](\eta)\times\exp\[\int_{\q,\k}\vph_\q\vph_\k\vph_{\k'}\F_{q,k,k'}(\eta) \],
\ea
where $\k'\equiv-\k-\q$, and we are now using the conformal time coordinate $\eta$ for the radiation era. 
As discussed in \S~\ref{sec:WF_nonlin_inflation}, the cubic kernel $\F_{k,k',q}(\eta)$ satisfies the equation of motion
\be
\F'(\eta) - i\alpha(\eta)\F(\eta) = -i \H^{\rm int}(\eta),
\ee
with solution
\be\label{F_gen_solution_rad}
\F_{k,k',q}(\eta) =-i \int_0^\eta d\eta' \H^{\rm int}_{\k,\k',\q}(\eta')\exp\[i\int_{\eta'}^\eta d\eta''\alpha_{k,k',q}(\eta'')\].
\ee
The source $\H^{\rm int}$ can be calculated by acting the interacting Hamiltonian $\H^{\rm int}=-\L_3$ (where $\L_3$ was defined in Eq. \eqref{cubic_model}) on the Gaussian wavefunction as in Eq. \eqref{H_int_source_def}, and is given by\footnote{The cotangents in Eq. \eqref{H_source_rad} come from factors of $\vph'\sim\pi=-i\d/\d\vph$ acting on the linear wavefunction and bringing down factors of ${\rm Im}A(k,\eta)\propto\cot(k\eta)$ out of the exponent of the wavefunction.}
\be\label{H_source_rad} 
\H^{\rm int}_{\k,\k',\q}(\eta) = \frac{1}{3}\frac{(k\eta)^2}{\tau_f^4}\[g+\tilde{g}\cot(k\eta)\cot(k'\eta)\] + \O(q/k,1/k\eta,k\tau_f).
\ee
We also have
\ba
\alpha_{k,k',q}(\eta) \equiv - \frac{1}{a^2(\eta)H^2}\(A_k(\eta)+A_{k'}(\eta)+A_q(\eta)\),
\ea
where $H$ is the inflationary Hubble rate. $A(k,\eta)$ is given in Eq.~\eqref{A_rad} for modes that crossed the horizon during inflation ($|k\tau_f|\ll1$), and leads to
\be\label{ImAlpha}
\Im\alpha_{k,k',q}(\eta) = \( k\cot(k\eta) - \eta^{-1} \) + \text{2 perms.} + \O(k\tau_f)^4.
\ee
Using  %
Eq.~\eqref{ImAlpha} in Eq.~\eqref{F_gen_solution_rad}, we obtain
\be
\Im\F_{k,k',q}(\eta) = -i\int_0^\eta d\eta' a(\eta')\H_{\k,\k',\q}^{(\rm int)}(\eta')\frac{\eta^3}{\eta'^3}\frac{\sin(k\eta')}{\sin(k\eta)}\frac{\sin(k'\eta')}{\sin(k'\eta)}\frac{\sin(q\eta')}{\sin(q\eta)},
\ee
up to corrections in $k\tau_f$.
With Eq.~\eqref{H_source_rad}, the solution for $\F_{k,k',q}(\eta)$ in the limit $q\ll k,k'$, at late times $k\eta\gg1$ such that the short modes have re-entered the horizon, is
\be\label{ImF_radiation}
\Im\F_{k,k',q}(\eta) = -\frac{g+\tg}{12} \frac{k^3(k\eta)^3}{(k\tau_f)^4}\frac{{\rm Si}((q+q_\parallel)\eta)+{\rm Si}((q-q_\parallel)\eta)}{\sin(k\eta)\sin(k'\eta)\sin(q\eta)} + \O(q/k,1/k\eta,(k\tau_f)^{-3}),
\ee
where ${\rm Si}(x) \equiv \int_0^x [\sin(y)/y]\dd y$ is the sine integral, and $q_\parallel\equiv \q\cdot\hat{\k}$.
We emphasive that Eq.~\eqref{ImF_radiation} is valid for all values of $q\eta$ (long modes both inside and outside the horizon), as long as the short mode is subhorizon, $k\eta\gg1$.
Note that Eq.~\eqref{ImF_radiation} is zero at $\eta=0$. Of course, $\Im\F_{k,k',q}(0)=\Im\F_{k,k',q}(\tau=\tau_f)\neq0$ if there were any interactions during inflation. In order to smoothly recover these initial conditions, we would have to track the corrections in $k\tau_f$ in Eq.~\eqref{ImF_radiation}. But we can ignore these (as we did for the Gaussian part of the wavefunction, $A(k,\eta)$) because we are interested in late times, $k\eta\gg1$.\footnote{Note that Eq.~\eqref{ImF_radiation} is enhanced by $1/k\tau_f$ relative to the the solution during inflation, Eq. \eqref{ImF_inflation}. This occurs for the same reason that the mode becomes exponentially more squeezed after inflation: the prolonged cosmological expansion increases the action, generating larger phases in the wavefunction.
In the absence of interactions after inflation, the cubic kernel $\F(\eta=0)$ set by inflation evolves without a source, $\F'=i\alpha\F$, which has the general solution
\be 
\F_{k,k',q}(\eta) = \F_{k,k',q}(0) \exp\[i\int_0^\eta d\eta' \alpha_{k,k',q}(\eta') \] \approx \F_{k,k',q}(0)\frac{k\eta}{\sin(k\eta)} \frac{k'\eta}{\sin(k'\eta)} \frac{q\eta}{\sin(q\eta)}.
\ee}

Note that the phase ${\rm Im}\F$ vanishes in the special case $g+\tg=0$. As a consequence, there will be no redundant records and consequently no wavefunction branching in this case. This is because the two cubic interactions each tilt the Wigner function as it rotates in phase space, either advancing or retarding its rotation. (This is reflected in the behavior of the conditional two-point functions calculated below in \S \ref{sec:records_decel}.) The two contributions cancel in the case where the shift to the quadratic Lagrangian is an overall rescaling, $\Delta\L_2\propto\L_2$. Classically, this only shifts the radius of a harmonic oscillator rotating in phase space, but not its phase. Quantum mechanically, the ellipticity of the Wigner function is perturbed, but not its angle of orientation.

\subsection{Long-wavelength Influence on Localized Short-wavelength Modes}
\label{sec:records_decel}

We now have all the pieces in place to calculate how the coarse-grained field is recorded in the conditional quantum states of short-wavelength localized modes.  Intuitively, we want to identify measurements that many localized observers could perform, from which they could reliably infer the same classical information.  More precisely
We will again condition the global wavefunction on a classical long-wavelength field configuration at $\eta=0$ and compute the corresponding states of localized field modes in nearby regions at later times $\eta >0$.  If the conditional states of a localized mode are orthogonal for sufficiently different values of a long-wavelength field variable, then that mode has a record of the background to the relevant precision.  If this is true for modes in many disjoint spatial regions, then there are redundant records of that long-wavelength variable.

The conditional two-point functions during the radiation era can be calculated in the same way as during the inflationary era.
Using Eq.~\eqref{WF_cubic_radiation} for $\Psi[\vph]$, with Eq.~\eqref{A_rad} for the Gaussian part and Eq.~\eqref{ImF_radiation} for the non-Gaussian part, and the definitions of $\Phi$ and $\Pi$ in Eq. \eqref{Phi_Pi_rad}, we find the $\O(g,\tg)$ contribution \footnote{We ignore various higher order corrections, which are denoted below in Eq.~\eqref{Phi_Pi_rad_final}.}
\ba 
\<\overline{\Phi_\k\Pi_{\k'}}\>_{\varphi_L}(\eta)  &=&  \vph_\q(\eta) \times (-3){\rm Im}\F_{k,k',q}(\eta)
\frac{1}{k^3} \frac{\sin(k\eta)}{k\eta} \frac{\sin(k'\eta)}{k'\eta},
\hspace{1cm} \text{where     } \q=-\k-\k'. \nn \\
&=& \frac{g+\tg}{4} \frac{k\eta}{(k\tau_f)^4} \times \vph_\q(\eta) \frac{{\rm Si}((q+q_\parallel)\eta)+{\rm Si}((q-q_\parallel)\eta)}{\sin(q\eta)}.  %
\label{Phi_Pi_rad_Fourier}
\ea
Convolving from Fourier modes to Gaussian wavepackets, now with support near the points $\x\pm\hat{\k}\eta$ (see \S~\ref{Phi_Pi_wavepacket}), we find that
\ba 
\<\overline{\Phi_{S}\Pi_{S}}\>_{\varphi_L}(k,\eta) &=& \frac{g+\tg}{4} \frac{k\eta}{(k\tau_f)^4} \int_\q\varphi_\q(\eta) e^{i\q\cdot\x-\frac{1}{4}\sigma^2q^2}\frac{1}{\sin(q\eta)}[{\rm Si}((q+q_\parallel)\eta)+{\rm Si}((q-q_\parallel)\eta)] + \O(1/\sigma k) \nn \\
&\equiv& \frac{g+\tg}{4} \frac{k\eta}{(k\tau_f)^4} \vph_{\hat{k}}(\x,\eta) + \O(\sigma/\eta,1/\sigma k,(k\tau_f)^{-3})
\label{Phi_Pi_rad_final}
\ea
where we have defined\footnote{We omit the cutoff scale $q^\star>1/\sigma$ here and in the following equations because, as we will explain, the suppression of sub-horizon modes removes (to a good approximation) the dependence on $q^\star$, allowing localized short modes to respond only to (conditioning on) sufficiently long-wavelength modes.}
\be\label{vph_khat_def}
\vph_{\hat{k}}(\x,\eta) \equiv \int_\q\vph_\q(\eta=0) e^{i\q\cdot\x} \[{\rm Si}((q+q_\parallel)\eta)+{\rm Si}((q-q_\parallel)\eta)\] / (q\eta)
\ee
Here, $\varphi_\q(0) = \varphi_\q(\eta)\frac{q\eta}{\sin(q\eta)}$ is the \textit{initial, scale-invariant} field at the end of inflation. Crucially, the sum of sine integrals reduces to $q\eta$ for $q\eta\ll1$, and asymptotes to $\pi$ for $q\eta\gg1$,\footnote{Since $-q\le q_\parallel\le q$, $\mathrm{Si}(0) = 0$, and $\lim_{x\to\infty} \mathrm{Si}(x) = \pi/2$, the sine integrals have the asymptotic behavior: $\lim_{q\eta\to \infty} [\mathrm{Si}(q\eta(1+\cos \theta)) + \mathrm{Si}(q\eta(1-\cos \theta))]=\pi$, unless $\q$ and $\k$ are exactly aligned or anti-aligned ($q_\parallel\equiv q\cos(\theta)=\pm q$), in which case the limit is $\pi/2$.}
so 
\be\label{vph_khat_split}
\vph_{\hat{k}}(\x,\eta) \sim \int_\q\Theta(q<\eta^{-1})\vph_\q(0)e^{i\q\cdot\x}+\pi\int_\q\Theta(q>\eta^{-1})\vph_\q(0)e^{i\q\cdot\x}\frac{1}{q\eta}. \nn
\ee
where the Heaviside theta functions enforce integration over scales either longer or shorter than $\eta$.
This is a consequence of the decay of modes for $k\eta\gg1$, as captured in the power spectrum $\<\vph\vph\>$ in Eq.~\eqref{power_spectrum_rad}: Because modes inside the horizon decay as $1/q\eta$, their contribution to $\vph_{\hat{k}}(\x,\eta)$ is increasingly suppressed at short scales.
From Eq.~\eqref{vph_khat_split} we see that $\vph_{\hat{k}}(\x,\eta)$ has a variance
\be
\<\(\vph_{\hat{k}}(\x,\eta)\)^2\> \sim \int_{q<1/\eta} \frac{d^3\q}{(2\pi)^3} \frac{1}{2q^3}+\int_{q>1/\eta} \frac{d^3\q}{(2\pi)^3} \frac{1}{2q^3}\frac{\pi^2}{(q\eta)^2} = (2\pi)^{-2} \ln(1/q_{\rm min}\eta)+\O(1),
\ee
with contributions only from superhorizon modes and from modes just re-entering the horizon ($q\eta\sim1$).
So $\vph_{\hat{k}}(\x,\eta)$ is the field smoothed on the horizon scale $\eta$, with some additional direction dependence.\footnote{Remember that the variable that is jointly recorded \emph{does not need need to be accessible in its original form}.  Likewise, following the measurement of an electron spin in the laboratory, there exists many stable records of that variable in macroscopically different and dispersed places even after the electron has been lost to the environment and has become scrambled.} 

The conditional two-point function $\<\Phi_\k\Phi_{\k'}\>_{\varphi_L}$ can be calculated in the same way, and is\footnote{Note that because $\<\Phi\Phi\>$ is $\O(g^2)$, it could potentially acquire corrections from the \textit{quartic} part of the wavefunction (which is dynamically sourced by the cubic part even in the absence of quartic interactions). However, we found that the quartic part does not contribute to $\<\Phi\Phi\>$ regardless of the cubic (or quartic) interactions, although other two-point functions are affected by it.}
\begin{align}\begin{split} 
\label{Phi_Phi_rad_Fourier}
\<\Phi_\k\Phi_{\k'}\>_{\varphi_L}(\eta) = \frac{(g+\tg)^2(k\eta)^2}{8k^3} \int_{\q,\q'} (2\pi)^3 &\d^3(\k+\k'+\q+\q') \vph_\q(\eta) \vph_{\q'}(\eta)\\
&\times
\frac{{\rm Si}((q+q_\parallel)\eta)+{\rm Si}((q-q_\parallel)\eta)}{\sin(q\eta)} \frac{{\rm Si}((q+q_\parallel)\eta')+{\rm Si}((q-q_\parallel)\eta')}{\sin(q\eta')},
\end{split}\end{align}
up to higher order corrections.
The resulting two-point function for Gaussian modes is
\be\label{Phi_Phi_rad_final}
\<\Phi_S^2\>_{\varphi_L}(k,\eta) = \frac{1}{2k^3} (k\tau_f)^8 + \frac{(g+\tg)^2 (k\eta)^2}{8k^3} \[ \( \vph_{\hat{k}}(\x,\eta) \)^2 + \O(\sigma^2/\eta^2,1/\sigma k,k\tau_f) \],\\
\ee
Finally, the rotated momentum $\Pi$ is unaffected by nonlinearities at leading order, and is equivalent to the linear theory result, Eq.~\eqref{Phi_Pi_2pt_linear_rad}, up to small corrections in $g,\tg$:
\be\label{Pi_Pi_rad_final}
\<\Pi_S^2\>_{\varphi_L}(k) \approx \frac{1}{2} \frac{1}{(k\tau_f)^8} k^3.  %
\ee
From Eqs.~\eqref{Phi_Pi_rad_final}, \eqref{Phi_Phi_rad_final}, and \eqref{Pi_Pi_rad_final}, we see that
\be\label{two_point_mixedness} 
\<\Phi_S^2\>_{\varphi_L}\<\Pi_S^2\>_{\varphi_L} - \<\overline{\Phi_{S}\Pi_{S}}\>^2_{\varphi_L} = \frac{1}{4}\[ 1 +  \frac{(g+\tg)^2(k\eta)^2}{4(k\tau_f)^8}\times\O(\sigma^2/\eta^2,1/\sigma k,k\tau_f) \] .
\ee

The conditional state of the mode $S$ is mixed (due to entanglement with other modes\footnote{The conditional state is not exactly Gaussian, due to the non-Gaussian entanglement between modes in the wavefunction. This is a small effect, suppressed by the interaction strengths, so we can still represent the Wigner function as an approximately Gaussian function in phase space, characterized by approximately elliptical regions of support.}) insofar as this quantity differs from the lower bound dictated by the uncertainty principle, $1/4 = (\hbar/2)^2$.
This mixing blurs out the Wigner function in phase space, increasing the thickness of its support in the direction with minimal uncertainty. (In the case of conditioning on $\vph_{\hat{k}}(\x,\eta)=0$, so that $\<\overline{\Phi_{S}\Pi_{S}}\>_{\varphi_L}=0$, this direction is simply $\Phi_S$, and it is clear that the mixing which causes the LHS of Eq.~\eqref{two_point_mixedness} to exceed $1/4$ is equivalent to an increase in $\<\Phi_S^2\>$ in Eq.~\eqref{Phi_Phi_rad_final}.) For $(g+\tg)\sigma k \gtrsim (k\tau_f)^4$, the entanglement between modes generates a large amount of mixing, and a correspondingly large increase of the area of support of the Wigner function.

However, the conditional state of a given short mode is still highly squeezed (although less so than in the absence of interactions), and can thus still act as a clock recording the value of $\vph_{\hat{k}}(\x,\eta)$. The range of angles at which the squeezed state can be oriented is captured by the product $\<\Phi_S^2\>_{\varphi_L}\<\Pi_S^2\>_{\varphi_L}$, which, for a typical background fluctuation, $\vph_{\hat{k}}(\x,\eta)=\O(1)$, is roughly the (square of the) area of phase space collectively covered by all conditional states. This area exceeds the area of support of a given conditional state by a factor of $\eta^2/\sigma^2$. That is,
\be\label{twopoint_inequality}
\[ \<\Phi_S^2\>_{\varphi_L}\<\Pi_S^2\>_{\varphi_L} - \<\overline{\Phi_{S}\Pi_{S}}\>^2_{\varphi_L} \] = \O\(\frac{\sigma^2}{\eta^2}\) \times \<\Phi_S^2\>_{\varphi_L}\<\Pi_S^2\>_{\varphi_L}
\hspace{1cm} \text{for} \ \ \varphi_L\sim\sqrt{\vph_L^2}, \ \ (g+\tg) k\eta \gg (k\tau_f)^4.
\ee
Since the conditional states are narrow ellipses that only rotate around the origin as a linear function of $\varphi_L$, the support of two conditional states become essentially non-overlapping when the difference in $\varphi_L$ exceeds the ratio of the conditioned and unconditioned supports.  (Cf.~the situation during inflation as discussed at the end of \S~\ref{sec:no_records_inflation}.)
In other words, a short mode can resolve the long-wavelength background with precision
\be
\Delta \varphi_{\hat{k}}(\x,\eta) \sim\sigma/\eta
\ee
despite its entanglement with other modes. Again, this is simply because modes shorter than the horizon scale have decayed as $1/\eta$ since re-entering the horizon. As a consequence, the interaction strength effectively becomes more suppressed for modes that have evolved further inside the horizon, and correlations between modes deep inside the horizon (e.g. shorter than $\sigma$) are suppressed relative to correlations with longer modes.

From Eq. \eqref{twopoint_inequality} we see that a mode with $\sigma\ll\eta$ contains a record with at least one bit of information (with at least minimal uncertainty, $\Delta\vph^2_{\hat{k}}(\x,\eta)\lesssim\<\vph^2_{\hat{k}}(\x,\eta)\>$), as long as
\be
(g+\tg) k\eta \gtrsim (k\tau_f)^4.
\ee
This is analogous to the inflationary condition $|k\tau|\lesssim g$ for a super-horizon mode $\vph_\k$ to contain information about a constant background $\vph_L$ (neglecting its entanglement with other modes).  In both cases, the mode must be sufficiently squeezed, and must be sufficiently entangled with long-wavelength modes, as encoded in the phase of the wavefunction, ${\rm Im}\F$.

In the absence of a minimum wavenumber $q_{\rm min}$, all modes that crossed the horizon during inflation will eventually reach this regime, and carry records.
On the other hand, given a finite $q_{\rm min}$, the condition that at least some modes acquire records is
\be\label{records_condition_1}
(g+\tg)q_{\rm min}\eta \gg (q_{\rm min}\tau_f)^4.
\ee
This ensures that modes several times shorter than $q_{\rm min}$ will acquire some information about the longest-wavelength fluctuations. This must occur by the time when the longest mode re-enters the horizon, $q_{\rm min}\eta\sim1$. Since the combination $|q_{\rm min}\tau_f|$ is simply the ratio $a_i/a_f$ of scale factors at the beginning and end of inflation, we can write this condition as
\be
\boxed{(g+\tg)^{1/4}\(\frac{a_f}{a_i}\)_{\rm inflation} \gg 1}.
\ee
We will now confirm that this condition further ensures that the many records (of a given long-wavelength variable) which form will occupy spatially disjoint regions.

\textit{Spatial Redundancy.}

We have seen that there are records of long-wavelength variables captured in the orthogonality of different conditional reduced states in the $(\Phi_S,\Pi_S)$ plane of phase space (specifically, for both $(\Phi_{S,r},\Pi_{S,r})$ and $(\Phi_{S,i},\Pi_{S,i})$, defined as in Eq.~\eqref{vph_S_def}).
Crucially, these records propagate into spatially disjoint regions.
As we saw in \S~\ref{Phi_Pi_wavepacket}, the degrees of freedom of the field $\vph$ and its conjugate momentum $\pi$ from which $\Phi_S=\Phi_S(\k,\x,\eta)$ and $\Pi_S=\Pi_S(\k,\x,\eta)$ are constructed are localized in regions of size\footnote{As long as $k\gtrsim \eta/\sigma^2$, so that the spreading of the Gaussian wavepackets is negligible -- see \S~\ref{Phi_Pi_wavepacket}.} $\sigma$ at the points $\x\pm\hat{\k}\eta$, describing the propagation of EPR pairs in opposite directions from their initial location $\x$.
As illustrated in Figure \ref{fig:records_propagate}, the regions of spatial support for different modes propagating in directions $\k$ and $\k'=\k+\Delta\k$ become disjoint once $\sigma/\eta\lesssim|\Delta\k|/k$.

Now, as long as Eq.~\eqref{records_condition_1} holds, with $g,\tg\ll1$, we are free to look for records in short modes with $\sigma\ll\eta$ (which guarantees that the record will be very precise), and $1/k\lesssim\sigma^2/\eta$ (which -- as noted in \S~\ref{Phi_Pi_wavepacket} -- guarantees that the record remains localized in the EPR-entangled pair of $\sigma$-sized regions\footnote{This condition is not essential, but we work in this regime for simplicity.}).
Combining the latter condition (no spreading) with the disjointness condition $\sigma/\eta\lesssim|\Delta\k|/k$, we have the hierarchy
\be
\frac{|\Delta\k|}{k}\gtrsim\frac{\sigma}{\eta}\gtrsim\frac{\sigma^{-1}}{k}.
\ee
Thus, while we need to consider sufficiently large $|\Delta\k|\gg|\Delta\k|_{\rm min}\sim\sigma^{-1}$ for a given pair of short modes to propagate into disjoint regions, we can still choose $|\Delta\k|/k$ to be very small at late times when $\sigma/\eta$ can be very small, and $1/\sigma k$ even smaller.

The long-wavelength variables $\vph_{\hat{k}}$ and $\vph_{\hat{k}'}$ recorded by these two modes (note that $\hat{\k}'\approx\hat{\k}+\Delta\k/k$) are approximately equivalent, up to corrections of order $|\Delta\k|/k$, which need not be much larger than $\sigma/\eta$. Thus, the difference in the variables recorded is no larger than the uncertainty in the records.
Consequently, it is possible to identify many spatially non-overlapping short modes which all contain the same information, up to differences and uncertainty of order $\sigma/\eta$.

\textit{Late-time behavior.}

We have assumed that the wavefunction is only perturbatively non-Gaussian, with the leading non-Gaussianity captured by the cubic kernel, or equivalently, three-point functions. This assumption breaks down at sufficiently late times, when the interactions have had enough time to generate $\O(1)$ deviations from Gaussianity. Here, we simply point out that this occurs at a later time than the formation of redundant records described above. A straightforward calculation shows that the three-point functions, normalized relative to the two-point functions, scale as
\be\label{3pt_late}
\frac{\<X_\q Y_\k Z_{\k'}\>(\eta)}{\sqrt{\<|X_\q|^2\>\<|Y_\k|^2\>\<|Z_{\k'}|^2\>}} \sim \O(g,\tg) \times k\eta \ \ \ \text{for} \ \ k\eta\gg1, \ q\eta\ll 1,
\ee
where $X,Y,Z=\{\vph,\pi\}$.
Here, we have omitted oscillatory functions and just written the overall scalings with the relevant parameters. A linear scaling with $a(\eta)\propto\eta$ is exactly what we should expect, since the cubic interaction with a superhorizon mode acts like a time-independent shift to the quadratic Hamiltonian, and the Schr\"{o}dinger equation determines the time-evolved wavefunction in terms of the time integral of the Hamiltonian, $\sim\int^\eta d\eta'(\rm constant)$. Once all modes re-enter the horizon, on the other hand, the interaction decays as $\sim1/\eta$ and the non-Gaussian entanglement ceases to grow.

We leave a full exploration of the late-time regime to future work, and emphasize that spatially redundant records proliferate already in the perturbative, nearly Gaussian regime, before modes evolve to be shorter than a fraction $\lesssim g,\tg$ of the horizon scale, at which point Eq.~\eqref{3pt_late} becomes $\O(1)$.

\section{Discussion}
\label{sec:discussion}

In this paper we analyzed the spatial entanglement arising in a simple model of a single self-interacting real scalar field.  This model can be interpreted in terms of cosmological perturbations produced during an initial inflationary epoch which freely propagate afterwards in a decelerating epoch.  Minimal interactions, such as those that must exist merely on account of gravity, generate entanglement between different scales that is  carried to distant regions.  In particular, the long-wavelength configuration of the field is recorded in the states of many spatially localized short-wavelength modes in the sense that conditioning on nearby but distinct long-wavelength configurations yields orthogonal states of the localized modes.  Thus, the localized modes redundantly record the configuration of the long wavelengths in many spatially disjoint regions.  We interpret this long-range spatial entanglement as a signature of objective wavefunction branching, with each branch corresponding to a potential macroscopic classical field-configuration outcome.  This demonstrates that the field decoheres itself in a preferred basis without assuming any preferred decomposition into subsystems, suggesting that the theory of decoherence may eventually be reformulated to rest on the more basic axiom of spatial locality.

The presence of redundant records means that no experiment can detect a phase between different branches without applying a unitary coherently coupling distant regions of space.  The difference between the coherent superposition of branches and their incoherent mixture is thus effectively undetectable, which is interpreted as the origin of apparent wavefunction collapse.  This holds even if individual records are in fact difficult to access because they require a joint measurement on extended EPR pairs, as depicted in Figure~\ref{fig:records_propagate}.  The background field configuration will nevertheless act as a ``latent'' classical variable, which can be traced out, for the purposes of predicting any other feasible measurement.  In this restricted sense, the branches are non-interacting.

In the rest of this section we consider possible avenues for future work.

\textit{Many-body phase information.} The presence of redundant records (and hence branches) in the state is captured by the phase of the field's wavefunction in the basis of field configurations.  This can be seen from the fact that the conditional two-point functions, which determine the amount of redundant information carried -- Eqs.~\eqref{Phi_Pi_kspace} and \eqref{Phi_Phi_kspace} in the inflationary era and Eqs.~\eqref{Phi_Pi_rad_Fourier} and \eqref{Phi_Phi_rad_Fourier} in the decelerating era -- depend \textit{only} on the phase $\rm Im \F$. Modestly different choices of interactions that change only the phase but still generate records will therefor still decohere the same distribution of long-wavelength field configurations.  It might be valuable to look for a similar relation between phase information, redundant records, and wavefunction branching in other many-body systems undergoing a quantum-to-classical transition.

\textit{Coherence and stochastic dynamics.} It would be interesting to quantify the redundancy and precision of records of long-wavelength field variables, identify the coherence of the field $\vph$ and field velocity $\vph'$, and ask under what conditions decoherence is sufficiently strong to generate stochastic behavior of long-wavelength fluctuations at late times in the form of a random walk.

\textit{Conditions for branching into classical fields.} The formation of redundant records is determined by the behavior of two- and three-point functions of the canonical variables $(\vph,\pi)$. It should be possible to identify more general conditions on correlation functions for nearly Gaussian wavefunctions that imply spatially redundant information, or on dynamics for nearly quadratic local Hamiltonians that generate branching into classical fields from the vacuum. 

\textit{Gravitational back-reaction.} In this paper, we have treated $\vph$ as a spectator field in an expanding space with scale factor $a(t)$ fixed by hand. In a fully self-consistent picture, the gravitational effects of the scalar mode $\vph$ would be accounted for. This could be done, for example, by (i) exploring a scalar field cosmology in which the background expansion is determined by a potential $V(\phi)$, with slow-roll inflation followed by a reheating epoch, and/or (ii) by studying tensor modes (primordial gravitational waves) in a similar context. In both of these cases, the relevant modes evolve outside the horizon and later re-enter the horizon in the presence of highly excited short-scale degrees of freedom, so we expect the qualitative features of wavefunction branching to be similar to those in our model.

\textit{Quantum cosmology.} We naturally desire to extend this study to wavefunction branching in quantum cosmology, treating the scale factor $a(t)$, as well as the zero mode of the field, as a quantum degree of freedom. In this case, we expect branching into classical backgrounds as a result of the squeezing of the modes, through which they acquire redundant information about the background.

\textit{Slow-roll Eternal Inflation.} In the regime of slow-roll eternal inflation, the post-inflationary spacetime cannot be described perturbatively. In this regime, we expect that branching of the wavefunction will occur in regions of space that have stopped inflating (allowing information to propagate spatially). It would be very interesting to explore the structure of redundant information and the emergence of classical geometry in this context.

\begin{acknowledgments}
	We thank Niayesh Afshordi, Cliff Burgess, Sean Carroll, Robin Blume-Kohout, Daniel Gottesman, Matthew Johnson, 
	Jason Pollack, and Guifre Vidal for discussion and feedback. Research at Perimeter Institute is supported by the Government of Canada through Industry Canada and by the Province of Ontario through the Ministry of Economic Development and Innovation.
\end{acknowledgments}

\appendix

\section{Records in momentum space}
\label{sec:momentum-records}

Here we discuss how the existence of records, like all entanglement structure, depends on the choice of preferred tensor decomposition into subsystems (or, equivalently, a preferred partitioning of observables \cite{zanardi2004quantum,viola2007entanglement}).  The basic issue appears already in the discrete case of a complex scalar ``field'' that is really just three harmonic oscillators in a line.  Suppose we have a state
\be
|\psi\rangle = (|0\rangle + |1\rangle)\otimes|0\rangle\otimes|0\rangle = |000\rangle + |100\rangle
\ee
where $|0\rangle$ and $|1\rangle$ are configuration states of the oscillator.  This state clearly has no entanglement, and hence no records, with respect to position space.  Performing a discrete Fourier transform 
\be
\tilde{\psi}_m = \sum_{n=0}^{N-1} \psi_n e^{2\pi i n m/N}
\ee
with N=3, this becomes the GHZ state
\be
|\tilde\psi\rangle = |0,0,0\rangle+ |1,1,1\rangle
\ee
in momentum space.  This state has redundant records in momentum space.  Furthermore, by simply appending new qubits in position space in the state $|0\rangle$, we can make the momentum space redundancy as high as we like.  Clearly, this sort of redundancy is not sufficient for declaring the associated wavefunction components $|0,0,0,\ldots\rangle$ and $|1,1,1,\ldots\rangle$ to be branches.

It's simple to extend this to the continuum.  Let $\xmin = \kmax^{-1}$ and $\xmax = \kmin^{-1}$ be the minimum and maximum length scales that can be probed in the field.  This means that two position excitations can't be distinguished if they are separated in space by less than $\xmin$ and two momentum excitations can't be distinguished if they are separated in momentum by less than $\kmin$.  For a scalar field with mass $m$, the field is effectively massless (ultra-relativistic) when $m \ll \kmin$ and non-relativistic when $m \gg \kmax$.

Consider a state of the field that is a super position of two field-configuration eigenstates given, in position space, by
\begin{align}
\vert\Psi\rangle = \vert \varphi(x) = 0 \rangle + \vert \varphi(x) = \varphi_0 e^{-x^2/2\sigma^2} \rangle
\end{align}
for some spatial width $\sigma$ and some arbitrary amplitude $\varphi_0$.  We can re-express this in momentum space as
\begin{align}
\vert\Psi\rangle = \vert \tilde\varphi(k) = 0 \rangle + \vert \tilde\varphi(x) = \tilde\varphi_0 e^{-k^2\sigma^2/2} \rangle
\end{align}
Our issue with redundancy appears clearly when $\sigma \ll \xmin = \kmax^{-1} < \kmin^{-1}$ implying that $\sigma^{-1} \gg \kmin$.  In this case, the two branches can be easily distinguished in momentum space by measuring several distinguishable modes in the range $-\sigma^{-1} \lesssim k \lesssim \sigma^{-1}$.  On the other hand, the two branches are identical outside the spatial range $-\sigma \lesssim x \lesssim \sigma$, and no more than one position mode can be accessed there.  Therefore, there is no redundancy, $R=1$, in real space, but potentially high redundancy, $R \sim (\sigma k_{min})^{-1}$, in momentum space.\footnote{Note that even though $\sigma \ll \xmin$, we always have at least one record in real space for sufficiently large $\varphi_0$.  The restrictions on the spatial resolution of measurements will simply mean we are unable to localize the excitation as tightly as it could be with better measurement equipment.}

The fact that records and their redundancy is potentially dependent on the scale at which the system can be probed has been discussed as a key remaining challenge for finding an objective definition of branch structure \cite{riedel2017classical}.

\section{Continuous branching}\label{sec:continuous-branching}

In this appendix we study simple examples of wavefunction branching for continuous degrees of freedom, which has several subtleties not present in the discrete case.

\subsection{Continuous branching of one degree of freedom}\label{sec:continuous-branching-single-degree}

\def\elll{\ell}

Consider a toy model where a preferred system $\Sys$ described by continuous degree of freedom $\nu$ has small transient interactions with a sequence of environmental piece $\Env^{(i)}$ respectively described by a continuous degree of freedom $y_i$.  Each $\Env_i$ couples to $\Sys$ only from time $t_i$ to $t_{i+1} = t_n+ \Delta t$ through the effective interaction Hamiltonian 
$\H_i = g \hat{\nu}\hat{y}_i$.  (In this section, we emphasize the distinction between the operators $\hat{\nu}$ and $\hat{y}$ and the scalar eigenvalues $\nu$ and $y_i$ by putting hats on the former.) The system and environment are initialized in states satisfying $\<\hat{\nu}\> = 0 = \<\hat y_i \>$ and $\<\hat y_i^2 \> = D$ for some constant $D$.  The pure global state after $N$ interactions is
\begin{align}
\ket{\Psi} = \int \! d\nu dy_1 \cdots dy_N\, \psi(\nu)\psi^{(y)}(y_1)\cdots\psi^{(y)}(y_1) 
\exp \[ i \nu g \Delta t  \(y_1 + \cdots + y_N\)\] 
\ket{\nu}\ket{y_1}\cdots\ket{y_N}
\end{align}
where $\psi$ and $\psi^{(y)}$ are the initial uncorrelated wavefunctions for $\Sys$ and the $\Env^{(i)}$ respectively.
Now set $g = \Delta t^{-1/2}$, fix the total time $T \equiv N\Delta t$, and take the limit $\Delta t \to 0$, $N\to\infty$.  Then the reduced dynamics of the density matrix $\rho$ for $\Sys$ become Markovian, and are described by a Lindblad equation\footnote{There is a precise sense in which this is the simplest model of pure decoherence of a continuous degree of freedom \cite{alicki2007quantum,riedel2016quantum}, and it well approximates the nonrelativistic evolution of a mass $m$ decoherenced by a bath of scattering particles on timescales for which the kinetic term $\hat{p}^2/2m$ of the object is negligible \cite{joos1985emergence,schlosshauer2008decoherence}.}
\begin{align}
\label{eq:decoh}
\partial_t \hat\rho &= D\(\hat{\nu} \hat\rho \, \hat{\nu} - \frac{1}{2}\{\hat{\nu}^2, \hat\rho \}\) \\
&= -D\[\hat{\nu},\[\hat{\nu},\hat\rho\]\]
\end{align}
After time $T$, the central object has been decohered over distances larger than the characteristic scale $\elll = 1/\sqrt{DT}$: 
\begin{align}
\rho(\nu,\nu') &= \bra{\nu}\Tr_{y}\big[\dyad{\Psi}\big]\ket{\nu'} = \psi(\nu)^* \psi(\nu') e^{-D (\nu-\nu')^2 T /2} \\
&= \rho_0(\nu,\nu')e^{-(\nu-\nu')^2 /2\elll^2}
\end{align}
In the infinite time limit, the coordinate $\nu$ becomes infinitely decohered, and the ``asymptotic branches'' are exact eigenstates of $\hat{\nu}$.

If we are to define branches by a high degree of redundancy after finite $T$, we could not find branches by partitioning the range of $\nu$ into bins of width $\elll$, because only two systems would contain a complete records of this discrete variable: $\Sys$ itself, and the \emph{entire} environment $\bigotimes_i \Env^{(i)}$.  Rather, there is a trade off between the lengthscale over which $\nu$ is ``irreversibly classical'' and the amount of redundancy quantifying the degree of irreversibility.  For any amount of redundancy $R$ we might target, branches for different values of $\nu$ are only redundantly recorded when separated by a distance $\delta \nu(R) \equiv \elll\sqrt{R} = \sqrt{R/DT}$.  To see this, we consider the overlap between the state of $N/R$ environmental parts $\Frag = \bigotimes_{i=1}^{N/R} \Env^{(i)}$ conditional on a different values $\nu$ and $\nu'$: 
\begin{align}
\Tr_{\Frag}\[\rho^{\Frag}_{\Sys:\nu} \rho^{\Frag}_{\Sys:\nu'}\] 
&\propto e^{-(\nu-\nu')^2/R\elll^2}
\end{align}
where $\rho^{\Frag}_{\Sys:\nu} = \bra{\nu}\Tr_{\,\overline\Frag}\[ \dyad{\Psi}\]\ket{\nu}$.
We will find it more convenient to describe this (highly symmetric) branch structure by writing the redundancy as a function of desired precision: 
\begin{align}
R(\delta \nu) \approx \delta \nu^2 DT.  
\end{align}
This equation is a compact way to summarize the following statement: ``At time $T$, the environment can be broken into $R(\delta \nu)$ disjoint fragments $\Frag^{(r)} = \bigotimes_{i \in I_r} \Env_i$ (with $r=1,\dots,R(\delta \nu)$ and $\{I_r\}$ a partition of $\{1,\ldots,N\}$) such that 
\ba
\langle  (\hat{\nu}-\nu_0)^2 \rangle_{\Frag^{(r)}:\nu_0} \equiv \frac{\langle \Psi\vert\Pi^{\Frag^{(r)}}_{\nu_0}  (\hat{\nu}-\nu_0)^2 \Pi^{\Frag^{(r)}}_{\nu_0}\vert\Psi \rangle}{\langle \Psi\vert\Pi^{\Frag^{(r)}}_{\nu_0}  \vert\Psi \rangle} \le \delta \nu^2
\ea
for all values of $\nu_0$, where $\Pi^{\Frag^{(r)}}_{\nu_0}$ is a projector on $\Frag^{(r)}$.''  Note that disjointness is defined \emph{with respect to} the tensor-product decomposition $\bigotimes_i \Env^{(i)}$, which in this toy model is put in by hand, but which ultimately we will ground in spatial locality.

One can speak about a ``number'' of branches with redundancy $R$ given by $L_\psi/\delta \nu(R)$, where $L_\psi$ is the support of the initial wavefunction $\psi(\nu)$.  But this is a fuzzy idea, since the branches bleed smoothly into each other; for $\delta \nu \ll (d\abs{\psi(\nu)}^2/d\nu)^{-1/2}$, the joint wavefunction looks locally spatially homogeneous, and the value of $\nu$ at which one branch ends and another begins is arbitrary.   Also, there may be many branches that are recorded with high redundancy but have little probability mass.  The entropy of the branches 
\begin{align}
S_\psi^{\delta \nu} = \int\! d\nu\abs{\psi(\nu)}^2 \ln \frac{\delta \nu}{\abs{\psi(\nu)}^2},
\end{align}
is a more elegant quantity than the absolute number of them.  It is translationally invariant and naturally accounts for the relative branch weights.

This toy model describes the branching of a continuous variable to arbitrary precision, which in principle may continue indefinitely so long as there is a supply of new environmental pieces.  However, it accomplishes this by neglecting any unitary dynamics that would induce transitions between different values of $\nu$.  If $\nu$ is a nonrelativistic test mass, the full Hamiltonian will also include a kinetic terms $\hat{p}^2/2m$.  Even if the initial wavepacket $\psi(\nu)$ is very wide and smooth compared to the mass, so that the dispersion of the wavefunction in space is negligible in the absence of decoherence, the kinetic term nevertheless becomes important on the timescale $\tilde{t} \sim \sqrt{m /D}$.  Intuitively this is because once $\Sys$ has been decohered over the scale $\tilde{\elll} = 1/\sqrt{D\tilde{t}} = (Dm)^{-1/4}$, the resulting momentum uncertainty $\tilde{\Delta p} = 1/\tilde{\elll} = (Dm)^{1/4}$ is producing strong wavefunction dispersion over that same distance scale: $\tilde{t}\tilde{\Delta p}/m \sim \tilde{\elll}$ on the same timescale (see, e.g., Ref.~\cite{diosi2000robustness}).

It is well known from the study of open quantum systems that this induces a random walk, with the Markovian quantum Brownian motion\footnote{Equations \eqref{eq:decoh-kinetic} and \eqref{eq:decoh-kinetic-potential}, as well as a large class of generalizations, have exact solutions in terms of the Wigner function \cite{brodier2004symplectic,robert2012time,riedel2016quantum}. Note, though, that most concrete models of quantum Brownian motion, the most well-known being the Caldiera-Leggett model \cite{caldeira1983path}, are only approximately Markovian.  Completely Markovian dynamics can only be obtained by integrating out a fast (microscopic) timescale, which in our case is achieved by taking the limit $\Delta t \to 0, N \to \infty$.  For a discussion of redundancy in the Caldiera-Leggett model, see Ref.~\cite{blume-kohout2008quantum}.} given by
\begin{align}
\label{eq:decoh-kinetic}
\partial_t \hat\rho &= 
-i \frac{1}{2m}[\hat{p}^2,\hat\rho] 
-D\[\hat{\nu},\[\hat{\nu},\hat\rho\]\]
\end{align}
Now the mass is moving before it can be decohered with arbitrary precision, so branches are formed corresponding to \emph{trajectories} $\nu(t)$ through time \cite{feynman1963theory,diosi1995decoherent,halliwell2007two}. One sensible way\footnote{A clean definition would best be done within the consistent histories formalism, where $R(\delta \nu, \delta t)$ would be a sort of generalization of the decoherence functional.  See Ref.~\cite{riedel2016objective} for intial work in this direction.} to describe the trade off between redundancy and precision is with the redundancy function $R(\delta \nu, \delta t)$, where a branch is defined by a trajectory with respective temporal and spatial precisions $\delta \nu$ and $\delta t$.

In fact, these dynamics can be represented stroboscopically as a iterated sequence of identical complete (entanglement-breaking) POVM measurements \cite{riedel2016quantum}, each separated exactly by a time $\tilde{t} = (12)^{1/4} \sqrt{m /D}$.  If we choose this for our time resolution, $\delta t = \tilde{t}$, then the object has decohered over distances larger than $\tilde{\elll}$. In order to get redundancy $R$ we must choose $\delta \nu = \tilde{\elll}\sqrt{R}$, i.e., $R(\delta \nu,\tilde{t}) \approx (\delta \nu/\tilde{\elll})^2$.  On the other hand, if we allow for a more coarse-grained time resolution, then redundancy is proportional to time (by construction of the environment) so our general expression is 
\begin{align}
R(\delta \nu,\delta t) \approx \(\frac{\delta \nu}{\tilde{\elll}}\)^2 \frac{\delta t}{\tilde{t}}.  
\end{align}

\subsection{Continuous branching of a field}

\def\xxx{\x}
\def\kkk{\k}
\def\chicord{\nu}
\def\chiconj{\mu}

Let us generalize our toy model to a field $\chicord(\xxx)$ (conjugate momentum $\chiconj(\xxx)$) in three dimensions, which will bring us closer to the continuous branch structure we expect for the scalar perturbation.  We likewise introduce a new dimension on the environment to include the spatial dimensions:
$y_{i}(\xxx)$, with the interaction Hamiltonian between time $t_i$ and $t_{i+1} = t_i + \Delta t$ set to $\H_i = g \int\! d\xxx \, \chicord(\xxx)y_{i}(\xxx)$.  Then assuming the field is an otherwise-free massless scalar, the complete Hamiltonian is 
\begin{align}
\H &= \frac{1}{2} \int\! d\xxx \[ \jstrut\chiconj^{2}(\xxx) + (\nabla \chicord)^2(\xxx) + 2g \Theta_i(t)\chicord(\xxx)y_{i}(\xxx) \]\\
&= \frac{1}{2} \int_\kkk \[ \jstrut \chiconj^2_\kkk + k^2 \chicord^2_\kkk + 2g \Theta_i(t)\chicord_\kkk\, y_{i,-\kkk} \]
\end{align}
where $\Theta_i(t) \equiv \Theta(t-t_i)\Theta(t_{i+1}-t)$ and $\Theta(\cdot)$ is the Heaviside step function.
So, for each value of $\kkk$, we have single continuous degree of freedom obeying similar Markovian dynamics  except for the addition of a quadratic potential term:
\begin{align}
\label{eq:decoh-kinetic-potential}
\partial_t \hat\rho_{\k} &= 
-i \[\frac{1}{2}{\hat{\chiconj}}_\k^2 + \frac{k^2}{2}{\hat{\chicord}}_\k^2\, , \, \hat\rho_\k\] 
-D\[{\hat{\chicord}}_\k,\[{\hat{\chicord}}_\k,\hat\rho_\k\]\]
\end{align} 
This reduces to \eqref{eq:decoh-kinetic} for small $k$ (weak potential) with $m=1$, and to \eqref{eq:decoh} for short times.  Thus, in these limits, we may consider the branches to be field configurations, each composed of a choice of configuration for every $\kkk$-mode.  Just like it is possible for a fraction $1/R$ of the environment to record the position of a single continuous degree of freedom $\nu$ to some precision $\delta \nu$, implying a redundancy of $R$, this larger environment can record a field configuration to some precision $\delta \chicord(\kkk) \sim \sqrt{R}\, \ell_\chicord(\kkk)$ after the field has been decohered on scales larger than $\ell_\chicord(\kkk)$. 

The branches of the primordial scalar perturbations $\vph$ will have at least two key differences from this toy model.  First, the tensor structure (indexed by $i$) with respect to which we define the redundancy in the toy model is put in by hand, but the objectiveness of the branch structure we seek in the main text comes from the natural but approximate tensor structure associated with spatial locality; that is, distinct records are to be found in disjoint spatial regions.  Second, instead of having the artificial system-environment distinction as for the toy model, the primordial scalar perturbations will act as their own environment; this is similar to the GHZ state, where all qubits are on equal footing, and where the coarse-grained variable (the average qubit value, analogous to the long-wavelength modes) plays the role of the central system.

\section{Alternative mode decompositions}

At an informal level, the Hilbert space of the field theory can be broken up into the tensor product of a continuum of independent momentum modes
\begin{align}
\Glob = \bigotimes_{\k} \Glob_\k
\end{align}
Each mode is spanned by the action of the field and conjugate-momentum operators ($\vph_\k$ and $\pi_\k$), or alternatively by the creation/annihilation operators ($a_\k$ and $a_\k^\dagger$).  Independence is expressed as the commutation of these operators for distinct values of $\k, \k'$.  Momentum modes are common and useful because of their relation to the translational symmetry of the system, 
but other orthonormal bases for the vector space of classical field configurations are permitted so long as they are sufficiently smooth. 

For instance, the eigenstate of the one-dimensional harmonic oscillator are the Hermite polynomials with Gaussian envelope,
\begin{align}
H_n(x) = (-1)^n \frac{d^n}{dx^n} e^{x^2/2}, \qquad n=0,1,2,\dots.
\end{align}
These can be combined to produce the corresponding basis functions in three dimensions, $H_{\mathbf{n}}(\x) = H_{n_1}(x)H_{n_2}(y)H_{n_3}(z)$, which are indexed by the triplet $\mathbf{n} = (n_1,n_2,n_3)$.
We can confirm using the inner product $(f,g) \equiv \int\! d\x \, f(\x)^*g(\x) = \int_\k \hat{f}(\k)^* \hat{g}(\k)$, where hats denote the Fourier transform, that $\{H_{\mathbf{n}}\}$ is an orthonormal basis for normalizable real functions in three dimensions.\footnote{We're working with real fields, so $f(\x) = f^*(\x)$ and $\hat{f}(\k) = \hat{f}^*(-\k)$.}  

Using any such orthonormal set $\{f_n\}$, we can define an alternative tensor decomposition $\Glob = \bigotimes_{n} \Glob_n$ where $\Glob_n$ is built up from the operators
\begin{align}
\label{eq:framemode}
\vph_n = \int_\k \hat{f}_n(\k) \vph_\k, \qquad \qquad \pi_n = \int_\k \hat{f}_n(\k) \pi_\k
\end{align}
This automatically induces the canonical commutation relations $[\vph_n,\pi_m] = i\delta_{nm}$ by virtue of the orthnormality of the $\{f_n\}$.  Note that since the index $n$ is discrete, there is no delta function like $\delta(\k+\p)$.

\section{Inflationary fluctuations and gravitational interactions}
\label{sec:inflation_theory}

\subsection{Review: metric fluctuations during inflation}
\label{app:review_metric}

The scalar curvature perturbation $\zeta$ in a quasi de Sitter background is defined by using the ADM formalism to decompose the spacetime into spatial hypersurfaces, and choosing surfaces of constant energy density.
(While the matter density is spatially uniform in this gauge, $\zeta$ describes the same dynamical degree of freedom as, for example, scalar field fluctuations in the spatially flat gauge.\footnote{See, for example: Andrew R. Liddle and David H. Lyth, \textit{The Primordial Density Perturbation} (2009), ch. 5.})
The scalar and tensor modes can then be described in the spatial metric components as
\be
g_{ij}(\x,t) = a^2(t) e^{2\zeta(\x,t)}(e^\gamma)_{ij} \approx a^2(t) \big[(1+2\zeta(\x,t))\d_{ij}+\gamma_{ij}(\x,t)\big]. \label{g_ij},
\ee
where the tensor modes satisfy $\nabla_i\gamma_{ij}=\gamma_{ii}=0$.
The scalar curvature perturbation effectively describes fluctuations in the amount of expansion at each point: $a_{\rm eff}(\x,t)\approx a(t)(1+\zeta(\x,t))$.  We will take the initial state for the fluctuations -- at very early times when the modes are much shorter than the Hubble scale -- to be the ground state, i.e., the usual Bunch-Davies vacuum \cite{Bunch:1978yq}.  

The action for $\zeta$ can be obtained by expanding the Einstein-Hilbert action using the ADM formalism to quadratic order in perturbations \cite{Maldacena:2002vr} (see also \cite{Brandenberger:1993zc,Garriga:1999vw}), and is\footnote{To have a nontrivial speed of sound, $c_s \neq 1$, we send $\dot\zeta^2 \to \dot\zeta^2/c_s^2$ in the action.}
\begin{align}
\label{eq:zeta-action}
\SS = \frac{1}{2}\int\! dt\, d^3\x \,\(2\eps \Mp^2\)a^3 \(\dot\zeta^2 - \frac{1}{a^2}(\nab\zeta)^2\).
\end{align}
where $\Mp\equiv1/\sqrt{8\pi G}$ is the reduced Planck mass, and $\eps\equiv-(dH/dt)/H^2$ is assumed to be constant up to higher order corrections in slow roll parameters. 
The resulting dynamics describe a massless scalar field in de Sitter space as long as $\eps$ is small and slowly varying, and are therefore encompassed in our model of a massless mode $\vph$.
The late-time power spectrum determined by Eq. \eqref{eq:zeta-action} is approximately scale-invariant:
\be
\<\zeta_\k\zeta_{\k'}\>|_{t\rarr\infty} = (2\pi)^3\d^3(\k+\k')\frac{2\pi^2}{k^3}\Delta_\zeta^2(k),
\ee
where
\be
\Delta_\zeta^2\approx \frac{1}{(2\pi)^2}\frac{H^2}{2\eps\Mp^2}.
\ee

Similarly, the action for tensor modes is
\be
\SS = \frac{\Mp^2}{8}\int\! dt\, d^3\x a^3 \(\dot{\gamma}_{ij}\dot{\gamma}_{ij} - \frac{1}{a^2}\nab_l\gamma_{ij}\nab_l\gamma_{ij}\),
\ee
and the resultant primordial gravitational wave spectrum is
\be
\<\gamma^s_\k\gamma^{s'}_{\k'}\>'|_{t\rarr\infty} = (2\pi)^3\d^3(\k+\k')\d_{ss'}\frac{2\pi^2}{k^3}\Delta_\gamma^2(k),
\ee
for tensor polarizations $s=+,\times$, where
\be
\Delta_\gamma^2 \approx \frac{1}{\pi^2}\frac{H^2}{\Mp^2}.
\ee
Both scalar and tensor modes can be described as a rescaling of our (dimensionless) massless mode $\vph$. As shown in \S \ref{sec:long_modes_examples} and Appendix \ref{sec:cubic_zeta}, their gravitational interactions are equivalent to those in our model, Eq. \eqref{cubic_model}, for specific choices of $(g,\tg)$.

\subsection{Cubic interactions of the scalar curvature perturbation $\zeta$}
\label{sec:cubic_zeta}

In this Appendix, we describe the derivation of Eq.~\eqref{L2_shift_zeta}, which captures the effect of a long-wavelength fluctuation of the scalar metric perturbation $\zeta$.

The self-interactions of the scalar curvature $\zeta$ can be obtained by expanding the Einstein-Hilbert action with a generic single-field matter Lagrangian as in \cite{Maldacena:2002vr,Chen:2006nt}, or more generally, by writing down an effective field theory for metric fluctuations in a nearly de Sitter background with slightly broken time translation invariance \cite{Cheung:2007st}. In either approach, setting $\Mp^2\equiv1$, the cubic action at leading order in $\eps$ is proportional to the equation of motion, $\d\L_2/\d\zeta$:
\be\label{L3_eom}
\L_3(\x) = \eps a^3 \[ \frac{2}{H} \zeta\dot{\zeta} + \frac{1}{2a^2 H^2}\(-(\nab\zeta)^2+\nab^{-2}\nab_i\nab_j(\nab_i\zeta\nab_j\zeta) \) \] \(\ddot{\zeta} + 3H\dot{\zeta} - a^{-2}\nab^2\zeta \) + \O(\eps^2).
\ee
These terms are not slow-roll suppressed relative to the quadratic action, and only affect correlation functions involving the conjugate momentum. They can be removed by a field redefinition which, at leading order in $\eps$, is trivial in the late-time limit, so they do not affect the late-time correlation functions of the field such as the three-point function $\<\zeta\zeta\zeta\>|_{t\rarr\infty}$.  Late-time correlation functions of the field arise from the $\O(\eps^2)$ terms which we have omitted from Eq.~\eqref{L3_eom}.
Note that the $\O(\eps)$ terms in Eq.~\eqref{L3_eom} arise, starting from the action in terms of inflaton fluctuations $\delta\phi$ in the spatially flat gauge \cite{Maldacena:2002vr}, by changing variables to  $\zeta$:
\be
-\frac{1}{\Mp\sqrt{2\eps}} \delta\phi = \zeta - \frac{1}{H} \zeta\dot{\zeta} - \frac{1}{4a^2H^2}\(-(\nab\zeta)^2+\nab^{-2}\nab_i\nab_j(\nab_i\zeta\nab_j\zeta) \) + \O(\eps\zeta^2).
\ee
Only the first term in brackets has a factor of $\zeta$ without derivatives, and can therefore couple short modes to a very long-wavelength mode. The remaining derivative terms in brackets only couple the gradient of the field. We will be interested in correlations between a very long mode with wavenumber $q$ and a shorter mode with wavenumber $k$, for which derivative terms will give contributions suppressed as $q/k$.  We will drop any such derivative-suppressed interactions.

After integrating by parts\footnote{Integration by parts generates boundary contributions to the action. In fact, there are additional boundary terms which we did not write in Eq.~\eqref{L3_eom}, which are cancelled by the integration by parts, leaving Eq.~\eqref{L3_2} without boundary contributions at $\O(\eps)$ \cite{Burrage:2011hd}.} \cite{Burrage:2011hd,Agarwal:2013rva} we can remove the $\ddot{\zeta}$ terms and write Eq.~\eqref{L3_eom} as
\be\label{L3}
\L_3(\x) = \eps a^3 \(\frac{-1}{H}\dot{\zeta}^3 + 3\zeta\dot{\zeta}^2 \) -\eps a \frac{2}{H}\zeta\dot{\zeta}\nab^2\zeta + \O(\eps^2) + ...,
\ee
where the dots include the derivative terms we have dropped.
Finally, integrating the $a\zeta\dot{\zeta}\nab^2\zeta$ term by parts first with respect to time, and then spatially, we have:
\be\label{L3_2}
\L_3(\x) = 3\eps a^3\zeta\dot{\zeta}^2 - \eps a \zeta(\nab\zeta)^2 - \eps a^3 \frac{1}{H}\dot{\zeta}^3 + \eps a\frac{1}{H}\dot{\zeta}(\nab\zeta)^2 + \O(\eps^2) + ... %
\ee
The $\dot{\zeta}^3$ and $\dot{\zeta}(\nab\zeta)^2$ terms in $\L_3$ have derivatives on all three factors of $\zeta$, so they will be suppressed as described above, and we drop them as we did the additional terms in Eq.~\eqref{L3_eom}.
We are then left with\footnote{These are the leading ``bulk'' interactions in the formulation of the action given in Eq.~(3.1) of \cite{Burrage:2011hd}.}
\be\label{L3_no_derivs}
\L_3(\x) = 3\eps a^3\zeta_L\dot{\zeta}^2 - \eps a \zeta_L(\nab\zeta)^2 + ...,
\ee
as the cubic interactions relevant for decoherence, where the dots indicate all terms that can be neglected, and the $L$ subscript indicates that we are just interested in very long-wavelength modes (with the remaining configurations absorbed into the dots).  This can be combined with the quadratic action in the simple form \cite{Creminelli:2011rh}
\be\label{L2_shift}
\L_2(\x)+\Delta \L_2(\x)= \eps a^3 \(1+3\zeta_L\) \dot{\zeta}^2 - \eps a(1+\zeta_L)(\nab\zeta)^2,
\ee
stated in the main text in Eq.~\eqref{L2_shift_zeta}.

\bibliographystyle{JHEP}
\bibliography{inflation_branches_bib}

\end{document}